\documentclass{JHEP3}
\usepackage{axodraw}
\usepackage{amsmath}
\usepackage[utf8]{inputenc}
\usepackage{longtable}

\usepackage{cite}
\usepackage{epsfig}
\usepackage{xspace}
\usepackage{graphics}
\usepackage{relsize}
\usepackage{amsfonts}
\usepackage{ifthen}
\usepackage[normalem]{ulem}	

\def\hide#1{}

\newcommand{\asme}{\alpha_{\textnormal{s\tiny{ME}}}}
\newcommand{\asps}{\alpha_{\textnormal{s\tiny{PS}}}}

\newcommand{\tms}{t_{\textnormal{\tiny{MS}}}}

\newcommand{\mz}{\textnormal{M}_{\textnormal{\tiny{Z}}}}
\newcommand{\mw}{\textnormal{M}_{\textnormal{\tiny{W}}}}
\newcommand{\mh}{\textnormal{M}_{\textnormal{\tiny{H}}}}

\newcommand{\Kf}{\ensuremath{K}}

\newcommand{\muf}{\mu_F}
\newcommand{\mur}{\mu_R}
\newcommand{\powhegbox}{P{\smaller OWHEG}-B{\smaller OX}\xspace}

\newcommand{\drad}{d\Phi_{\textnormal{rad}}}

\newcommand{\wckkwl}[1]{w_{#1}}
\newcommand{\wumeps}[1]{w_{#1}^\prime}
\newcommand{\Oas}[1]{\mathcal{O}(\as^{#1})}
\newcommand{\Oasof}[2]{\mathcal{O}\left(\as^{#1}\left(#2\right)\right)}
\newcommand{\ordms}{\ensuremath{\rho_{\textnormal{\tiny MS}}}}

\newcommand{\mesqof}[2]{\left|\mathcal{M}_{#1}\left(#2\right)\right|^2}

\newcommand{\Bornev}[1]{\textnormal{B}_{#1}}
\newcommand{\Bbarev}[1]{\overline{\textnormal{B}}_{#1}}
\newcommand{\Btilev}[1]{\widetilde{\textnormal{B}}_{#1}}

\newcommand{\Virtev}[1]{\textnormal{V}_{#1}}

\newcommand{\Dipev}[1]{\textnormal{D}_{#1}}
\newcommand{\Insev}[1]{\textnormal{I}_{#1}}

\newcommand{\Iev}[2]{\int_s\widehat{\textnormal{B}}_{#1\rightarrow #2}}
\newcommand{\Tev}[1]{\widehat{\textnormal{B}}_{#1}}

\newcommand{\termX}[2]{\left[\vphantom{\sum}#1\right]_{#2}}

\newcommand{\tree}[1]{\mathbb{T}_{#1}}
\newcommand{\virt}[1]{\mathbb{V}_{#1}}
\newcommand{\subt}[1]{\mathbb{S}_{#1}}

\newcommand{\untree}[1]{\mathbb{B}_{#1}}
\newcommand{\unvirt}[1]{\mathbb{L}_{#1}}
\newcommand{\unsubt}[1]{\mathbb{I}_{#1}}

\newcommand{\hugeint}{\mathop{\mathlarger{\mathlarger{\mathlarger{\int}}}}}
\newcommand{\hugesum}{\mathop{\mathlarger{\mathlarger{\mathlarger{\sum}}}}}
\newcommand{\hugercurly}{\mathop{\mathlarger{\mathlarger{\Bigg\}}}}}
\newcommand{\hugelcurly}{\mathop{\mathlarger{\mathlarger{\Bigg\{}}}}
\newcommand{\hugersquare}{\mathop{\mathlarger{\mathlarger{\Bigg]}}}}
\newcommand{\hugelsquare}{\mathop{\mathlarger{\mathlarger{\Bigg[}}}}

\newcommand{\powheg}{P{\smaller OWHEG}\xspace}
\newcommand{\Bbar}{\ensuremath{\bar{B}}}
\newcommand{\sherpa}{S{\smaller HERPA}\xspace}

\newcommand{\pytppp}{P{\smaller YTHIA}8\xspace}

\newcommand{\as}{\ensuremath{\alpha_{\mathrm{s}}}}

\newcommand{\ECM}{\ensuremath{E_{\mathrm{CM}}}}

\newcommand{\particle}[1]{\ensuremath{\mathrm{#1}}}
\newcommand{\antiparticle}[1]{\ensuremath{\bar{\mathrm{#1}}}}
\newcommand{\el}{\particle{e}}
\newcommand{\eplus}{\ensuremath{\el^+}}
\newcommand{\eminus}{\ensuremath{\el^-}}

\newcommand{\g}{\particle{g}}
\newcommand{\p}{\particle{p}}
\newcommand{\q}{\particle{q}}

\newcommand{\W}{\particle{W}}
\newcommand{\Higgs}{\particle{H}}

\newcommand{\Wm}{\ensuremath{\W^-}}

\newcommand{\qu}{\particle{u}}

\newcommand{\qc}{\particle{c}}

\newcommand{\qbar}{\antiparticle{q}}

\newcommand{\ubar}{\antiparticle{u}}
\newcommand{\dbar}{\antiparticle{d}}

\newcommand{\cbar}{\antiparticle{c}}

\newcommand{\tee}{\ensuremath{\el^+\el^-}\xspace}

\newcommand{\ord}{\ensuremath{\rho}}

\newcommand{\state}[1]{\ensuremath{S_{+#1}}}
\newcommand{\splitP}{\ensuremath{P}}

\newcommand{\done}[1]{}

\def\mrm#1{\mathrm{#1}}
\newcommand{\cf}{c.f.\xspace}
\newcommand{\ie}{i.e.\xspace}
\newcommand{\eg}{e.g.\xspace}

\def\sup#1{\ensuremath{^{\mrm{#1}}}}

\def\Pnoem{\ensuremath{\Pi}}
\def\noem#1{\ensuremath{\Pnoem_{\state{#1}}}}

\def\f2d3{\ensuremath{F_2^{\mrm{D}3}}}

\providecommand{\eqref}[1]{eq.~(\ref{#1})\xspace}
\renewcommand{\eqref}[1]{eq.~(\ref{#1})\xspace}
\newcommand{\eqsref}[1]{eqs.~(\ref{#1})\xspace}

\newcounter{aenumct}

\newcounter{ienumct}

{\end{list}}

\newcounter{enumct}
\renewenvironment{enumerate}{\begin{list}{\arabic{enumct}.}%
{\usecounter{enumct}\setlength{\topsep}{1mm}%
\setlength{\partopsep}{1mm}\setlength{\itemsep}{0mm}%
\setlength{\parsep}{1mm}}}{\end{list}}

\def\showcommentsflag{0}
\newcommand{\showcomments}{\def\showcommentsflag{1}}

\newcounter{commentcounter}%
\newcommand{\comment}[1]{\ifnum\showcommentsflag > 0%
\addtocounter{commentcounter}{1}%
\Red{\ensuremath{\ddagger^{\arabic{commentcounter}}}}%
\marginpar{\raggedright\tiny\it\Red{\ensuremath{\ddagger^{\arabic{commentcounter}}} #1}}
\fi%
}
\newcommand{\commentdel}[2]{\ifnum\showcommentsflag > 0%
\Red{\sout{#1}}\comment{#2}%
\fi
}
\newcommand{\commentadd}[2]{\ifnum\showcommentsflag > 0%
\comment{#2}\Red{#1}%
\else
#1
\fi
}
\newcommand{\commentchange}[3]{\ifnum\showcommentsflag > 0%
\Red{\sout{#2}}\comment{#3}\Red{#1}%
\else
#1
\fi
}
\newcommand{\nocomment}[1]{\ifnum\showcommentsflag > 0%
{\tiny\it\Red{\{#1}\}}
\fi%
}
\newcommand{\nocommentdel}[1]{\ifnum\showcommentsflag > 0%
\Red{\sout{#1}}%
\fi
}
\newcommand{\nocommentadd}[1]{\ifnum\showcommentsflag > 0%
\Red{#1}%
\else
#1
\fi
}
\newcommand{\nocommentchange}[2]{\ifnum\showcommentsflag > 0%
\Red{\sout{#2}}\Red{#1}%
\else
#1
\fi
}

\showcomments

\preprint{LU-TP 12-43\\MCnet-12-17\\arXiv:1211.7288 [hep-ph] (December 3, 2012)}

\skip\footins = 1\bigskipamount plus 2pt minus 4pt                              

\title{Merging Multi-leg NLO Matrix Elements with Parton Showers\footnote{Work supported in parts by
    the Swedish research council (contracts 621-2009-4076 and
    621-2010-3326).}}

\author{Leif Lönnblad and Stefan Prestel\\
  Dept.~of Astronomy and Theoretical Physics, Lund University, Sweden\\
  E-mail: \email{Leif.Lonnblad@thep.lu.se}
    and \email{Stefan.Prestel@thep.lu.se}}
  
  \abstract{We discuss extensions the CKKW-L and UMEPS tree-level matrix 
    element and parton shower merging approaches to next-to-leading order
    accuracy.

    The generalisation of CKKW-L is based on the NL$^3$ scheme
    previously developed for \tee-annihilation, which is extended to
    also handle hadronic collisions by a careful treatment of parton
    densities. NL$^3$ is further augmented to allow for more readily
    accessible NLO input.

    To allow for a more careful handling of merging scale dependencies
    we introduce an extension of the UMEPS method. This approach,
    dubbed UNLOPS, does not inherit problematic features of CKKW-L,
    and thus allows for a theoretically more appealing definition of
    NLO merging.

    We have implemented both schemes in \pytppp, and present
    results for the merging of $\W$- and Higgs-production events, where
    the zero- and one-jet contribution are corrected to next-to-leading
    order simultaneously, and higher jet multiplicities are described by
    tree-level matrix elements. The implementation of the procedure
    is completely general and can be used for higher jet
    multiplicities and other processes, subject to the availability of
    programs able to correctly generate the corresponding partonic
    states to leading and next-to-leading order accuracy.

  }

\begin{document}

\sloppy

\section{Introduction}

Particle physics phenomenology has been awed by the accuracy of LHC 
analyses. The precision at which, for example, the Higgs candidate mass has
been measured could only be achieved through a very detailed understanding of the
structure of collision events in an environment that can safely be called 
messy. Remnants of single collisions alone give rise to large numbers of 
hadronic jets, leptons and photons -- even before pile-up events are taken 
into account. In order to separate and determine the characteristics of rare 
signal events, highly accurate methods have evolved to describe background
processes.

Precise theoretical calculations for scatterings with multiple jets in
particular are necessary for reliable background estimates.  For
generic processes this, until recently, meant that multi-jet
tree-level matrix element and parton shower merging (MEPS) techniques
were the method of choice, with CKKW-inspired prescriptions
\cite{Catani:2001cc,Lonnblad:2001iq,Lavesson:2005xu,Hoeche:2009rj,Lonnblad:2011xx}
being widely used.  These methods impose a weight containing the
parton shower resummation on tree-level-weighted $n$-parton phase
space points. Phase space points with soft and/or collinear partons in
the matrix element (ME) event generation are removed by a
jet-resolution cut, leaving only $n$-parton phase space points that
contain exactly $n$ resolved jets. The same cut is also used to
restrict the parton shower (PS) to only produce unresolved partons as
long as tree-level calculations for the resulting state are
available. The combination of reweighting and phase space slicing (by
the jet separation cut) allows to add tree-level samples with
different jet multiplicity without introducing any phase space
overlap.

This method has an evident drawback, even if we would be content with
a tree-level prescription of multiple jets: Simply adding
$n$-resolved-jet states cannot guarantee a stable inclusive
(lowest-multiplicity) cross section. In particular, the inclusive
cross will depend on the jet separation parameter, $\tms$, so that
choosing $\tms$ unwisely may result in significant cross section
uncertainties. This problem is remedied by the UMEPS method
\cite{Lonnblad:2012ng}, which infers the notion of parton shower
unitarity to derive an add-subtract scheme to safeguard a fixed
inclusive cross section.

However, MEPS methods only improve the description of the shape of
multi-jet observables, and cannot describe overall normalisations or
decrease theoretical uncertainties due to scale variations.  This
requires predictions of next-to-leading order (NLO) QCD.  Through
formidable efforts of the fixed-order community, we have recently
witnessed an NLO revolution\footnote{We cannot do justice to all results
of these intricate calculations, so that we limit ourselves to the
more conceptual papers \cite{Bern:1994zx,Britto:2004nc,Denner:2005nn,Ossola:2006us,Ellis:2007br,Ossola:2008xq,Ellis:2008ir,Berger:2008sj,Becker:2010ng,Cascioli:2011va}, which
made this progress possible.}, meaning that today, the automation of NLO
calculations is practically a solved problem.  Such calculations become directly
comparable to LHC data by incorporating the NLO results into
General-Purpose event generators.  NLO matrix element and parton
shower matching methods like \powheg \cite{Nason:2004rx,Frixione:2007vw,Alioli:2010xd,Platzer:2011bc} and MC@NLO \cite{Frixione:2006gn,Hoeche:2011fd,Hoeche:2012ft,Hirschi:2011pa} have -- in parallel with
the NLO revolution -- become robust tools that allow a coalescence of
resummation, low-scale effects and hadronisation with NLO
calculations.

The latest step in these developments are multi-jet NLO merging prescriptions \cite{Lavesson:2008ah,Gehrmann:2012yg,Hoeche:2012yf,Frederix:2012ps}.
These address the problem of simultaneously describing observables for any 
number of (additional) jets with NLO accuracy, and are thus direct successors
of the tree-level schemes. The problem in NLO multi-jet merging is twofold. 
It is firstly mandatory -- as in tree-level merging -- to ensure that 
configurations with $n$ hadronic jets are described by the $n$-jet ME. 
If we have a better calculation at hand, we do 
not want to predict rates for $n$ hadronic jets by adding a parton shower 
emission to the $(n-1)$-jet NLO calculation. 
This problem has already been solved in tree-level merging 
methods.
Secondly, each $n$-jet observable has to be described with NLO
accuracy (if an NLO calculation of the $n-$jet cross section was used as 
input), while all higher orders in $\as$ should be given by the parton shower
resummation (possibly with improvements). This problem can be overcome by 
\begin{enumerate}
\item[(1)] Using tree-level matrix elements only as seeds for higher-order 
      corrections, \ie\ including the full resummation in tree-level events, 
      and safeguarding 
that the weighting of $n$-jet tree-level configurations 
      does not introduce $\Oas{n+1}$-terms.
\item[(2)] Defining an NLO cross section for $n$-parton states that does not 
      include $n+1$ resolved jets, and making sure that no (uncontrolled)
      $\Oas{n+2}$-corrections are introduced by the NLO calculation. 
\item[(3)] Adding the corrected tree-level and the NLO events.
\end{enumerate}
This means that we have to decide how to define NLO cross section for
$n$-parton states that do not include $n+1$ resolved jets. We call an
NLO calculation ``exclusive'' if it contains weights for $n$-jet phase
space points that include Born, virtual and unresolved real
corrections, where the resolution criterion is defined by exactly the
same function as the merging scale. If \emph{all} real emission corrections
are projected onto $n$-jet kinematics, the calculation will be called
``inclusive" instead. It is feasible to make an inclusive calculation
exclusive by introducing explicit counter-events that are distributed
according to the resolved-emission contribution, and subtracting these
events from events generated according to the inclusive NLO cross
section.

\noindent
The points (1) - (3) will schematically lead to an algorithm of the form
\begin{itemize}
\item Reweight $n$-resolved-jet tree-level events with weight used in the 
      tree-level merging scheme.
\item Subtract the $\Oas{n}$ and $\Oas{n+1}$ terms introduced by this 
      prescription from the tree-level events.
\item Add NLO-weighted $n$-resolved-jet events.
\end{itemize}
Many variations of this basic form are possible. The first conceptual
paper on NLO merging \cite{Nagy:2005aa} for example advocated
subtracting $\Oas{n+1}$-terms from the NLO cross section. This is also
the case in the MINLO NLO matching scheme \cite{Hamilton:2012np}, and
the NLO merging scheme introduced for aMC@NLO
\cite{Frederix:2012ps}. MEPS@NLO \cite{Gehrmann:2012yg,Hoeche:2012yf}
exerts full control over the NLO calculation to avoid
point (1). 
We hope that the near future will bring detailed comparisons
of all these schemes.

Moving beyond tree-level merging prescriptions has long been regarded
the next crucial step in background simulations for the LHC.  The aim
of this article is to present a comprehensive guide to NLO merging
schemes in \pytppp \cite{Sjostrand:2007gs}. We will present two
different NLO merging schemes, choosing to generalise both the CKKW-L
and UMEPS methods.  We will refer to the NLO version of CKKW-L as
NL$^3$, since this is an extension of the NLO merging scheme for
$\eplus\eminus$-collision presented in \cite{Lavesson:2008ah} to
hadronic collisions, now also allowing for \powheg input.  The virtue of
this method is its relative simplicity in combining NLO accuracy with
PS resummation. However, it inherits violations of the inclusive cross
section from CKKW-L. Cross section changes are the result of adding
higher-multiplicity matrix elements -- containing logarithmic terms that
are beyond the accuracy of the parton shower, and can thus not be
properly cancelled. At tree-level, this issue was resolved by the
UMEPS method. Thus, we believe that a NLO generalisation of UMEPS,
which also cancels new logarithmic contributions appearing at NLO, is
highly desirable. This method will be coined UNLOPS for unitary
next-to-leading-order matrix element and parton shower merging\footnote{While 
finishing this article, a conceptual publication \cite{Platzer:2012bs} was
presented, which discusses similar methods.}.

This publication is split into a main text and a large appendix section.
The main text should be regarded as an introduction of the methods, while all
technical details and derivations are collected in appendices. Sections in the
main body are intended to give an overview of NL$^3$ and UNLOPS, and explain 
some benefits 
with simple examples. The appendices
are aimed at completeness, and in principle allow an expert reader to 
implement our methods in detail. As such, the appendices can also be 
considered a technical manual of the \pytppp implementation.

We begin by reviewing the CKWW-L method (section \ref{sec:ckkwl}) and the 
UMEPS improvement (section \ref{sec:umeps}). Then, we move to NLO merging 
methods (section \ref{sec:nlo-merging-intro}), discuss NL$^3$ in section 
\ref{sec:nl3}, and describe UNLOPS in section \ref{sec:unlops-math}. After
these, we show the feasibility of the NLO merging schemes by presenting 
results for $\W$-boson production (sections
\ref{sec:w-results} and \ref{sec:w-results-data}) and for 
$\Higgs$-boson production in gluon fusion (section \ref{sec:h-results}). Then,
we end the main text by concluding in section \ref{sec:conclusions}.

In appendix \ref{sec:nlo-prereq} we discuss some of the prerequisites
that we need in order to derive our merging schemes, such as our
choice of merging scale (\ref{sec:pythia-evolution-pt}), the form of
NLO input events that is required for NLO merging in \pytppp
(\ref{sec:exclusive-cross-sections}), a detailed description of the
notation we use (\ref{sec:notation}) and an outline of how the
\powhegbox program can be used to produce the desired NLO input
(\ref{sec:powheg-usage}).

All technical details
on how weights and subtraction terms are generated is deferred to appendix
\ref{sec:weight-generation}, ending in a summary 
(appendix \ref{sec:weight-generation-summary}). From there, we move to a
motivated derivation of the general NL$^3$ method in 
appendix \ref{sec:nl3-derivation}, which can also be understood as a validity 
proof. 
The corresponding derivation of UNLOPS is given in appendix 
\ref{sec:unlops-derivation}, to which a comment on pushing this method to
NNLO is attached (section \ref{sec:unlops-nnlo}). We finally discuss the 
addition of multiparton interactions to NLO-merged results in 
appendix \ref{sec:mpi-in-nlo}.

Before moving to the main text, we would like to apologise for the inherent 
complexity of
NLO merging methods. Also, we would like to affirm that in \pytppp,
intricacies are handled internally, so that with reasonable input, producing
NLO-merged predictions should not be difficult. The schemes described in this 
paper will be part of the next major \pytppp release.

\section{Tree-level multi-jet merging}
\label{sec:multi-jet-merg}

Since the methods presented in this article are heavily indebted to tree-level
merging methods, we would like to start with a brief discussion 
of CKKW-inspired schemes. Let us first introduce some technical jargon.

We think of the results of fixed-order calculations as matrix element (ME)
weights, integrated over the allowed phase space of outgoing particles. 
In the following, we will refer to a phase space point (\ie\ a set of
momentum-, flavour- and colour values for a configuration with $n$ 
\emph{additional} partons) as a $n-$parton event, $n-$parton state or simply 
$\state{n}$. Let us use the term ``$n$-resolved-jet phase space point" 
(often shortened to $n$-jet point, $n$-jet configuration or $n$-jet state) for a 
point in the integration region for which all $n$ partons have a jet 
separation larger than a cut $\tms$. We further classify any configuration 
of hadronic jets by the number of resolved jets from which it emerged. The 
goal of merging schemes is to describe configurations with $n$ hadronic jets 
with $n$-jet fixed-order matrix elements, meaning that the distribution of 
hadronic jets is governed by the ``seed" partons in the $n-$jet phase space 
point, while parton showers only dress the seed partons in soft/collinear 
radiation. We will often use the notation 
$\mathcal{O}\left(\state{nj}\right)$ to indicate that the observable 
$\mathcal{O}$ has been evaluated on configurations containing $n$ resolved 
jets.

Tree-level merging schemes have to ensure that an $n$-resolved-jet
configuration never evolves into a state with $n+1$ well-separated
hadronic jets. This can be achieved by applying Sudakov factors to the
ME input events and by vetoing emissions above $\tms$. To
capture the full parton shower resummation, it is common to also
reweight the input events with a running coupling. Because no
$n$-resolved-jet event evolves to an $(n+1)$-resolved-jet state, it is
possible to add all contributions, and combine the tree-level
description of well-separated jets with the resummation of parton
showers, which can then be processed by hadronisation models.

To summarise, tree-level merging is realised by calculating
tree-level-weighted $n$-jet phase space points for up to $N$
additional jets, reweighting these events, guaranteeing that ME events
do not fill overlapping regions of phase space, and combining the
different event samples for predictions of observables. It is not
reasonable to limit observable predictions to only include
configurations with up to $N$ additional resolved jets. Instead, the
parton shower is used to generate resolved jets for all multiplicities
$n>N$, for which no tree-level calculation is available. This is
accomplished by \emph{not} restricting the emissions off the highest
multiplicity ($N$-jet) ME state to unresolved partons only.

There are in principle different ways how to combine the reweighted
samples in tree-level merging. CKKW-inspired methods use an additive
scheme, while unitarised MEPS opts for an add-subtract
prescription. In the following, we will briefly discuss CKKW-L and
UMEPS.


\subsection{CKKW-L}
\label{sec:ckkwl}

The CKKW-L method
\cite{Lonnblad:2001iq,Lavesson:2005xu,Lonnblad:2011xx} imposes
tree-level accuracy on the parton shower description of phase space
regions with $n\le N$ well-separated partons. To this purpose,
tree-level-weighted phase space point are generated in the form of Les
Houches Event files \cite{Alwall:2006yp}. The cross section of
producing a state $\state{n}$ with $n$ partons has to be regularised
by a cut $\tms$ on the momenta of the partons. In CKKW-L, any
(collection of) cuts that regularise the calculation are
allowed. $\tms$ is commonly called merging scale.

The events with $n=1,\dots,N$ additional partons will then be reweighted to 
incorporate parton shower resummation. The parton shower off states 
$\state{n<N}$ is further forbidden to generate radiation that passes the 
cut $\tms$. Reweighting with no-emission 
probabilities, and ensuring that parton shower emissions do not fill phase 
space regions for which 
events are available from other ME multiplicities,
will guarantee that 
there is no double-counting between events with different number of partons 
in the ME calculation. In this publication, we will use the minimal parton 
shower evolution variable $\ord$ as regularisation cut, and hence denote the 
merging scale by $\ordms$. This choice is discussed in appendix 
\ref{sec:pythia-evolution-pt}.

The full CKKW-L weight to make $n$-parton events exclusive, and minimise the 
dependence on $\ordms$, is given by
\begin{eqnarray}
\wckkwl{n} &=& 
 \tfrac{x_0^+f_0^+(x_0^+,\ord_0)}{x_n^+f_n^+(x_n^+,\muf)}
 \tfrac{x_0^-f_0^-(x_0^-,\ord_0)}{x_n^-f_n^-(x_n^-,\muf)}
 \times \left(\prod_{i=1}^{n}
  \tfrac{x_i^+f_{i}^+(x_i^+,\ord_i)}{x_{i-1}^+f_{i-1}^+(x_{i-1}^+,\ord_{i-1})}
 ~\tfrac{x_i^-f_{i}^-(x_i^-,\ord_i)}{x_{i-1}^-f_{i-1}^-(x_{i-1}^-,\ord_{i-1})}
  \right)\nonumber\\
&&\times\left(\prod_{i=1}^{n}\tfrac{\as(\ord_i)}{\as(\mur)}\right)
  \times\left(\prod_{i=1}^{n}\noem{i-1}(x_{i-1},\ord_{i-1},\ord_i)\right)
  \times\noem{n}(x_n,\ord_n,\ordms)\label{eq:ckkwl-wgt-old}\\
&=&  \tfrac{x_{n}^+f_{n}^+(x_{n}^+,\ord_n)}{x_{n}^+f_{n}^+(x_{n}^+,\muf)}
     \tfrac{x_{n}^-f_{n}^-(x_{n}^-,\ord_n)}{x_{n}^-f_{n}^-(x_{n}^-,\muf)}\nonumber\\
&&\times 
   \prod_{i=1}^{n} \Bigg[\frac{\as(\ord_i)}{\as(\mur)}
     \tfrac{x_{i-1}^+f_{i-1}^+(x_{i-1}^+,\ord_{i-1})}
          {x_{i-1}^+f_{i-1}^+(x_{i-1}^+,\ord_i)}
     \tfrac{x_{i-1}^-f_{i-1}^-(x_{i-1}^-,\ord_{i-1})}
          {x_{i-1}^-f_{i-1}^-(x_{i-1}^-,\ord_i)}
\noem{i-1}(x_{i-1},\ord_{i-1},\ord_i)\Bigg]\nonumber\\
&&\times\,\,\noem{n}(x_n,\ord_n,\ordms)
  ~,\label{eq:ckkwl-wgt}
\end{eqnarray}
where $\ord_i$ are reconstructed emission scales. The
first PDF ratio in \eqref{eq:ckkwl-wgt-old} means that the total cross
section is given by the lowest order Born-level matrix element, which
is what non-merged \pytppp uses. The PDF ratio in brackets comes
from of the fact that shower splitting probabilities are products of
splitting kernels and PDF factors. The running of $\as$ is correctly
included by the second bracket. Finally, the event is made exclusive
by multiplying no-emission probabilities. The \pytppp implementation reorders
the PDF ratios according to \eqref{eq:ckkwl-wgt}, so that only PDFs of fixed 
flavour and x-values are divided, thus making the weight piecewise 
numerically more stable. This will also later be useful when expanding the 
CKKW-L weight. For the highest multiplicity, the last no-emission
probability $\noem{N}(x_N,\ord_N,\ordms)$ is absent to not suppress 
well-separated emissions for which no ME calculation is available.

The calculation of the CKKW-L weight is made possible by using a parton shower
history. Parton shower histories are crucial for all merging methods, so it
is necessary to elaborate. The matrix element state, $\state{n}$, (read from a 
LHE file) is interpreted as the result of a sequence of PS splittings, 
evolving from a zero-jet state, $\state{0}$, to a one-jet state, $\state{1}$, 
etc.\ until the the state $\state{n-1}$ splits to produce the input 
$\state{n}$. All splittings occur at associated scales $\ord_1, \ldots,
\ord_{n}$. A parton shower history (short PS history) for an input state 
$\state{n}$, \ie\ a sequence of states, $\state{0},\dots,\state{n}$, and 
scales $\ord_{1}, \dots,\ord_{n}$, is constructed from the input event by 
inverting the parton shower phase space on $\state{n}$. This means that in a 
first step, we identify partons in $\state{n}$ that could have resulted from 
a splitting, recombine their momenta, flavours and colours, and iterate this
procedure on $\state{n-1},\dots,\state{0}$. Clearly, there can be many ways
of constructing such a history ``path". Indeed, we construct all possible 
parton shower histories for each input state $\state{n}$, and then choose
one history path probabilistically, using the product of PS branching 
probabilities as discriminant. More on this matter can be found in 
\cite{Lonnblad:2011xx}.

A CKKW-L merged prediction for and observable $\mathcal{O}$ is obtained by 
adding all contributions for fixed numbers of resolved jets 
$\mathcal{O}({\state{nj}})$, given by the reweighted (\ie\ exclusive) the 
$\state{n}$ events, for all multiplicities $n=1,\dots,N$. Using the symbol
$\Bornev{n}$ for the fully differential $n$-parton tree-level hadronic cross
section, the prediction for $\mathcal{O}$ is given by
\begin{eqnarray}
\langle \mathcal{O} \rangle
 &=& \int d\phi_0\Bigg\{ \mathcal{O}({\state{0j}}) \Bornev{0}\wckkwl{0}
 + \int \mathcal{O}({\state{1j}}) \Bornev{1}\wckkwl{1}
 + \int\!\!\!\int \mathcal{O}({\state{2j}})\Bornev{2}\wckkwl{2} + \\
 \nonumber
&& \qquad\qquad
 \dots
 ~+~ \int\!\dots\!\int \mathcal{O}({\state{Nj}})\Bornev{N}\wckkwl{N}~
\Bigg\} \nonumber\\
&=&
  \sum_{n=0}^N \int d\phi_0 
  \int\!\dots\!\int
  \mathcal{O}({\state{nj}})~
  \Bornev{n}\wckkwl{n}\label{eq:ckkwl}
~,
\end{eqnarray}
where we have used the symbol $\state{nj}$ to indicate states with $n$ 
resolved jets, resolved meaning above the cut $\ordms$ as defined by the 
merging scale definition. The contribution of states with more than $N$ 
resolved jets is included by allowing the parton shower to produce emissions
above the merging scale when showering the $N$-jet ME events.


\subsection{UMEPS}
\label{sec:umeps}

The idea of unitarised matrix element + parton shower (UMEPS) merging
\cite{Lonnblad:2012ng} is to supplement CKKW-L merging with
approximate higher orders for low-multiplicity states, in order to
exactly preserve the $n$-jet inclusive cross sections. In UMEPS,
events with additional jets, which are simply added in CKKW-L, are
also subtracted, albeit from lower-multiplicity states. This
subtraction is motivated by the mechanism for how non-corrected parton
showers would preserve the inclusive cross section. The contribution
for a jet being emitted off $\state{0}$ at scale $\ord$, for example,
is cancelled with contributions for no jet being emitted between
$\ord_{max}$ and $\ord$. UMEPS makes this cancellation explicit by
constructing subtraction terms through integration over the phase
space of the last emitted jet. The guiding principle is ``subtract
what you add": If $n-$parton events are added, those events should, in
an integrated form, be subtracted from $(n-1)-$parton states.
Improvements in multi-jet observables are retained, since integrated
$n$-parton events and ``standard" events contribute to different jet
multiplicities.

\noindent
In UMEPS, Les Houches events with (initially) $n-$partons are reweighted with
\begin{eqnarray}
\wumeps{n}
&=&  \tfrac{x_{n}^+f_{n}^+(x_{n}^+,\ord_n)}{x_{n}^+f_{n}^+(x_{n}^+,\muf)}
     \tfrac{x_{n}^-f_{n}^-(x_{n}^-,\ord_n)}{x_{n}^-f_{n}^-(x_{n}^-,\muf)}
   \nonumber\\
&&\times
     \prod_{i=1}^{n} \Bigg[\tfrac{\as(\ord_i)}{\as(\mur)}
     \tfrac{x_{i-1}^+f_{i-1}^+(x_{i-1}^+,\ord_{i-1})}
          {x_{i-1}^+f_{i-1}^+(x_{i-1}^+,\ord_i)}
     \tfrac{x_{i-1}^-f_{i-1}^-(x_{i-1}^-,\ord_{i-1})}
          {x_{i-1}^-f_{i-1}^-(x_{i-1}^-,\ord_i)}
  \noem{i-1}(x_{i-1},\ord_{i-1},\ord_i)\Bigg]
  ~.\label{eq:umeps-wgt}
\end{eqnarray}
This is the CKKW-L weight $\wckkwl{n}$ without the last no-emission 
probability $\noem{n}(x_n,\ord_n,\ordms)$ (\ie\ for the highest 
multiplicity $N$: $\wumeps{N} = \wckkwl{N}$).
As before, we make use of a PS history to calculate this weight.
Denoting the tree-level differential $n$-parton cross
section by $\Bornev{n}$, and introducing the notation
\begin{eqnarray}
\label{eq:umeps-b-i-event-def}
\Bornev{n}\wumeps{n} ~=~ \Tev{n} \qquad\textnormal{and}\qquad
\int d^{n-m}\phi~\Bornev{n}\wumeps{n}
 ~=~ \Iev{n}{m}~,
\end{eqnarray}
we can write the UMEPS n-jet merged prediction for an observable 
$\mathcal{O}$ as
\begin{eqnarray}
\langle \mathcal{O} \rangle
 &=& \int d\phi_0\Bigg\{ \mathcal{O}({\state{0j}})\left[ \Tev{0}
 ~-~ \Iev{1}{0}
 ~-~ \Iev{2}{0}
 ~-~
\dots
 ~-~ \Iev{N}{0}~\right]\nonumber\\
&&\qquad\quad
 + \int \mathcal{O}({\state{1j}}) \left[
     \Tev{1}
 ~-~ \Iev{2}{1}
 ~-~
\dots
 ~-~ \Iev{N}{1}~\right]\nonumber\\
&&\qquad\quad
 +\phantom{\int} \dots\nonumber\\
&&\qquad\quad
 + \int\!\dots\!\int \mathcal{O}({\state{N-1j}}) \left[
     \Tev{N-1}
 ~-~ \Iev{N}{N-1}~\right]
\nonumber\\
&&\qquad\quad
 + \int\!\cdots\!\int \mathcal{O}({\state{Nj}})~\Tev{N} ~\Bigg\}\nonumber\\
&=&
  \sum_{n=0}^N \int d\phi_0 
    \int\!\dots\!\int
  \mathcal{O}({\state{nj}})~\left\{
  \Tev{n}
 ~-~ 
  \sum_{i=n+1}^{N}
  \Iev{i}{n}
  ~\right\}
~.
\end{eqnarray}
Many parts of standard CKKW-L implementations can be recycled to
construct UMEPS predictions. The letter $s$ on the integrals in the
samples $\Iev{n+i}{n}$ indicates that the integrated states can
directly be read off from intermediate states in the parton shower
history. If a one-particle integration includes revoking the effect of
recoils, it is possible that the state after performing one
integration contains unresolved partons. In this case, we decide to
perform further integrations (as indicated by the integration measure
in \ref{eq:umeps-b-i-event-def}), until the reconstructed
lower-multiplicity state involves only resolved jets. Multiple
integrations will include the effect of $\ordms$-unordered emissions
into the description of lower-multiplicity states. We think of these
($\ordms$-unordered, sub-leading) contributions as improvements to a strictly ordered parton
shower.

It might however not always be possible to find any parton shower histories
that will permit at least one integration. If the flavour and
colour configurations of a $+n$-parton phase space point cannot be projected
onto an ``underlying Born" configuration with $n-1$ partons, we call the 
parton shower history of the phase space point incomplete \cite{Lonnblad:2011xx}. The existence of
configurations with incomplete histories is reminiscent of which particles are
consider radiative partons, meaning that if $\W$-radiation were allowed, a 
history
\begin{eqnarray*}
\qc\cbar\to\qu\dbar\Wm \qquad\Longrightarrow\qquad \qc\cbar\to\qu\ubar
\end{eqnarray*}
is possible, while otherwise, the history of $\qc\cbar\to\qu\dbar\Wm$ is 
incomplete. Note that the effect of incomplete 
configurations on the cross section is minor, since such contributions are 
related to flavour changes of fermion lines through radiation. 
Configurations with incomplete histories are not regarded as corrections to 
the lowest-multiplicity process, and will be treated as completely new process.
Therefore, we will not (and cannot) subtract configurations with incomplete
histories from lower-multiplicity states, which leads to marginal changes in 
the inclusive cross section.


\subsection{Getting ready for NLO merging}
\label{sec:getting-ready}

Following appendix \ref{sec:pythia-evolution-pt}, we define the
merging scale in terms of the shower evolution variable, thus putting
$\tms = \ordms$.  We further rescale the weights $\wckkwl{n}$ and
$\wumeps{n}$ by a \Kf-factor, $K=\int\Bbar/\int B$, to arrive at a
better normalisation of the total cross section.
This means the introduction of additional
$\mathcal{O}(\as(\mur))-$terms, which have to be removed later
on. Appendix \ref{sec:weight-generation} discusses the generation of
these \Kf-factors, which were introduced in \cite{Lavesson:2008ah} to
avoid dicontinuities across the merging scales.  Note that we include
\Kf-factors only because we do not see a formal reason against
rescaling. In this publication, we attempt to provide a general
definition of our new NLO merging schemes, and thus include these
factors.

Parton showers make $\as$ a tunable parameter, so that \eg\ $\as(\mz)$ is 
chosen to fit data as closely as possible. This means $\asps(\mz)$ used in 
the parton shower might not be the value $\asme(\mz)$ used in the matrix 
element calculation. We can recover a uniform $\as$-definition 
by shifting 
\begin{equation}
\label{eq:asscaleb}
\asps(\rho) = \asme(b_{i}\rho),
\end{equation}
where $b_i$ might be take different values, $b_I$ or $b_F$, if $\asps(\mz)$ is 
different for initial and final state splittings. If $\asme(b_{I/F}\rho)$ 
would then be used instead of $\as$ everywhere, a uniform $\as$ definition 
would be recovered. For this paper, we choose $\asps(\mz) = \asme(\mz) 
= \alpha_{s,\textnormal{\tiny{PDF}}}(\mz)$, \ie\ fix the value of
$\as(\mz)$ to the one used in the parton distributions. In the future, when 
developing a NLO tune, we will interpret $\asps(\mz)$ as a tuning parameter, 
so that we can check the influence of NLO merging on the (rather high) parton 
shower $\as$ value. For the results in this publication, we will drop the
index "ME" on $\as$, and understand $b_i=1$. Our starting point
for NLO merging are $n-$parton samples reweighted by
\begin{eqnarray}
\wumeps{n}
&=&  K \cdot \tfrac{x_{n}^+f_{n}^+(x_{n}^+,\ord_n)}{x_{n}^+f_{n}^+(x_{n}^+,\muf)}
     \tfrac{x_{n}^-f_{n}^-(x_{n}^-,\ord_n)}{x_{n}^-f_{n}^-(x_{n}^-,\muf)}\nonumber\\
&&\times
   \prod_{i=1}^{n} \Bigg[\tfrac{\as(\ord_i)}{\as(\mur)}
     \tfrac{x_{i-1}^+f_{i-1}^+(x_{i-1}^+,\ord_{i-1})}
          {x_{i-1}^+f_{i-1}^+(x_{i-1}^+,\ord_i)}
     \tfrac{x_{i-1}^-f_{i-1}^-(x_{i-1}^-,\ord_{i-1})}
          {x_{i-1}^-f_{i-1}^-(x_{i-1}^-,\ord_i)}
  \noem{i-1}(x_{i-1},\ord_{i-1},\ord_i)\Bigg]
  \label{eq:wt-umeps-k}\\
\wckkwl{n} &=& \wumeps{n}\noem{n}(x_n,\ord_n,\ordms)
  \label{eq:wt-ckkwl-k}~,
\end{eqnarray}
in the case of UMEPS and CKKW-L, respectively. When referring to the weight in
UMEPS and CKKW-L we will from now on always allude to the weights including a
\Kf-factor.

Since we aim at interfacing two different program codes -- NLO matrix
element generators and parton shower event generators -- we need to
make sure that the output of one stage (\ie\ the NLO ME generator) is
completely understood, before using it as input for the event
generation step.  Thus, we require that all fixed-order calculations
are performed with fixed factorisation and renormalisation scale,
since dynamic scale choices in the fixed-order calculation result in
subtle changes in higher orders\footnote{The preparation of output of
  the \powhegbox program \cite{Alioli:2010xd} is outlined in appendix
  \ref{sec:powheg-usage}}.  All higher-order terms due to
$\as-$running and PDF evolution will be carefully taken into account
in the merging algorithm by reweighting with \eqref{eq:wt-umeps-k} (or
\eqref{eq:wt-ckkwl-k}).


\section{Next-to-leading order multi-jet merging}

\label{sec:nlo-merging-intro}

Before sketching the NLO merging schemes we want to present here, we
apologise that the discussion is (even after shifting most technical details
into appendices) unfortunately very notation-heavy. 

Multi-jet merging schemes act on exclusive fixed-order input. This,
for example, means that all phase space points that are allowed in the
evaluation of tree-level matrix elements with $n$ outgoing partons
correspond to configurations with exactly $n$ resolved jets, and no
unresolved jets.
The resolution criterion is given by the minimal separation of
jets, with the relative transverse momentum used as shower evolution variable
defining the separation\footnote{See appendix \ref{sec:pythia-evolution-pt} for 
details.}.

The idea of using exclusive inputs is adopted for NLO merging, 
however, the notion of exclusive cross section needs to be refined for 
next-to-leading order calculations: We consider an $n-$jet NLO calculation 
exclusive if the output consists of $n-$parton phase space points with weights
that correspond to the sum of Born, virtual and unresolved real radiation
terms, where by unresolved real emission, we mean that the additional 
emission does not produce an additional resolved jet.
It is possible to amend the NLO merging scheme if the requirement 
that all real emission terms are unresolved is not met
(see discussion about exclusive vs.\ inclusive NLO calculations 
in appendix \ref{sec:exclusive-cross-sections} for details).

For an NLO merging scheme it is however crucial that virtual and unresolved
real contributions contribute to the same phase space points, since otherwise,
it is not possible to guarantee an implementation that is independent of the
infrared regularisation in the NLO calculation. This problem is solved in 
\powheg and MC@NLO, where real-emission contributions are projected onto 
$n-$jet phase space points by integrating over the radiative phase space. In 
this article, we use the \powhegbox program \cite{Alioli:2010xd} as NLO matrix element 
generator\footnote{See appendix \ref{sec:powheg-usage} for details.}.

Note that we do not require any change in the NLO matrix element
generator. It is acceptable to produce LHE output with only minimal cuts. The
merging scale jet separation will then be enforced internally in \pytppp, 
meaning that after reading the input momentum configuration from LHE file, any
event not passing the cut will be dismissed. \pytppp itself can decide if 
the required number of resolved jets are found, thus rendering the input 
exclusive. 

The aim of this section is to briefly describe two NLO merging algorithms. 
Each description will be split into a more formal part, and an algorithmic 
section, with the goal of presenting an overview of the NLO merging 
prescriptions coined NL$^3$ and UNLOPS. So that the flow of the narrative is 
not overly cluttered with technicalities, we have shifted all details into 
appendices. We however wish to introduce the reader to the 
symbols\footnote{See appendix \ref{sec:notation} for details.}

\vspace*{1ex}
\begin{longtable}{lp{0.83\textwidth}}
$\Bornev{n}$: &   Tree-level matrix element for $n$ outgoing partons.\\
$\int_s\Bornev{n\rightarrow m}$: & Sum of tree-level cross sections with
                  $n$ outgoing partons in the input ME events,
                  after integration over the phase space of $n-m$ partons.\\
$\Bornev{n+1|n}$: & Sum of tree-level configurations with $n+1$ partons with 
                  a definite correspondence to a $n$-parton tree-level matrix 
                  element.\\
$\Virtev{n}$: &   Virtual correction matrix element for $n$ outgoing partons.\\
$\Dipev{n+1|n}$:  & Sum of infrared regularisation terms for $n$ resolved and
                  one unresolved parton.\\
$\Insev{n+1|n}$:  & Sum of integrated infrared regularisation terms for 
                  $n$ resolved and one unresolved parton.\\ 
$\Bbarev{n}$: &   Inclusive NLO matrix element for $n$ outgoing partons,
                  \ie\ sum of Born, virtual and all real contributions as 
                  weight of $n-$parton phase space points.\\
$\Btilev{n}$: &   Exclusive NLO matrix element for $n$ outgoing partons,
                  \ie\ sum of Born, virtual and unresolved real contributions
                  as weight of $n-$parton phase space points.\\
$\int_s\Bbarev{n\rightarrow m}$: & Inclusive NLO cross sections with
               $n$ outgoing partons in the input ME events,
               after integration over the phase space of $n-m$ partons.\\
$\int_s\Btilev{n\rightarrow m}$: & Exclusive NLO cross sections with
               $n$ outgoing partons in the input ME events,
               after integration over the phase space of $n-m$ partons.\\
$\Tev{n}$:    & UMEPS-processed $n$-resolved-jet tree-level events.\\
$\Iev{n}{m}$: & UMEPS-processed tree-level cross sections with initially
               $n$ resolved jets in the input ME events, 
               after integration over the phase space of $n-m$ partons.\\
$\termX{A}{-a,b}$: & Contribution $A$, with terms of powers $\as^a$ and $\as^b$ 
               removed.\\
$\termX{A}{c,d}$: & Contribution $A$, with only terms of power $\as^c$
and $\as^d$
               retained.
\end{longtable}
\vspace*{1ex}

\noindent
Appendix \ref{sec:notation} is intended to give more thorough explanations
of the notation. Particularly the last two short-hands are helpful when 
isolating orders in $\as$. For example, we have
\begin{align}
&\termX{\Bornev{2}}{-2} &=&~ 0 \nonumber\\
&\termX{\Btilev{0}}{1} &=&~ 
         \Virtev{n}
       + \Insev{n+1|n}
       +
         \int \drad
         \Big[~ \Bornev{n+1|n} \Theta\left( \ordms - t(\state{n+1},\ord) \right)
              - \Dipev{n+1|n}~\Big]
\nonumber\\
&\termX{\Bornev{0}\wckkwl{0}}{-0,1} &=&~ 
  \Bornev{0}
  \Big\{
  \wckkwl{0}
 - \termX{\wckkwl{0}}{0}
 - \termX{\wckkwl{0}}{1}
  \Big\}\nonumber\\
& &=&~ 
  \Bornev{0}
  \hugelcurly
  \noem{0}(x_0,\ord_0,\ordms)
 - 1
\nonumber\\
&& &
 +\int^{\ord_{0}}_{\ordms}\!\!
  d\ord~dz~
  \frac{\as(\mur)}{2\pi}
  \left[
  \sum_{a\in\{ \textnormal{\tiny{outgoing}} \}}\!\!\!\sum_{j}
  \splitP_j^a\left(z\right)
~+\!\!\!
  \sum_{a\in\{ \textnormal{\tiny{incoming}} \}}\!\!\!\sum_{j}
  \frac{f_{j}^a(\frac{x_i^a}{z},\muf)}{f_{i}^a(x_{i}^a,\muf)}
  \splitP_j^a\left(z\right)
  \right]
  \hugercurly\nonumber
\end{align}
All details on the expansion of the tree-level weights can be found in 
appendix \ref{sec:weight-generation}.
We are now equipped for extending CKKW-L and 
UMEPS tree-level merging to next-to-leading order accuracy.


\subsection{NL$^3$: CKKW-L at next-to-leading order}
\label{sec:nl3}

The NL$^3$ prescription \cite{Lavesson:2008ah} in principle starts from the CKKW-L-weighted 
tree-level cross sections $\Bornev{n}\wckkwl{n}$, adds events weighted 
according to the exclusive NLO cross sections $\Btilev{n}$, and removes 
approximate $\mathcal{O}(\as^0(\mur))$ and $\mathcal{O}(\as^1(\mur))$
terms in the CKKW-L weight $\wckkwl{n}$.
Since exclusive NLO samples are rarely accessible, we instead use the inclusive 
NLO cross section $\Bbarev{n}$, and generate explicit subtraction events
by using 
higher-multiplicity tree-level matrix elements. For details of this 
choice, we refer to appendix \ref{sec:powheg-usage}.

All details about the derivation of the NL$^3$ method can be found in appendix 
\ref{sec:nl3-derivation}. Here, let us assume the construction of NLO 
accuracy + parton shower higher orders is possible for configurations with 
exactly $m$ resolved jets, and that the desired accuracy is achieved 
for any number of resolved jets $m \in \{0,\dots,M\}$. On top of these 
NLO-correct multiplicities, NL$^3$ allows the inclusion of tree-level matrix
elements with $n \in \{ M+1,\dots,N\}$ additional partons. The 
highest-multiplicity tree-level sample further allows the
generation of more than $N$ resolved jets, by allowing parton shower 
emissions to produce resolved partons. The complete result is
then obtained by simply adding the partial results for each jet multiplicity.
This means that the NL$^3$ result for an observable $\mathcal{O}$, when 
merging $N$ tree-level, and $M<N$ next-to-leading order calculations, is 
\begin{eqnarray}
\label{eq:nl3-full-inclusive-main-text}
\langle \mathcal{O} \rangle
 &=& \sum_{m=0}^M \int d\phi_0 \int\!\cdots\!\int 
     \mathcal{O}(\state{mj})
     ~\Bigg\{ \termX{\Bornev{m}\wckkwl{m}}{-m,m+1} + \Bbarev{m} 
 -\int_s \Bornev{m+1\rightarrow m} \quad\Bigg\}\nonumber\\
&+&
  \sum_{n=M+1}^N
  \int d\phi_0 \int\!\cdots\!\int 
  \mathcal{O}(\state{nj})
  \Bornev{n} \wckkwl{n}
\end{eqnarray}
where the crucial change from CKKW-L (\cf\ \eqref{eq:ckkwl}) is in the
first line, where we add the exclusive NLO events and remove the
corresponding \as-terms from the CKKW-L weight.
From a technical point of view, it is often convenient to think of this in
terms of processing the samples
\begin{align}
&\tree{m}^{\prime} &=&~\termX{\Bornev{m}\wckkwl{m}}{-m,m+1} = \Bornev{m}
          \left\{
          \wckkwl{m}
        - \termX{\wckkwl{m}}{0}
        - \termX{\wckkwl{m}}{1}
          \right\} &\qquad \textnormal{ for } m \leq M\\
&\virt{m} &=&~\phantom{-} \Bbarev{m} &\qquad \textnormal{ for } m \leq M \\
&\subt{m} &=&~-\int_s \Bornev{m+1\rightarrow m} &\qquad \textnormal{ for } m \leq M\\
&\tree{n} &=&~\phantom{-} \Bornev{n} \wckkwl{n} &\qquad \textnormal{ for } M < n \leq N
\end{align}
and writing simply
\begin{eqnarray}
\label{eq:nl3-full-inclusive-main-text-samples}
\langle \mathcal{O} \rangle
 &=& \sum_{m=0}^M \int d\phi_0 \int\!\cdots\!\int 
     \mathcal{O}(\state{mj})
     ~\Bigg\{ \tree{m}^{\prime} + \virt{m} + \subt{m} ~\Bigg\}
\nonumber\\
&+&
  \sum_{n=M+1}^N
  \int d\phi_0 \int\!\cdots\!\int 
  \mathcal{O}(\state{nj})
  \tree{n}
\end{eqnarray}
The prediction for a generic observable can be obtained by calculating the 
result $\mathcal{O}_{b}(\state{k})$, measured for $k-$jet phase space 
points, filling the histogram bin $\mathcal{O}_{b}$ with weight 
$\tree{k}$ (or $\tree{k}^{\prime}$/$\virt{k}$/$\subt{k}$, depending on the 
sample), and summing over all multiplicities $k$.

The construction of the necessary weights is done with the help of a
parton shower history and is detailed in appendix
\ref{sec:weight-generation}. Once the weights are calculated, further
parton showering is attached. The shower off inclusive NLO events and
phase space subtractions is started at the last reconstructed scale,
and all emissions above $\ordms$ vetoed. This means that all
higher-order terms above the merging scale are taken solely from the
reweighted tree-level matrix elements, thus ensuring that the
prescription preserves the parton shower description corrections
beyond the reach of the NLO calculation. All samples have to be added
to produce NLO-accurate $m=0,\dots,M$ jet observables, with
higher-order corrections given by CKKW-L. Details on how the weights
of different samples are motivated, as well as a proof of NLO + PS
correctness, are given appendix \ref{sec:nl3-derivation}.

Here, let us illustrate how NLO accuracy is achieved for one
particular jet multiplicity. For this, we examine the samples
contributing to $M-$jet observables (where M is the highest
multiplicity for which an NLO calculation is available). We start by
analysing the $\Oas{M}$ and $\Oas{M+1}$ contributions. We find
\begin{align}
&  \termX{\langle \mathcal{O} \rangle_{M}}{M}
 + \termX{\langle \mathcal{O} \rangle_{M}}{M+1}
 = \termX{\mathcal{O}(\state{Mj})\virt{M}}{M}
 +
 \termX{\mathcal{O}(\state{Mj})\Big\{\virt{M} + \subt{M}\Big\}}{M+1}\nonumber\\
&\quad
 =~ \mathcal{O}(\state{Mj})\Big\{\Bornev{M}
 + \Virtev{M}
 + \Insev{M+1|M}
 + \int \drad
   \left( \Bornev{M+1|M} - \Dipev{M+1|M}\right)
 - \int_s \Bornev{M+1\rightarrow M}\Big\}\nonumber\\
&\quad
 =~ \mathcal{O}(\state{Mj})~\Btilev{M}
\end{align}
Thus, the description of $M-$jet states is NLO-correct. For $M+1-$jet events,
we have
\begin{align}
&  \termX{\langle \mathcal{O} \rangle_{M+1}}{M+1}
 = \mathcal{O}(\state{Mj})~\Bornev{M+1}~,
\end{align}
providing tree-level accuracy. Both these facts mirror the NLO description
of observables. Keeping only the next-higher powers $\Oas{M+2}$ above 
$\ordms$, we see that
\begin{eqnarray}
&&  \termX{\langle \mathcal{O} \rangle_{M}}{M+2}
+  \termX{\langle \mathcal{O} \rangle_{M+1}}{M+2}
+  \termX{\langle \mathcal{O} \rangle_{M+2}}{M+2}\\
&&\qquad =  \mathcal{O}(\state{Mj})~\Bornev{M}\termX{\wckkwl{M}}{2}
 +  \mathcal{O}(\state{M+1j})~\Bornev{M+1}\termX{\wckkwl{M+1}}{1}
 +  \mathcal{O}(\state{M+2j})~\Bornev{M+2}\nonumber
\end{eqnarray}
For $M-$jet observables, only the reweighted $M-$parton LO matrix 
element contributes, while $M+1-$jet observables are described by reweighted 
$M+1-$parton tree-level states. $M+2-$jet observables are determined by the
$M+2-$parton tree-level prediction. These are the results of default CKKW-L.
Thus, the method is NLO accurate for 
$M-$jet observables, and also retains exactly the resummation of CKKW-L in 
higher orders for $M-$ and $M+1-$jet observables.

\subsubsection{NL$^3$ step-by-step}
\label{sec:nl3-step-by-step}

In the NL$^3$ algorithm, we have to handle three classes of event samples:
\begin{enumerate}
\item[A:] Inclusive next-to-leading order samples $\virt{m}$ for
  $m\leq M$ resolved jets.
\item[B:] Tree-level samples $\tree{m}^\prime$ for $m\leq M$ resolved jets, and 
      tree-level samples $\tree{n}$ for $M < n \leq N$ jets.
\item[C:] Tree-level samples $\subt{m}$ with initially $m+1$ partons, after 
      integration over the radiative phase space of the $(m+1)$'th parton
      (for $m\leq M$).
\end{enumerate}
Samples of class A are produced with the \powhegbox program, by
setting the minimal scale for producing radiation to $\ECM$. For
calculations that need to be regularised, we use minimal cuts in
\powhegbox, and reject events without exactly the number of required
jets internally in \pytppp.  The samples of class A are processed in the most
simple manner:

\begin{enumerate}
\item[A.I] Pick a jet multiplicity, $m$, and a state $\state{m}$, 
           according to the cross sections given by the (NLO) matrix element 
           generator. Reject any state with unresolved jets.
\item[A.II] Find all parton shower histories for 
            $\state{0},\dots, \state{m}$, and pick
            a parton shower history probabilistically.
\item[A.III] Do not perform any reweighting on $\state{m}$.
\item[A.IV] Start the shower off $\state{m}$ at the latest
  reconstructed scale $\ord_m$. Veto shower emissions resulting in an
  additional resolved jet.  $\ordms$.
\item[A.V] Start again from A.I.
\end{enumerate}

\noindent
To amend that we have used inclusive NLO cross sections where we should have 
used exclusive calculations, we have to introduce samples of class C. The 
first step in the construction of these samples is to generate tree-level 
weighted events with $1\leq m \leq M+1$ partons above $\ordms$. Then, 

\begin{enumerate}
\item[C.I] Pick a jet multiplicity, $m+1$, and a state $\state{m+1}$,
  according to the cross sections given by the (LO) matrix element
  generator. Reject any state with unresolved jets.
\item[C.II] Find all parton shower histories for $\state{0},\dots,
  \state{m},\state{m+1}$, and pick a parton shower history
  probabilistically. Replace $\state{m+1}$ with the $\state{m}$ of by
  the chosen history\footnote{We do not apply any further action if
  $\state{m}$ contains unresolved jets in NL$^3$, in contrast UMEPS 
  (or UNLOPS).}.
\item[C.III] Weight $\state{m}$ with $-1$.
\item[C.IV] Start the shower off $\state{m}$ at the latest
  reconstructed scale $\ord_{m}$. Veto shower
  emissions resulting in an additional resolved jet.
\item[C.V] Start again from C.I.
\end{enumerate}

\noindent
Higher orders in $\as$ (in the CKKW-L scheme) are introduced by including
events of class B. Again, tree-level weighted
events for $0\leq n \leq N$ partons are needed as input. Then, 

\begin{enumerate}
\item[B.I] Pick a jet multiplicity, $n$, and a state $\state{n}$,
  according to the cross sections given by the matrix element
  generator. Reject any state with unresolved jets.
\item[B.II] Find all parton shower histories for 
            $\state{0},\dots, \state{n}$, and pick
            a parton shower history probabilistically.
\item[B.III] Perform reweighting:
  \begin{enumerate}
    \item[B.III.1] If $n>M$, weight with $\wckkwl{n}$, as would be the case 
                   in CKKW-L.
    \item[B.III.2] If $n\leq M$, weight with $\left\{\wckkwl{n}
                   - \termX{\wckkwl{n}}{0} - \termX{\wckkwl{n}}{1} \right\}$.
  \end{enumerate}
\item[B.IV] Start the shower off $\state{n}$ at the latest 
            reconstructed scale $\ord_n$.
  \begin{enumerate}
    \item[B.IV.1] If $n=N$, allow any shower emission.
    \item[B.IV.2] If $n < N$, veto shower emissions resulting in an
      additional resolved jet.
  \end{enumerate}
\item[B.V] Start again from B.I.
\end{enumerate}

\noindent
All samples of all classes are finally added to produce the $M$-NLO-jet- and
$N$-LO-jet-merged prediction.
Both the samples B and C require tree-level input, \ie\
the input events for C-samples can be also be used as input for B-samples.
In total, the \pytppp implementation requires $M$ NLO-weighted Les Houches
event files, and $N$ tree-level-weighted files as input, but some of the
tree-level input files need to be processed twice.  

Due to the ubiquity of multiparton interactions (MPI) in hadronic collisions,
we are still far from a full event description at the LHC, even after 
combining multi-jet calculations and parton showers. How MPI can be 
attached to NL$^3$ is discussed in appendix \ref{sec:mpi-in-nlo}.


\subsection{UNLOPS: UMEPS at next-to-leading order}

\label{sec:unlops-math}

Although NL$^3$ accomplishes a merging of multiple NLO calculations to the 
specified accuracy, it inherits the merging scale dependence of the inclusive 
lowest multiplicity cross section from CKKW-L. For lack of a better term, we
will refer to changes in the inclusive cross section as ``unitarity violations".
When including additional jets in $\W$-boson production, unitarity violations
enter at the same order in $\as$ as \eg\ the NLO corrections to 
$\W+j-$production. Even if changes of the inclusive cross section are 
generally small as long as the merging scale is not set too small, it is not 
clear how much of the shape changes we observe are really due to not 
cancelling logarithms. Thus, we want to promote UMEPS, where these unitarity 
violations are absent \cite{Lonnblad:2012ng}, to NLO accuracy as well.

Extending UMEPS to include multiple NLO calculations is slightly more involved 
than the CKKW-L case. The complete method is derived in appendix 
\ref{sec:unlops-derivation}. In a sense, NL$^3$ and UNLOPS are complementary: 
NL$^3$ is, in the accuracy claimed by the method, easily applicable to any
 number of jets, while UNLOPS aims at higher accuracy for the dominant low 
multiplicities\footnote{In fact, the UNLOPS zero- and one-jet NLO merging 
presented here can easily be promoted to a NNLO matching scheme, as outlined 
in appendix \ref{sec:unlops-derivation}.}.
The strategy to extend UMEPS to NLO accuracy is similar to NL$^3$. We remove
any approximate $\Oasof{0}{\mur}$ and $\Oasof{1}{\mur}$ terms in the UMEPS 
weighting procedure, and simply add the correct NLO result. To disturb the 
description of higher order contributions as little as possible, we only 
cancel those terms of the UMEPS weight that would have a better description
in the NLO matrix element.

The UNLOPS method aims to move beyond UMEPS not only in terms of fixed-order 
accuracy for multiple exclusive $n-$jet observables, but also in the 
description of higher orders in low-multiplicity states. This is a direct 
consequence of requiring unitarity, \ie\ that the inclusive cross section be
fixed to the zero-jet NLO result. In the spirit of UMEPS, this means 
that once we
want to add a one-jet NLO calculation, we have to subtract its integrated 
version from zero-jet events.
Similarly we need to remove the $\Oasof{0}{\mur}$ and $\Oasof{1}{\mur}$ in 
the UMEPS tree-level weights, not only for the one-jet events but also for 
the corresponding subtracted zero-jet events. In this way we ensure that the 
inclusive zero-jet cross section is still given by the NLO calculation and 
we will also improve the $\Oas{2}-$term of exclusive zero-jet observables.

The UNLOPS prediction for an observable $\mathcal{O}$, when simultaneously
merging inclusive NLO calculations for m=$0,\dots,M$ jets, and including up 
to N tree-level calculations, is given by
\begin{eqnarray}
\label{eq:unlops-full-inclusive-main-text}
\langle \mathcal{O} \rangle
 &=&
 \sum_{m=0}^{M-1}~ \int d\phi_0 \int\!\cdots\!\int 
     \mathcal{O}(\state{mj})
     ~\Bigg\{
   \Bbarev{m}
 + \termX{\Tev{m}}{-m,m+1}
\nonumber\\
&&\qquad\qquad\qquad\quad~
 - \sum_{i=m+1}^{M} \int_s\Bbarev{i\rightarrow m}
 - \sum_{i=m+1}^{M} \termX{\Iev{i}{m}}{-i,i+1}
 - \sum_{i=M+1}^{N} \Iev{i}{m}
~\Bigg\}\nonumber\\
&+&
   \int d\phi_0 \int\!\cdots\!\int 
   \mathcal{O}(\state{Mj})\Bigg\{~
   \Bbarev{M}
 + \termX{\Tev{M}}{-M,M+1}
 - \sum_{i=M+1}^{N} \Iev{i}{M}~
   \Bigg\}\nonumber\\
&+&
  \sum_{n=M+1}^N
  \int d\phi_0 \int\!\cdots\!\int 
  \mathcal{O}(\state{nj})~\left\{ \Tev{n} - \sum_{i=n+1}^{N} \Iev{i}{n} ~\right\}
\end{eqnarray}

Here we see, in the first line, the addition of the $\Bbarev{m}$ and
the removal of the $\Oasof{m}{\mur}$ and $\Oasof{m+1}{\mur}$ terms of
the original UMEPS $\Tev{m}$ contribution. On the second line we see
the subtracted integrated $\Bbarev{m+1}$ term to make the $m$-parton
NLO-calculation exclusive and the corresponding $\Oasof{m}{\mur}$- and
$\Oasof{m+1}{\mur}$-subtracted UMEPS term together with subtracted
terms from higher multiplicities where intermediate states in the
clustering were below the merging scale. The third line is the special
case of the highest multiplicity corrected to NLO, and the last line
is the standard UMEPS treatment of higher multiplicities.

The full derivation of this master formula is given in appendix
\ref{sec:unlops-derivation}, where we also discuss the case of
exclusive NLO samples and explain the necessity for subtraction terms
from higher multiplicities.
We will limit ourselves to including only zero- and one-jet NLO 
calculations in the results section. For the sake of clarity we will thus only
discuss this special case here. For this case, the UNLOPS prediction
(when including only up to two tree-level jets) is 
\begin{eqnarray}
\label{eq:unlops-01-main-text}
\langle \mathcal{O} \rangle
 &=& \int d\phi_0 \Bigg\{ \mathcal{O}({\state{0j}}) \left( \Bbarev{0}
 ~-~ \int_s \Bbarev{1\rightarrow 0} 
 ~-~ \termX{\Iev{1}{0}}{-1,2}
 -~ \Iev{2}{0} ~\right)\nonumber\\
&&\qquad\quad
 + \int \mathcal{O}({\state{1j}}) \left(
\Bbarev{1} + \termX{\Tev{1}}{-1,2}
 -~ \Iev{2}{1} ~\right)\nonumber\\
&&\qquad\quad
 + \int\!\!\int \mathcal{O}({\state{2j}})
     \Tev{2}
\end{eqnarray}
In an implementation, this is conveniently arranged in terms of the samples
\begin{eqnarray}
\untree{2}
 &=&~
 \Bornev{2}\wumeps{2}\\
\untree{1}
 &=&~ \termX{\Tev{1}}{-1,2}
 ~=~
 \Bornev{1}\left\{
 \wumeps{1} - \termX{\wumeps{1}}{0} - \termX{\wumeps{1}}{1}
 \right\}\\
\virt{0}
 &=&~ \Bbarev{0}
 ~=~      \Bornev{0} + \Virtev{0} + \Insev{1|0}
        + \int \drad \left( \Bornev{1|0} - \Dipev{1|0} \right)\\
\virt{1}
 &=&~ \Bbarev{1}
 ~=~      \Bornev{1} + \Virtev{1} + \Insev{2|1}
        + \int \drad \left( \Bornev{2|1} - \Dipev{2|1} \right)\\
\unsubt{1}
 &=&~ - \int_s
       \Bornev{2\rightarrow1}\wumeps{2}
     ~-~ \int_s
       \Bornev{2\rightarrow0}\wumeps{2} \\
\unsubt{0}
 &=&~ - \termX{\Iev{1}{0}}{-1,2}
 ~=~
  - \int_s
        \Bornev{1\rightarrow0}
        \left\{
         \wumeps{1} - \termX{\wumeps{1}}{0} - \termX{\wumeps{1}}{1}
        \right\}
\\
\unvirt{0}
 &=&~  - \int_s
        \Bbarev{1\rightarrow0}
\end{eqnarray}
meaning that we have two tree-level samples ($\untree{1}$, $\untree{2}$),
two NLO samples ($\virt{0}$, $\virt{1}$), two subtractive samples 
($\unsubt{0}$, $\unsubt{1}$) and one integrated NLO sample ($\unvirt{0}$).
The prediction is formed by reading tree-level input events (for 
$\untree{1}$, $\untree{2}$, $\unsubt{0}$ and $\unsubt{1}$), or inclusive
NLO input (for $\virt{0}$, $\virt{1}$ and $\unvirt{0}$), generating the 
necessary merging weights, and filling histogram bins with the product of 
matrix element and merging weight. For technicalities on the generation of 
the weights, we refer to appendix \ref{sec:weight-generation}. 

In the inclusive cross section, it can immediately be 
checked that all contributions except zero-jet NLO terms cancel exactly, 
meaning that the inclusive cross section is given by the zero-jet NLO cross 
section. As in UMEPS however, we have to accept marginal changes of the 
inclusive cross section in the presence of incomplete histories, \ie\ when
it is not possible to regard $n$-jet states as corrections to $n-1$-jet 
states, because no underlying Born configuration exists.
(see discussion at the end of section \ref{sec:umeps}). The contribution from
such configurations is, for the results presented in this publication, 
numerically insignificant.

Let us turn to the UNLOPS description of exclusive observables. Only looking
at zero-jet observables in \eqref{eq:unlops-01-main-text}, we see
\begin{eqnarray}
\langle \mathcal{O} \rangle_0
 &=& \int d\phi_0
     \mathcal{O}(\state{0j})
     ~\Bigg\{~
     \Bornev{0} + \Virtev{0} + \Insev{1|0}
 + \int \drad \Big[~\! \Bornev{1|0}\Theta\left( \ordms - t(\state{1},\ord) \right) - \Dipev{1|0}~\! \Big]\nonumber\\
&&\qquad
 - \int_s \left[ \Virtev{1} + \Insev{2|1}
               + \int\drad \left( \Bornev{2|1} - \Dipev{2|1} \right) \right] \nonumber\\
&& - \int_s \Bornev{1\rightarrow0}
        \left\{
         \wumeps{1} - \termX{\wumeps{1}}{0} - \termX{\wumeps{1}}{1}
        \right\} 
 - \int_s \Bornev{2\rightarrow0}\wumeps{2}
\quad\Bigg\}~,
\end{eqnarray}
where we have, between the first and second lines, cancelled the tree-level 
contribution of 
$\int_s\Bbarev{1\rightarrow0}$ with the resolved real-emission term 
$\int_{\ordms} \drad \Bornev{1|0}$  appearing in $\Bbarev{0}$. When 
extracting only the $\Oasof{0}{\mur}$ and 
$\Oasof{1}{\mur}$ terms, this gives the contribution of the exclusive NLO 
matrix element, \ie\ of tree-level, virtual correction and unresolved real
contributions. At $\Oasof{2}{\mur}$, we have 
\begin{align}
\termX{\langle \mathcal{O} \rangle_0}{2}
 = \int d\phi_0
     \mathcal{O}(\state{0j})
     \Bigg\{
 - \int_s \left[ \Virtev{1} + \Insev{2|1}
               + \int\drad \left( \Bornev{2|1} - \Dipev{2|1} \right) \right]
 - \int_s \Bornev{2\rightarrow0} ~\Bigg\}
\end{align}
The first group of terms in the curly brackets gives an approximation of NNLO 
corrections, since in a NNLO calculation, all logarithmic terms in the NLO 
$+1-$jet calculation are removed by two-loop and double-real terms. 
Conversely, we should be able to include
the correct logarithmic terms of two-loop and double unresolved terms by 
integrating over the jet in the $+1-$jet NLO calculation. The last term in 
curly brackets is sub-dominant
and corresponds to emissions that are unordered in $\ordms$
\footnote{This term already appears in UMEPS.}. 

Examining the $\Oasof{3}{\mur}$ contributions, we are left with 
\begin{eqnarray}
\termX{\langle \mathcal{O} \rangle_0}{3}
 &=& \int d\phi_0
     \mathcal{O}(\state{0j})
     ~\Bigg\{~
 - \int_s \Bornev{1\rightarrow0}\termX{\wumeps{1}}{2}
 - \int_s \Bornev{2\rightarrow0}\termX{\wumeps{2}}{1}
\quad\Bigg\}
\end{eqnarray}
This is simply the parton shower approximation, amended with a term
corresponding emissions that are unordered in $\ordms$.

Let us move on to the discussion of one-jet observables. If we use the fact
that we can cancel the contribution of two resolved real-emission 
jets in $\Bbarev{1}$ by the $\Oasof{2}{\mur}$ term in 
$\Iev{2}{1}$, we find 
\begin{eqnarray}
&&  \termX{\langle \mathcal{O} \rangle_1}{1}
+ \termX{\langle \mathcal{O} \rangle_1}{2}\\
&&=~ \int d\phi_0
     \int
     \mathcal{O}(\state{1j})
     ~\Bigg\{~
   \Bornev{1} + \Virtev{1} + \Insev{2|1}
 + \int^{\ordms} \drad \left( \Bornev{2|1} - \Dipev{2|1} \right) \Bigg\}~.
\nonumber
\end{eqnarray}
Thus, the method describes one-jet observables with NLO accuracy. The 
$\Oasof{3}{\mur}-$term is given by
\begin{eqnarray}
\termX{\langle \mathcal{O} \rangle_1}{3}
 &=& \int d\phi_0
     \int
     \mathcal{O}(\state{1j})
     ~\Bigg\{~
        \Bornev{1}\termX{\wumeps{1}}{2}
 - \int_s \Bornev{2\rightarrow1}\termX{\wumeps{2}}{1}~\Bigg\}~,
\end{eqnarray}
which is simply the UMEPS-improved parton shower approximation. In conclusion,
we find the method is NLO-correct, improves the logarithmic behaviour of  
zero-jet observables, and otherwise includes the parton shower resummation
of the UMEPS procedure.


\subsubsection{UNLOPS step-by-step}
\label{sec:unlops-step-by-step}

As a complication on top of NL$^3$, UNLOPS requires four classes of events.
We will step-by-step formulate the method for including $M$ inclusive 
next-to-leading order calculations, combined with $N$ tree-level matrix 
elements. This means that we need to handle the samples
\begin{enumerate}
\item[A:] Inclusive next-to-leading order samples $\virt{m}$ for
  $m$ resolved jets.
\item[B:] Tree-level samples $\untree{n}$ for $n<N$ resolved jets.
  (There is no zero-jet tree-level sample.)
\item[C:] Tree-level samples $\unsubt{n}$ with initially $m$
  partons, after integration over the radiative phase space of
  the one or more emissions, as required by the UMEPS method.
\item[D:] Next-to-leading order samples $\unvirt{m}$ with initially $m$ 
  resolved jets, after integration over the radiative phase space of the 
  emission.
\end{enumerate}
Samples of class A are produced with the \powhegbox program, exactly as in 
NL$^3$.
The \powhegbox output files are then processed:
\begin{enumerate}
\item[A.I] Pick a jet multiplicity $m$, and a state $\state{m}$,
  according to the cross sections given by the (NLO) matrix element
  generator. Reject any state with unresolved jets.
\item[A.II] Find all parton shower histories for $\state{0},\dots,
  \state{m}$, and pick a parton shower history probabilistically.
\item[A.III] Do not perform any reweighting on $\state{m}$.
\item[A.IV] Start the shower off $\state{m}$ at the latest
  reconstructed scale $\ord_m$. Veto shower emissions resulting
  in an additional resolved jet.
\item[A.V] Start again from A.I.
\end{enumerate}

\noindent
This is exactly the treatment we already know from NL$^3$.
To ensure that the inclusive (lowest-multiplicity) cross section is not 
changed, we need to subtract the integrated variants $\unvirt{m+1}$ of the 
$(m+1)$-jet NLO calculation, \ie\ introduce the samples of class D. This will 
also remedy the fact that we have used the inclusive $m$-jet NLO cross 
sections, while we should have used exclusive $\Btilev{m}$ input for 
$\virt{m}$.
As starting point, we use the $\Bbarev{m+1}-$distributed event sample 
(\ie\ the same input as for $\virt{m+1}$). Then

\begin{enumerate}
\item[D.I] Reject any state with unresolved jets.
\item[D.II] Find all parton shower histories for
  $\state{0},\ldots,\state{m+1}$, and pick a parton shower history
  probabilistically. Replace $\state{m+1}$ with the $\state{m}$, or
  the first state $\state{l}$ with all $l\le m$ partons above the
  merging scale.  (lower multiplicity states are taken from the
  intermediate states of the chosen PS history)
\item[D.III] Weight $\state{m}$ with $-1$.
\item[D.IV] 
  Start the shower off $\state{m}$ at the latest reconstructed scale
  $\ord_{m}$. Veto shower emissions resulting in an additional
  resolved jet.
\item[D.V] Start again from D.I.
\end{enumerate}

\noindent
To add the UMEPS resummation to these samples (and correct that we have used
$\Bbarev{M}$ events rather than exclusive $\Btilev{M}$ input for $\virt{M}$), we
include samples of class C. These are generated from the $(n+1)$-jet 
tree-level samples, by following the steps

\begin{enumerate}
\item[C.I] Pick a jet multiplicity, $n+1$, and a state $\state{n+1}$,
  according to the cross sections given by the matrix element
  generator.  Reject any state with unresolved jets.
\item[C.II] Find all parton shower histories for $\state{0},\dots,
  \state{n}$, and pick a parton shower history probabilistically.
\item[C.III] Perform reweighting:
  \begin{enumerate}
    \item[C.III.1] If $n+1 > M$, weight with $\wumeps{n+1}$, as would be 
                   the case in UMEPS.
    \item[C.III.2] If $n+1 \leq M$, weight with $\left\{\wumeps{n+1}
                   - \termX{\wumeps{n+1}}{0} - \termX{\wumeps{n+1}}{1} \right\}$.
  \end{enumerate}
\item[C.IV] Replace $\state{n+1}$ with the $\state{n}$, or the first state 
  $\state{l}$ with all $l\le n$ partons above the merging scale (lower
  multiplicity states are taken from the intermediate states of the
  chosen history).  Start the shower off $\state{n}$ at the latest
  reconstructed scale $\ord_{n}$. Veto shower emissions resulting in an additional resolved jet.
\item[C.V] Start again from C.I.
\end{enumerate}

\noindent
The last contributions we have to include are reweighted tree-level samples,
\ie\ events of class B. There is no zero-jet tree-level contribution in 
UNLOPS, since the $\Oasof{0}{\mur}-$term is already included by $\virt{0}$. 
Samples for class B are generated very similar to events of class 
C, with no ``integration step" required for class B: 

\begin{enumerate}
\item[B.I] Pick a jet multiplicity, $n>0$, and a state $\state{n}$,
  according to the cross sections given by the matrix element
  generator.  Reject any state with unresolved jets.
\item[B.II] Find all parton shower histories for $\state{0},\dots,
  \state{n}$, and pick a parton shower history probabilistically.
\item[B.III] Perform reweighting:
  \begin{enumerate}
  \item[B.III.1] If $n>M$, weight with $\wumeps{n}$, as would be
    the case in UMEPS.
  \item[B.III.2] If $n \le N$, weight with $\left\{\wumeps{1} -
      \termX{\wumeps{1}}{0} - \termX{\wumeps{1}}{1} \right\}$.
  \end{enumerate}
\item[B.IV] Start the shower off $\state{n}$ at the latest 
            reconstructed scale $\ord_n$.
  \begin{enumerate}
    \item[B.IV.1] If $n=N$, allow any shower emission.
    \item[B.IV.2] If $n < N$, veto shower emissions resulting in an
      additional resolved jet.
  \end{enumerate}
\item[B.V] Start again from B.I.
\end{enumerate}

Note that although the UNLOPS procedure is more complicated than
NL$^3$, no additional user input is required: \pytppp only needs
$M$ inclusive NLO event samples and $N-1$ tree-level event files, since
$A$ and $B$ use the same NLO input, and the same tree-level input can
be employed in both C and D.
This concludes our discussion of NLO merging prescriptions. Information on how
underlying event is added to our prescription is given in appendix \ref{sec:mpi-in-nlo}.


\subsection{Short comparison}
Before presenting results, let us pause and recapitulate the last
section. We have presented two different NLO merging schemes, which
differ in several ways

\vspace*{1ex}
\begin{longtable}{p{0.45\textwidth}p{0.05\textwidth}p{0.45\textwidth}}
NL$^3$
    &&        UNLOPS \\
    
    && \\
$\bullet$ Generalisation of CKKW-L
    && $\diamond$ Generalisation of UMEPS\\
$\bullet$ Needs exclusive or inclusive NLO calculations as input.
    && $\diamond$ Needs exclusive or inclusive NLO calculations as input.\\
$\bullet$ Straight-forward when moving to high jet multiplicities.
    && $\diamond$ Less transparent when moving to high jet multiplicities.\\
$\bullet$ Changes the inclusive NLO cross sections.
    && $\diamond$ Preserves the NLO inclusive cross sections.\\
$\bullet$ Reproduces the logarithmic behaviour of the PS in zero-jet
observables. Does not fully cancel sub-leading logarithmic enhancements
of higher multiplicity NLO calculations.
    && $\diamond$ Explicitly cancels logarithmic enhancements, has 
       improved logarithmic behaviour in low-multiplicity jet observables.\\
$\bullet$ Produces negative weights.
    && $\diamond$ Produces even more negative weights.\\
\end{longtable}

\noindent
At this point, we will not make comparisons with other NLO merging
methods, but hope to be able to contribute to a thorough comparison in
a future publication. Here we will only make some brief remarks on the
formal accuracy of our methods, compared to the ones presented in
\cite{Gehrmann:2012yg,Hoeche:2012yf} (MEPS@NLO) and
\cite{Frederix:2012ps} (aMC@NLO). All of these methods rely on the
introduction of a merging scale and it is relevant to investigate how
the description of observables are affected by changes in this
scale. In particular it is interesting to make sure that the
NLO-correctness of the of the methods are not spoiled by large
logarithms involving the merging scale,
$L=\ln\left(\frac{\mu_{r/f}}{\ordms}\right)$. Even if the dependence on the merging scale
vanishes to the logarithmic approximation of the shower (normally at
best NLL), the sub-leading logarithmic dependence may become as large
as the NLO correction which we want include.

To exemplify (following the arguments of Bauer et
al. \cite{Bauer:2008qh,Bauer:2008qj}) we look at the inclusive $n$-jet
cross section, which in all methods have been corrected to reproduce
the NLO cross section, so it is exact to $\Oas{n}$ and
$\Oas{n+1}$. But if we look at the $\Oas{n+2}$-term there will be
dependencies on the merging scale, which we can symbolically expand
out in logarithms as $\as^{n+2}(L^4+L^3+L^2+\ldots)$, Even for a
NLL-correct parton shower where the both the $\as^{n+2}L^4$ and
$\as^{n+2}L^3$ terms will cancel exactly, we will have dependencies of
the order $\as^{n+2}L^2$. This means we that have to choose the
merging scale such that $\as L^2\ll 1$, to be sure we do not spoil the
effect of the $\Oas{n+1}$-correction of the NLO calculation.

For the MEPS@NLO method it was shown that it at most has a dependence
of order $\as^{n+2}L^3$ which is colour-suppressed, but certainly has
a dependence of order $\as^{n+2}L^2$. For the aMC@NLO method we do not
know of any formal analysis of the logarithmic correctness, but it is
difficult to see how it could have avoid dependencies of order
$\as^{n+2}L^3$. Also in our NL$^3$ method, where the dependence is
given by the precision of the shower, it cannot be claimed that the
dependence of order $\as^{n+2}L^3$ is absent, as it has not been
proven that the \pytppp shower is formally NLL-correct. However, for
our UNLOPS method, we explicitly conserve the inclusive NLO cross
section, and the merging scale dependence is cancelled almost
completely through our ``subtract everything that is added''
strategy. We say \textit{almost} cancelled, as this is clearly an
observable-dependent statement. From
\eqref{eq:unlops-full-inclusive-main-text} we see that in order for
the addition of a higher order matrix element $B_k$ to completely
cancel for an inclusive $n$-jet observable, we require (symbolically)
\begin{equation}
  \label{eq:observable-cancellation}
  \int d\phi_0 \int\!\cdots\!\int\mathcal{O}_n(\state{kj})B_k=
  \sum_{i=n}^{k-1}\int d\phi_0 \int\!\cdots\!\int
  \mathcal{O}_n(\state{ij})B_{k\rightarrow i},
\end{equation}
which is clearly never an exact cancellation. There is also an
implicit merging scale dependence here as, whether or not \eg\ $B_3$
is projected into $B_{3\to2}$ and measured with
$\mathcal{O}(\state{2j})$ or into $B_{3\to1}$ and measured with
$\mathcal{O}(\state{1j})$, depends on the merging scale. However, for
small enough merging scales, this should not matter for collinear- and
infrared-safe observables, and we do not expect any large logarithms
of the merging scale to appear. Also we note that there are some
$n$-parton states that cannot be projected down to a lower
multiplicity state using parton shower splittings (\textit{incomplete}
states in section \ref{sec:umeps}) as described in
\cite{Lonnblad:2012ng}, where we also found that such diagrams give
numerically very small contributions.

UNLOPS also shares features with the LoopSim method
\cite{Rubin:2010xp,Campanario:2012fk}, in particular the use of an
integrated version of one-jet NLO calculations. However, we cannot
cancel logarithms of the form $\ln\left(\frac{p_{\perp
      \textnormal{jet}}}{\mw}\right)$, which arise by soft/collinear
$\W$-radiation, because we do not allow an integration over the
(radiated) $\W$-boson. The study of such ``giant \Kf-factor effects''
is postponed until a full electroweak shower becomes available in \pytppp.

Finally it should be noted that NLO merging methods can be useful even
if only the NLO calculation for the lowest multiplicity is
available. Since an NLO merging scheme consistently splits the real
emission into unresolved and resolved parts by defining a merging
scale $\ordms$, and uses the same definition to separate states with
two resolved jets from states with one resolved and several unresolved
jets, any NLO calculation can be improved by merging further
tree-level calculations for additional jets. Such schemes go under the
name of MENLOPS
\cite{Hamilton:2010wh,Hoche:2010kg,Alioli:2011nr}. Promoting a NLO
calculation to a MENLOPS prediction is straight-forward with our
methods.

\section{Results}

The UNLOPS and NL$^3$ methods have been implemented in \pytppp, and will 
be included in the next major release version. 
In this section, we will present sample results for NLO merging with inclusive
NLO calculations. The aim of this section is to affirm that the implementation
in \pytppp is working smoothly. We do so by presenting results for $\W$-boson
production and Higgs ($\Higgs$) production in gluon fusion, when simultaneously merging
zero and one additional jet at next-to-leading order with two additional jets on tree-level.

All input matrix element configurations are taken from Les Houches Event 
Files. We use the following input:
\begin{itemize}
\item $\W+0$, $\W+1$ and $\W+2$ at tree-level generated by MadGraph/MadEvent.
\item $\W+0$, $\W+1$ at NLO \cite{Alioli:2008gx,Alioli:2010qp} generated by \powhegbox (see appendix \ref{sec:powheg-usage}).
\item $\Higgs+0$, $\Higgs+1$ and $\Higgs+2$ at tree-level generated by MadGraph/MadEvent.
\item $\Higgs+0$, $\Higgs+1$ at NLO \cite{Alioli:2008tz,Campbell:2012am} generated by \powhegbox (see appendix \ref{sec:powheg-usage}).
\item Fixed-order input was calculated with three values for fixed 
      renormalisation scales and factorisation scales, 
  \begin{itemize}
  \item Central scales: $\mur=\mz^2$ and
      $\muf=\mw^2$ for $\W$-production, $\mur=\mz^2$ and
      $\muf=\mh^2 = (125 \textnormal{ GeV})^2$ for $\Higgs$-production.
  \item Low scales: $\mur=(\mz/2)^2$ and
      $\muf=(\mw/2)^2$ for $\W$-production, $\mur=(\mz/2)^2$ and
      $\muf=(\mh/2)^2$ for $\Higgs$-production.
  \item High scales: $\mur=(2\mz)^2$ and
      $\muf=(2\mw)^2$ for $\W$-production, $\mur=(2\mz)^2$ and
      $\muf=(2\mh)^2$ for $\Higgs$-production.
  \end{itemize}
  In all Figures we will label curves generated from central scale input with
  $cc$, from low scale input with $ll$, and from high scale input with $hh$. 
\item CTEQ6M parton distributions and $\as(\mz^2) = 0.118$.
\item The merging scale $\ordms$ is defined by the minimal \pytppp evolution
      variable (see appendix \ref{sec:pythia-evolution-pt}).
\end{itemize}
The value of $\as(\mz^2)$ was set to match the $\as$-value obtained in the 
parton distributions used in the ME calculation. We use the same PDFs and
$\as(\mz^2)$-value in the parton shower evolution. 
For all internal analyses, we use \texttt{fastjet}-routines 
\cite{Cacciari:2011ma} to define jets. The momentum of the
intermediate $\W$-boson will, if required, be extracted directly from the
Monte Carlo event.

We will compare our results to the result of the \powhegbox program
for $\W$+jet production. For these comparisons, we have generated
default \powhegbox output, fixing the renormalisation and
factorisation scales, and regularising the Born configuration with a
cut $p_{\perp,\textnormal{parton}} = 5$ GeV. To determine a shower
starting scale for these \powhegbox output events, we reconstruct all
possible (including unordered) parton shower histories, choose one,
and start the shower from the last reconstructed scale. No visible
effects of using different options to choose history have been
found. This is not the default interface to the \powhegbox, which
requires truncated showers if the scale definition on the \powhegbox
and the parton shower do not match. Appropriately vetoed showers are
normally used instead in \pytppp, because no truncated showers are
available. Since the scale definition in the \powhegbox could change
depending on the details of the implementation (being different for
Catani-Seymour- and Frixione-Kunzt-Signer-based approaches), we do not
use vetoed showers, and rather choose starting scale by constructing a
parton shower history. For $\W+$jet production, we found only
insignificant differences between both methods.

When taking ratios to default \pytppp (often given by the
lower insets of figures), we rescale the \pytppp reference by 
$\Kf(\mur,\muf)=\frac{\int \Bbarev{0} (\mur,\muf) }{\int \Bornev{0} (\mur,\muf)}$. This
guarantees that we remove the variation of the normalisation of the 
inclusive cross section due to scale choices: 
\begin{eqnarray}
\frac{\frac{1}{\int \Bbarev{0}(\mur,\muf)}\langle\mathcal{O}\rangle_{\textnormal{NLO merged}}}
     {\frac{1}{\int \Bornev{0}(\mur,\muf)}\langle\mathcal{O}\rangle_{\textnormal{Pythia8}}}
=
\frac{\langle\mathcal{O}\rangle_{\textnormal{NLO merged}}}
     {\Kf(\mur,\muf) \cdot \langle\mathcal{O}\rangle_{\textnormal{Pythia8}}}
\end{eqnarray}
The variation of the overall normalisation will otherwise obscure interesting 
features.
For Higgs production in gluon fusion for example we will compare
merged curves generated with $\mur=2\mz$ and $\muf=2\mh$ (labelled
$hh$) to \pytppp, multiplied by with $\Kf(2\mz,2\mh)$. For central
scales, we would use $\Kf(\mz,\mh)$.

\FIGURE[t]{
\centering
  \includegraphics[width=0.47\textwidth]{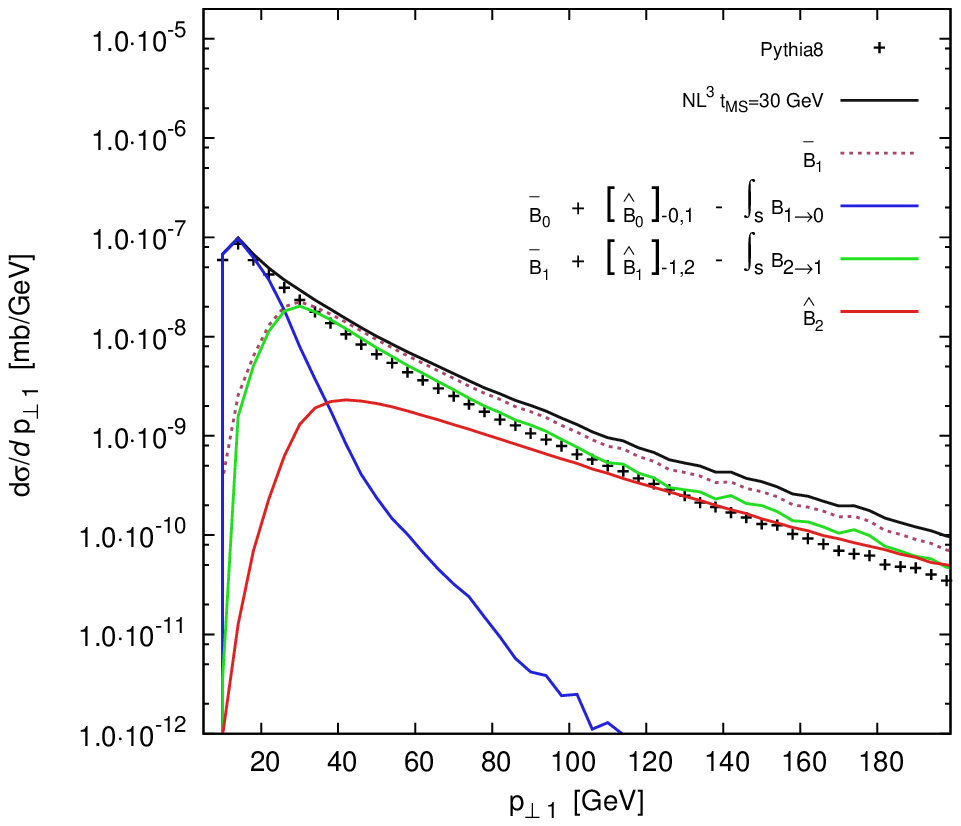}
  \includegraphics[width=0.47\textwidth]{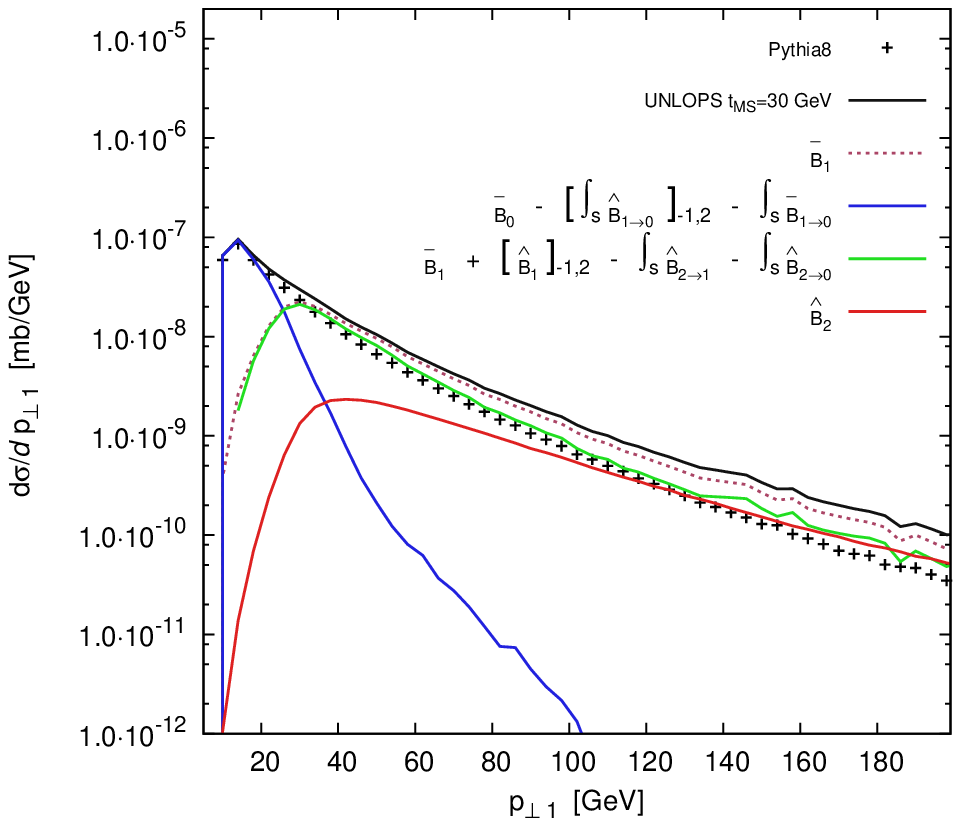}
  \vspace*{-5mm}
\caption{\label{fig:w-pT1-split} Transverse momentum of the hardest jet, for $\W$-boson production in $\p\p$ collisions at $\ECM=7000$ 
  GeV, when merging up to two additional partons at LO, and zero and one additional parton at NLO. Jets were defined with the $k_\perp$-algorithm, with $k_{\perp,min} = 10$ 
  GeV. Multi-parton interactions and hadronisation were excluded. 
  We choose to order the contributions by the jet multiplicity in the states
  on which further the showering is initiated.
  Left panel: Results of the NL$^3$ scheme.
  Right panel: Results of the UNLOPS scheme.}
}
\FIGURE[t]{
  \centering
  \includegraphics[width=0.47\textwidth]{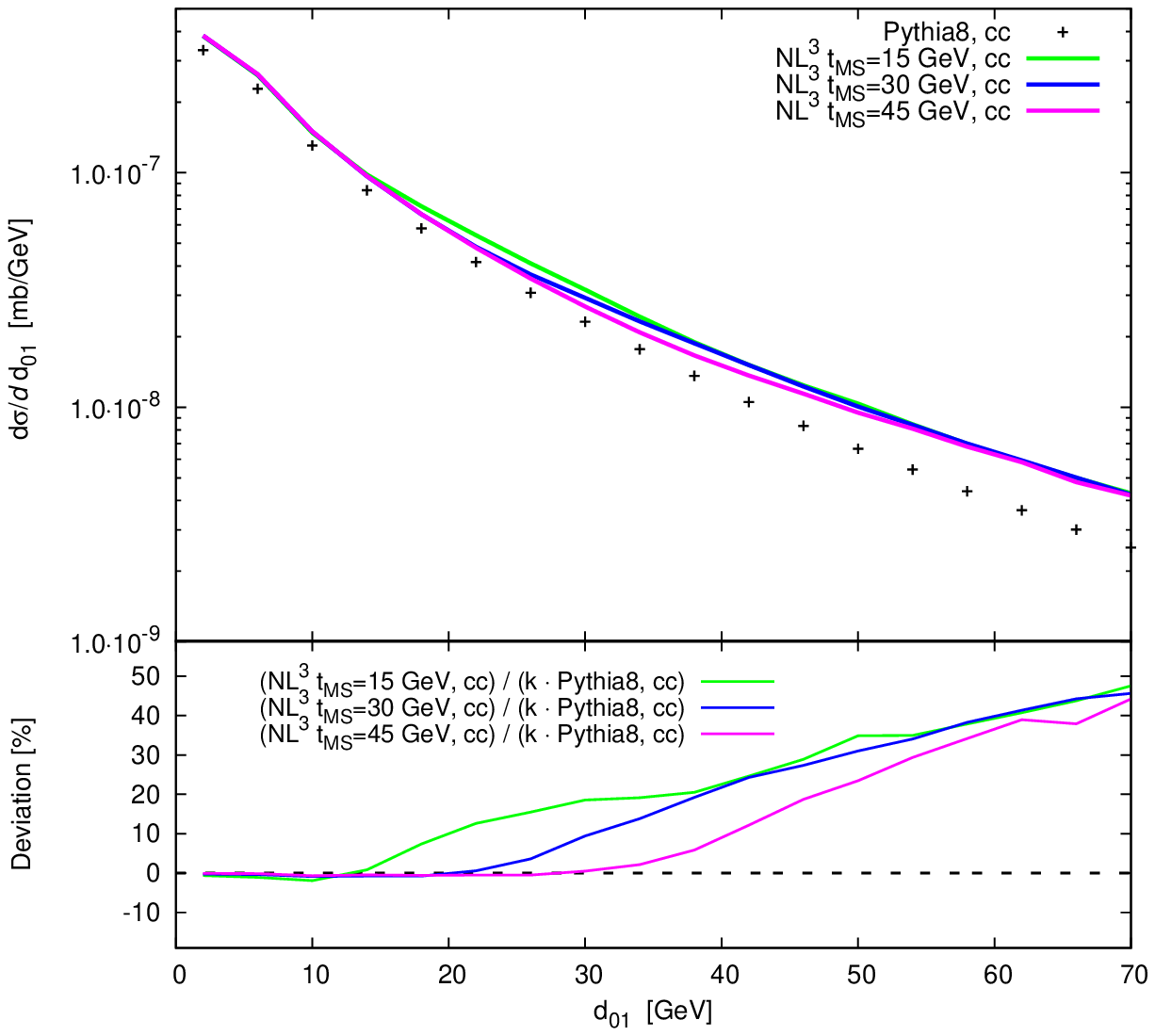}
  \includegraphics[width=0.47\textwidth]{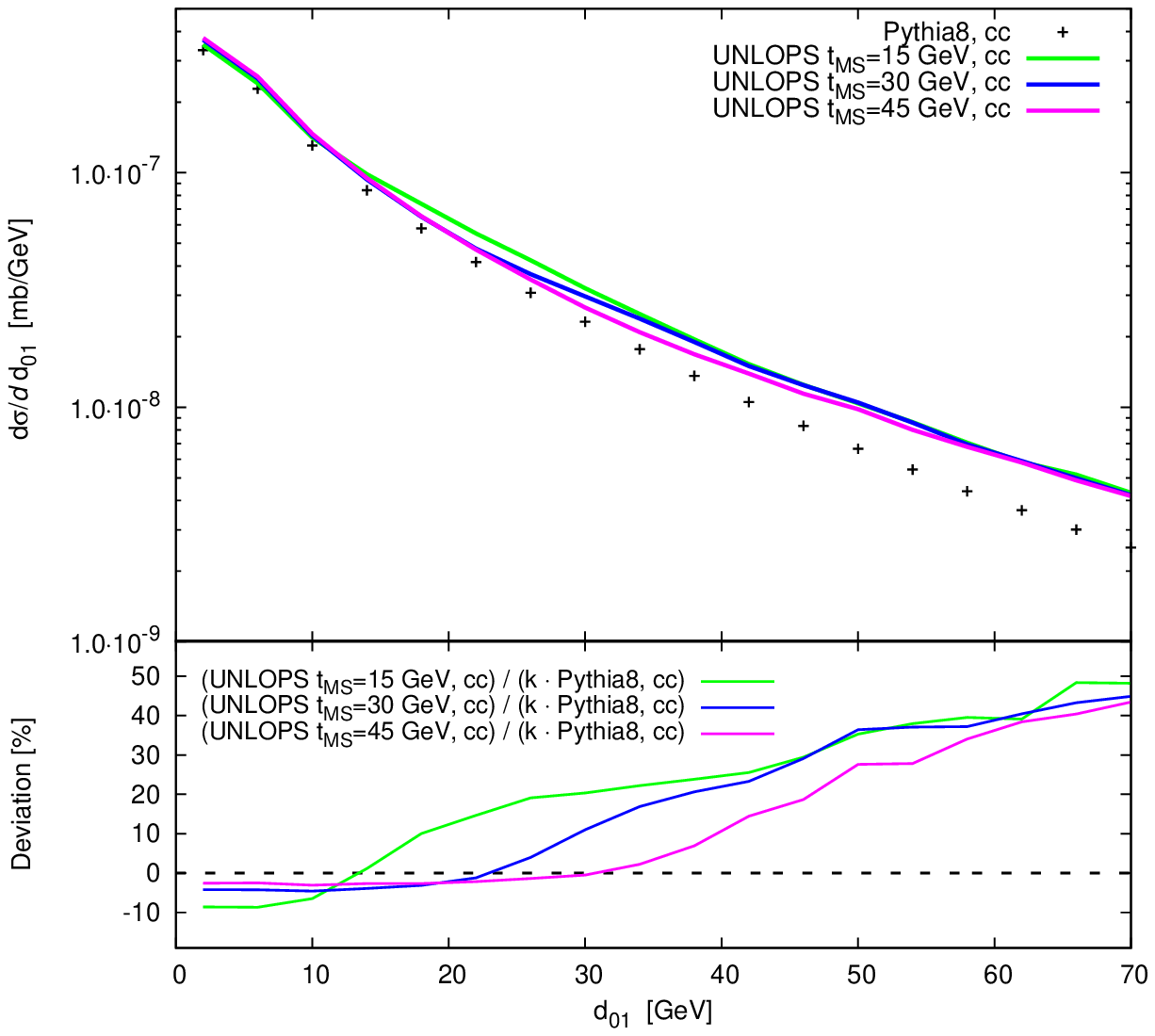}
  \vspace*{-5mm}
  \caption{\label{fig:w-pTj1-tmsvar} $k_\perp$- separation $d_{01}$ of the first jet and 
    the beam, for $\W$-boson production in $\p\p$ collisions at $\ECM=7000$ 
    GeV, when merging up to two additional partons at LO, and zero and one 
    additional parton at NLO, for three different merging scales. Jets were 
    defined with the $k_\perp$-algorithm, by clustering to exactly one jet. 
    Multi-parton interactions and hadronisation were excluded.
    Left panel: Results of the NL$^3$ scheme.
    Right panel: Results of the UNLOPS scheme.}
}
\FIGURE[t]{
\centering
  \includegraphics[width=0.47\textwidth]{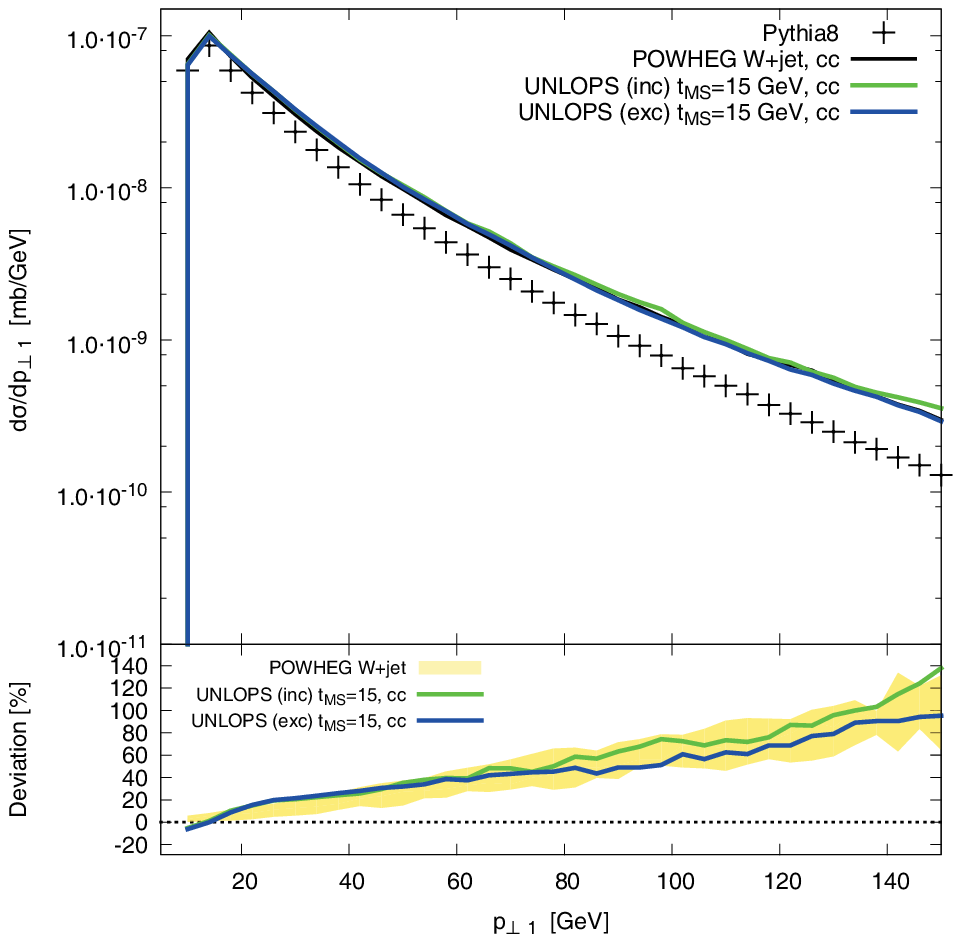}
  \includegraphics[width=0.47\textwidth]{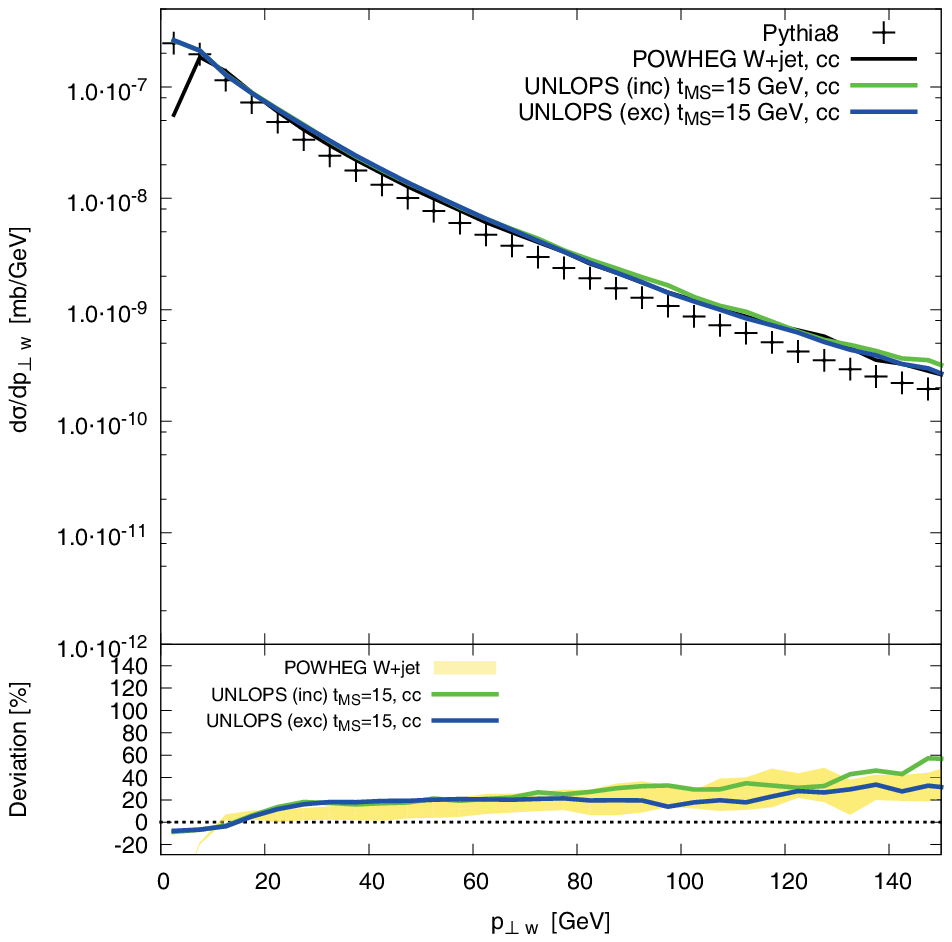}
\caption{\label{fig:w-pts-btilde-vs-bbar} Comparison of using exclusive and 
  inclusive NLO input for $\W$-boson production in $\p\p$ collisions at $\ECM=7000$ 
  GeV, when merging up to two additional partons at LO, and zero and one 
  additional parton at NLO. Curves labelled ``inc" are produced with the
  $\Bbarev{}$-prescription, while ``exc" indicate a generation with $\Btilev{}$-input.
  The lower inset shows the deviation from \pytppp. The band labelled 
  ``POWHEG W+jet" is given by the envelope of varying the renormalisation 
   scale in the \powhegbox program between $\frac{1}{2}\mz,\dots,2\mz$, and 
  the factorisation scale between $\frac{1}{2}\mw,\dots,2\mw$.}
}
\FIGURE[t]{
\centering
  \includegraphics[width=0.47\textwidth]{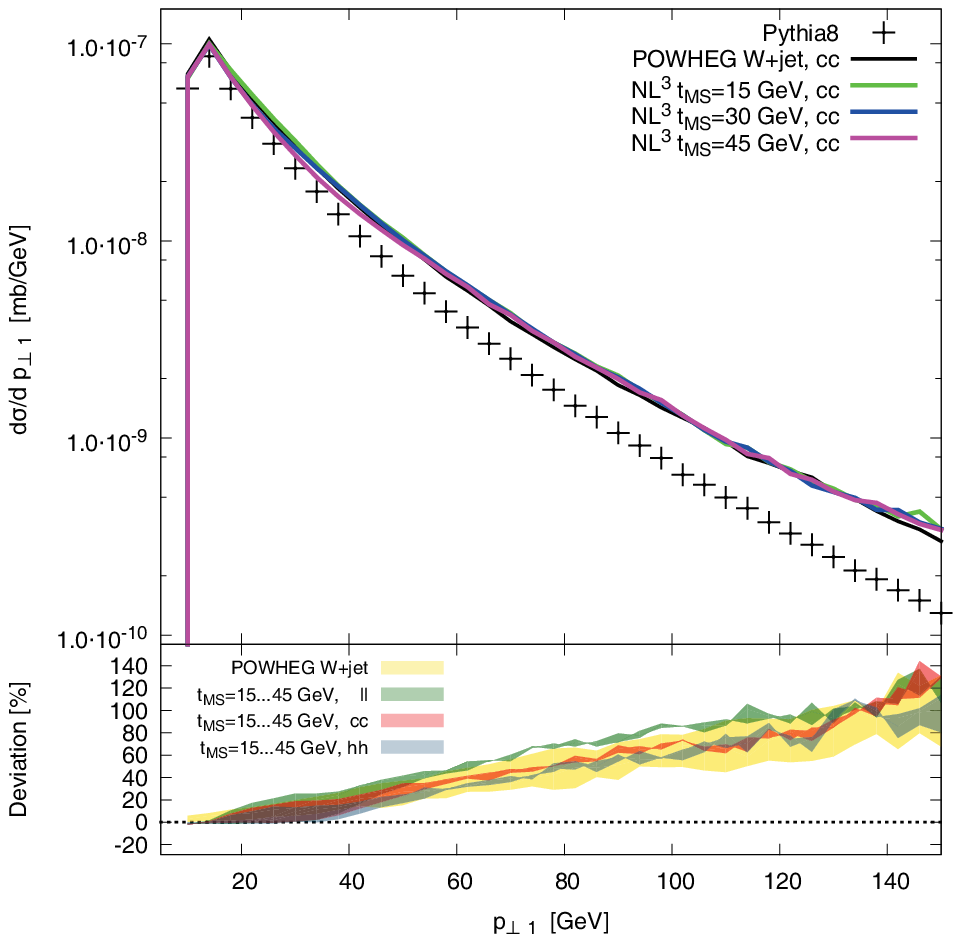}
  \includegraphics[width=0.47\textwidth]{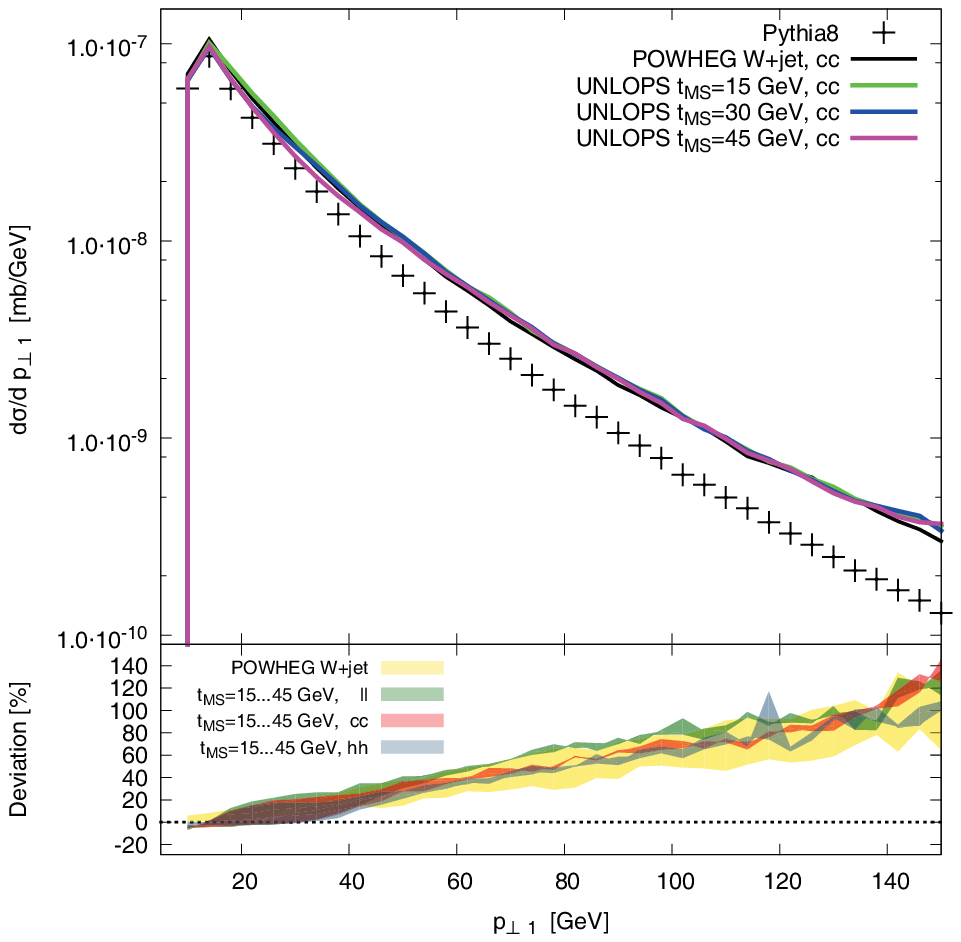}
\caption{\label{fig:w-pT1-vs-powheg}Transverse momentum of the hardest jet, 
  for $\W$-boson production in $\p\p$ collisions at $\ECM=7000$ GeV, when 
  merging up to two additional partons at LO, and zero and one additional 
  parton at NLO. Jets were defined with the $k_\perp$-algorithm, with 
  $k_{\perp,min} = 10$ GeV. Multi-parton interactions and hadronisation were 
  excluded. 
  The lower inset shows the deviation from \pytppp. The band labelled 
  ``POWHEG W+jet" is given by the envelope of varying the renormalisation 
   scale in the \powhegbox program between $\frac{1}{2}\mz,\dots,2\mz$, and 
  the factorisation scale between $\frac{1}{2}\mw,\dots,2\mw$. Left panel:
  Results of NL$^3$. Right panel: Results of UNLOPS.}
}

\subsection{$\W$-boson production}
\label{sec:w-results}

Let us start by discussing results for $\W$-boson production, when combining
inclusive NLO calculations for $\W+0$- and $\W+1$ parton with the \pytppp
event generator. This section is intended mainly for validation, and we will
thus switch off multi-parton interactions and hadronisation. We present
results for both NL$^3$ and UNLOPS. Our preferred method is UNLOPS, since 
the inclusive cross sections are there handled more consistently. By showing 
the results for both NL$^3$ and UNLOPS, we hope to convey a 
rudimentary idea of the effect of potentially problematic logarithmic 
enhancements in standard observables.

Figure \ref{fig:w-pT1-split} shows the transverse momentum of the hardest jet
in $\W$-production at the LHC, for both NL$^3$ and UNLOPS. The sum of the 
solid, coloured curves gives the full NLO merged result, \ie\ the black line.
The dashed curve is included only to illustrate that the hard tail of the 
$p_{\perp 1}$-spectrum is dominated by the the $\W+$jet NLO
sample (labelled $\Bbar_{1}$), both in NL$^3$ and UNLOPS, which is of course desired. This fact makes
the $p_{\perp 1}$-spectra of NL$^3$ and UNLOPS very similar.

Differences between NL$^3$ and UNLOPS are expected in the 
intermediate- / low-scale regions. This is illustrated by 
Figure \ref{fig:w-pTj1-tmsvar}, which shows the $d_{01}$-distribution of the
first jet\footnote{The observable $d_{01}$ is very closely related to 
$p_{\perp 1}$, but avoids a $k_{\perp,min}$-cut in defining the jet, by clustering to exactly one
jet. This allows to show the lowest scale features.}.
Since UNLOPS explicitly preserves the 
$\W$-production NLO cross section, the increase in the tail has to be 
compensated by decreasing contributions below the merging scale. The 
description at low $d_{01}$ in NL$^3$ is, by construction, completely 
governed by the \pytppp result.

Before continuing, we would again like to stress that we are using
inclusive NLO cross section as input in this publication, as discussed
in appendix \ref{sec:exclusive-cross-sections}. There, it was found
that making inclusive cross sections exclusive by constructing
explicit phase space subtractions (through the phase space mapping of
\pytppp) will produce slightly harder partons in the core process (see
Figure \ref{fig:exc-xsec-btilde}). This tendency persists after
showering, as shown in Figure \ref{fig:w-pts-btilde-vs-bbar} for the
UNLOPS case. Clearly, this is a non-negligible effect, although the
differences are contained in the NLO scale variation band. We believe
that using exclusive input is conceptually superior. However, this
section is intended to give uncertainty estimates for NLO merged
parton showers, and in particular to sketch merging scale uncertainty,
and there is no reason to assume that the $\Bbarev{}$- and the
$\Btilev{}$-prescription differ in this respect. Using inclusive
input, however, makes merging scale variations much simpler and
quicker and avoids having to tamper with the internals of the
\powhegbox. Because of this speed factor, we chose to use inclusive
input for the results of this publication.

In the following, we will often include merging scale variations in the ratio
plots. So that the plots become less cluttered, we will give the envelope 
of curves for merging scales between $\ordms=15$ GeV and $\ordms=45$ GeV as 
uncertainty band, rather than show the actual curves.

Figure \ref{fig:w-pT1-vs-powheg}
shows that the transverse momentum of the
hardest jet is heavily affected by NLO merging. This is due to the $\W+$jet NLO
calculation, as already seen in Figure \ref{fig:w-pT1-split}. The merging 
scale variations, as well as the $\mur$/$\muf$-variation for NLO merged 
results lie within the scale variation band of the NLO calculation, but the
combined variation is not significantly smaller. The NLO merged 
predictions touch the upper limit of the NLO scale variation band, because of
 the use of inclusive NLO input, as discussed earlier. Merging scale 
variations alone are minor.   

\FIGURE[t]{
\centering
  \includegraphics[width=0.47\textwidth]{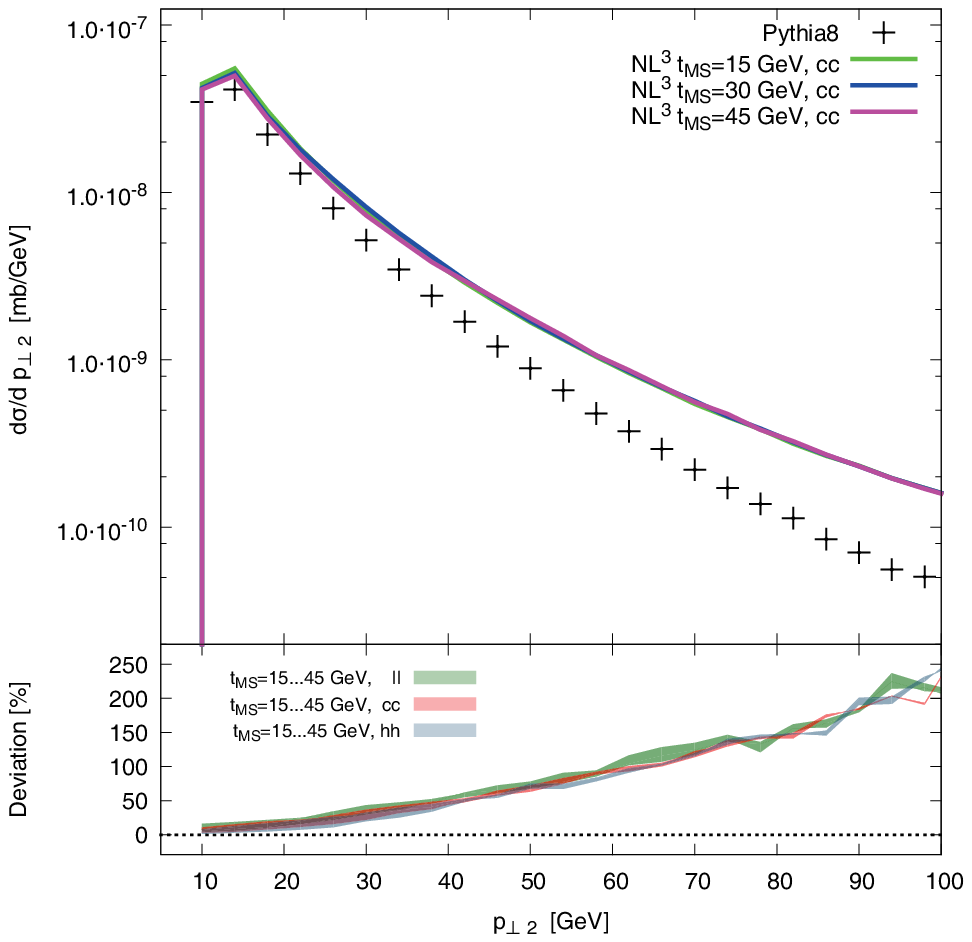}
  \includegraphics[width=0.47\textwidth]{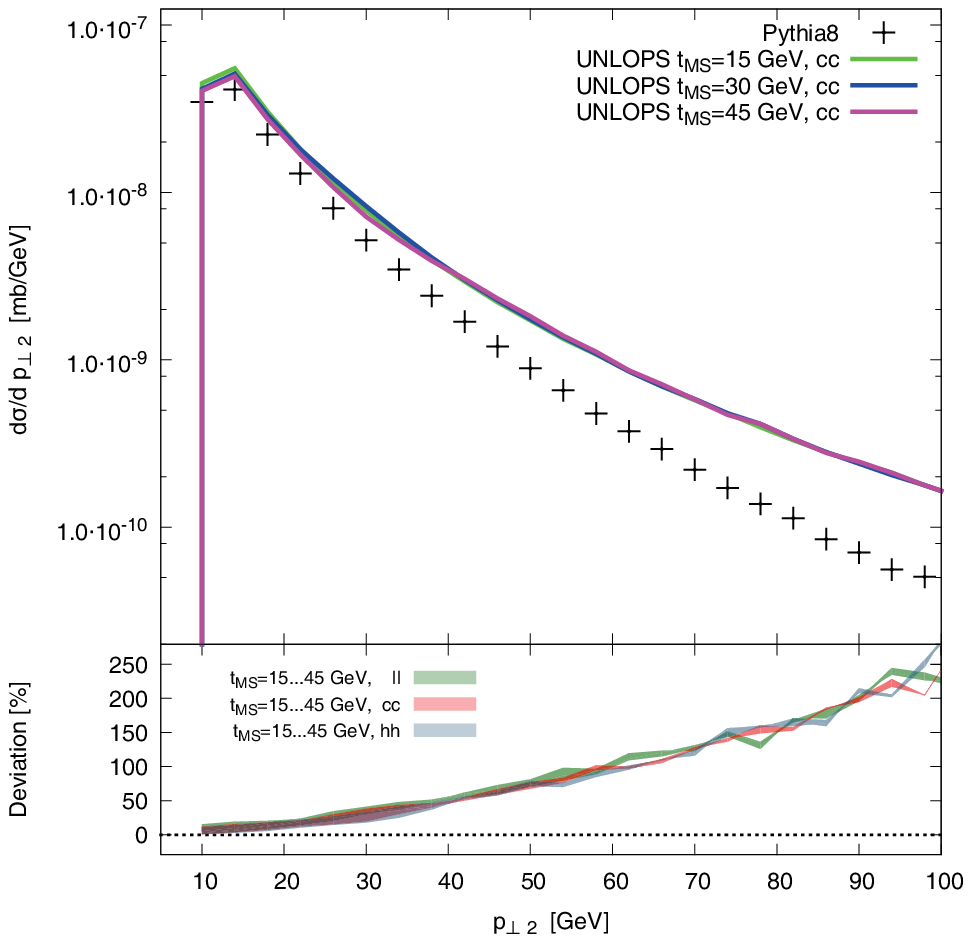}
\caption{\label{fig:w-pT2}Transverse momentum of the second hardest jet, 
  for $\W$-boson production in $\p\p$ collisions at $\ECM=7000$ GeV, when 
  merging up to two additional partons at LO, and zero and one additional 
  parton at NLO. Jets were defined with the $k_\perp$-algorithm, with 
  $k_{\perp,min} = 10$ GeV. Multi-parton interactions and hadronisation were 
  excluded. 
  The lower inset shows the deviation from \pytppp. Left panel:
  Results of NL$^3$. Right panel: Results of UNLOPS.}
}

Merging scale uncertainties are also small in Figure \ref{fig:w-pT2}, which
shows the transverse momentum of the second hardest jet. It is particularly
reassuring that even combined with $\mur$/$\muf$-variation, the bands are
smaller than in CKKW-L and UMEPS \cite{Lonnblad:2012ng}

We would like to conclude this section by noting that differences
between UNLOPS and NL$^3$ are hardly noticeable for the displayed
observables. This is true for all observables we have investigated in
$\W$-boson production, which can be interpreted to mean that the
logarithmic improvements in UNLOPS do not result in major changes in
$\W$-boson production for the merging scale we have chosen. We
anticipate larger effects once scale hierarchies become larger, \ie\
if the merging scale is significantly decreased. For now, we will
instead investigate if the introduction of a slightly larger mass
scale and incoming gluons in the lowest order process leads to visible
effects.

\subsection{$\Higgs$-boson production in gluon fusion}
\label{sec:h-results}

This section is intended to demonstrate that the \pytppp
implementation is not specific to $\W+$jets, and that different
processes can be used to guide algorithmic choices. We have chosen to
investigate Higgs-boson production in gluon fusion, mainly because of
the presence of incoming gluons in the lowest order process and the
very large NLO corrections.

\FIGURE[t]{
\centering
\begin{minipage}{1.0\linewidth}
  \includegraphics[width=0.47\textwidth]{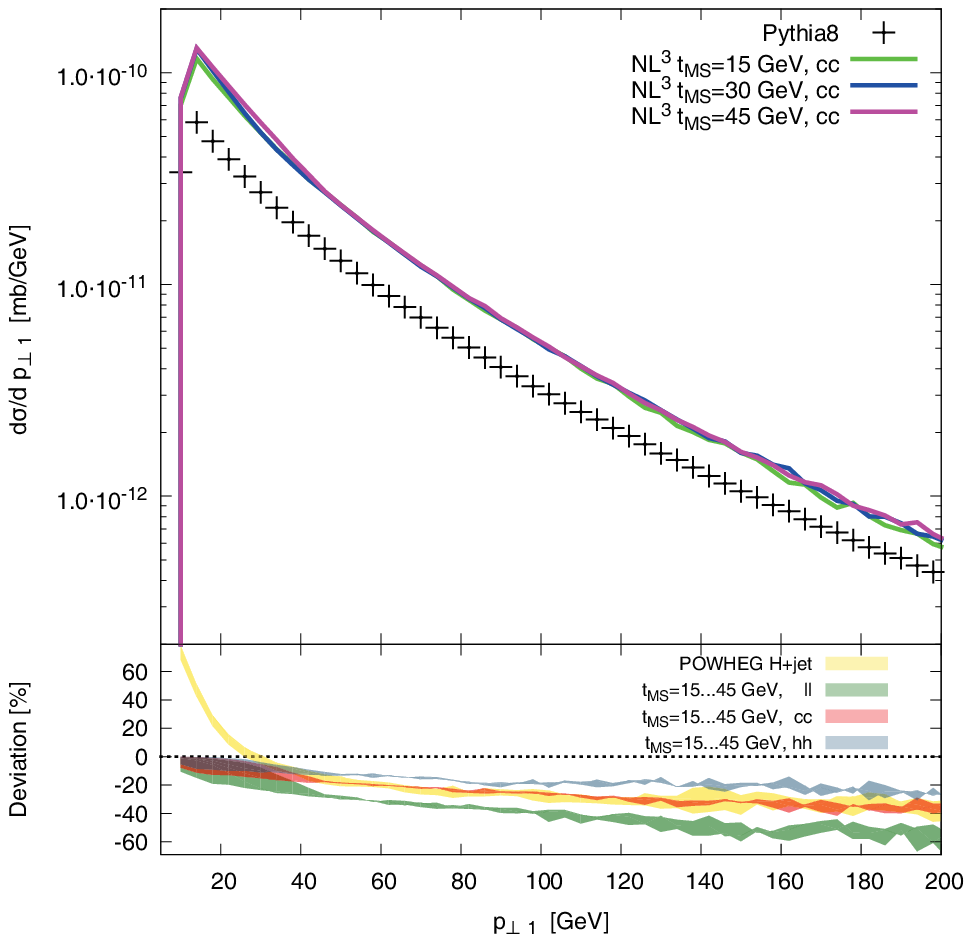}
  \includegraphics[width=0.47\textwidth]{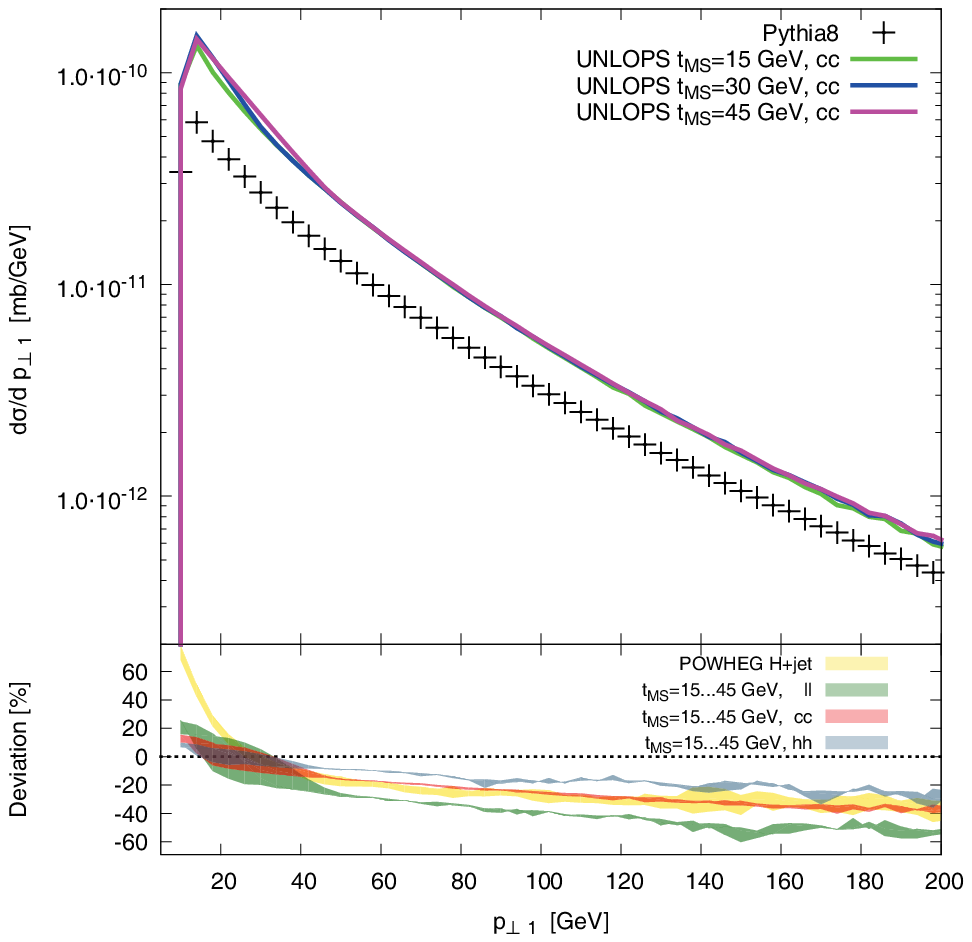}

  \includegraphics[width=0.47\textwidth]{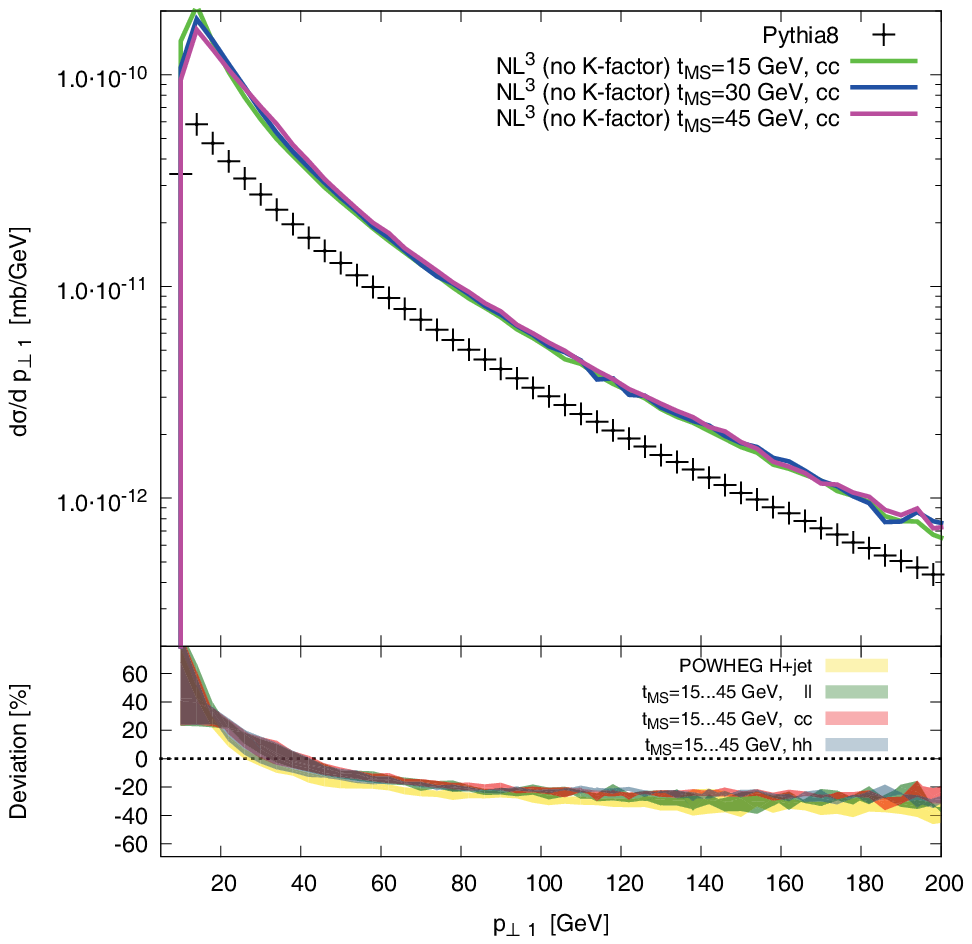}
  \includegraphics[width=0.47\textwidth]{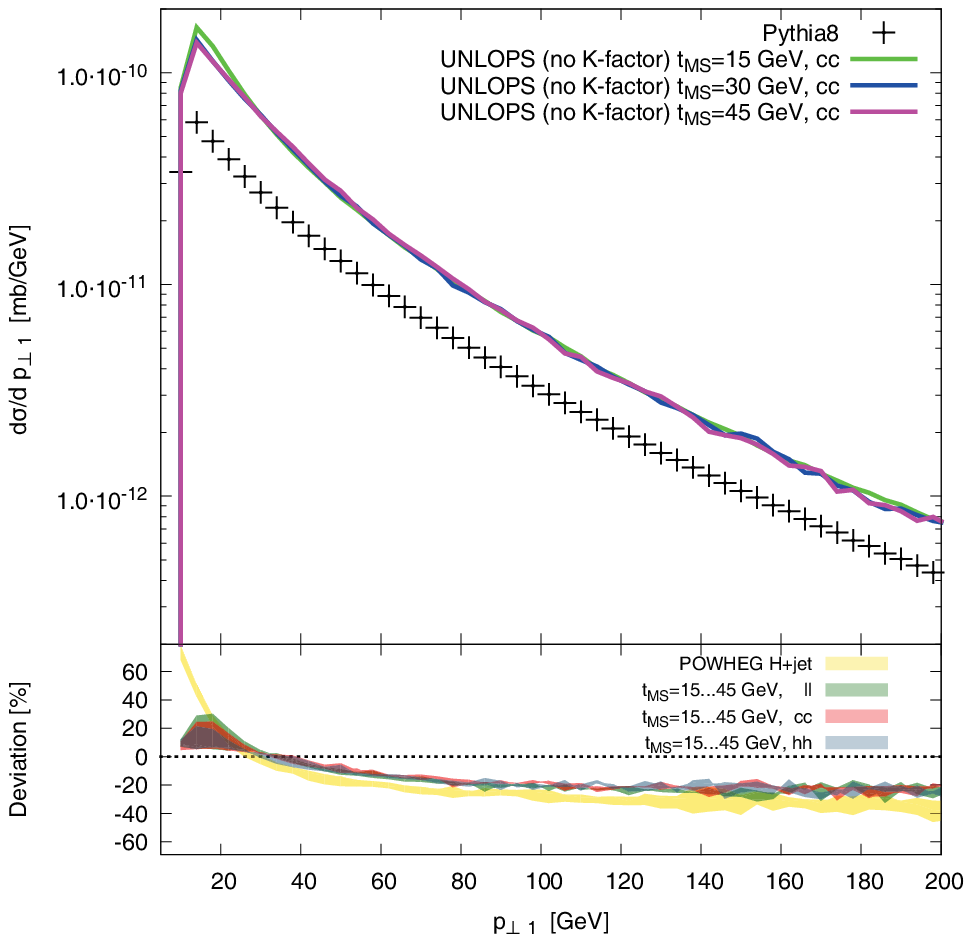}
\end{minipage}
\caption{\label{fig:h-pT1-vs-powheg}Transverse momentum of the hardest jet, 
  for $\Higgs$-boson production in $\p\p$ collisions at $\ECM=7000$ GeV, when 
  merging up to two additional partons at LO, and zero and one additional 
  parton at NLO. Jets were defined with the $k_\perp$-algorithm, with 
  $k_{\perp,min} = 10$ GeV. Multi-parton interactions and hadronisation were 
  excluded. 
  The lower inset shows the deviation from \pytppp. The band labelled 
  ``POWHEG W+jet" is given by the envelope of varying the renormalisation 
   scale in the \powhegbox program between $\frac{1}{2}\mz,\dots,2\mz$, and 
  the factorisation scale between $\frac{1}{2}\mh,\dots,2\mh$.
  The upper panels show the results for using the zero-jet 
  \Kf-factor (i.e.\ $K = \frac{\int \Bbarev{0}}{ \int \Bornev{0}}$) throughout
  the NLO merging procedures. The lower panel show the result when \emph{not}
  using any \Kf-factor (i.e.\ $K = 1$). Left columns: Results of NL$^3$. Right
  columns: Results of UNLOPS.}
}

\FIGURE[t]{
  \centering
  \begin{minipage}{1.0\linewidth}
    \begin{center}
      \includegraphics[width=0.5\textwidth]{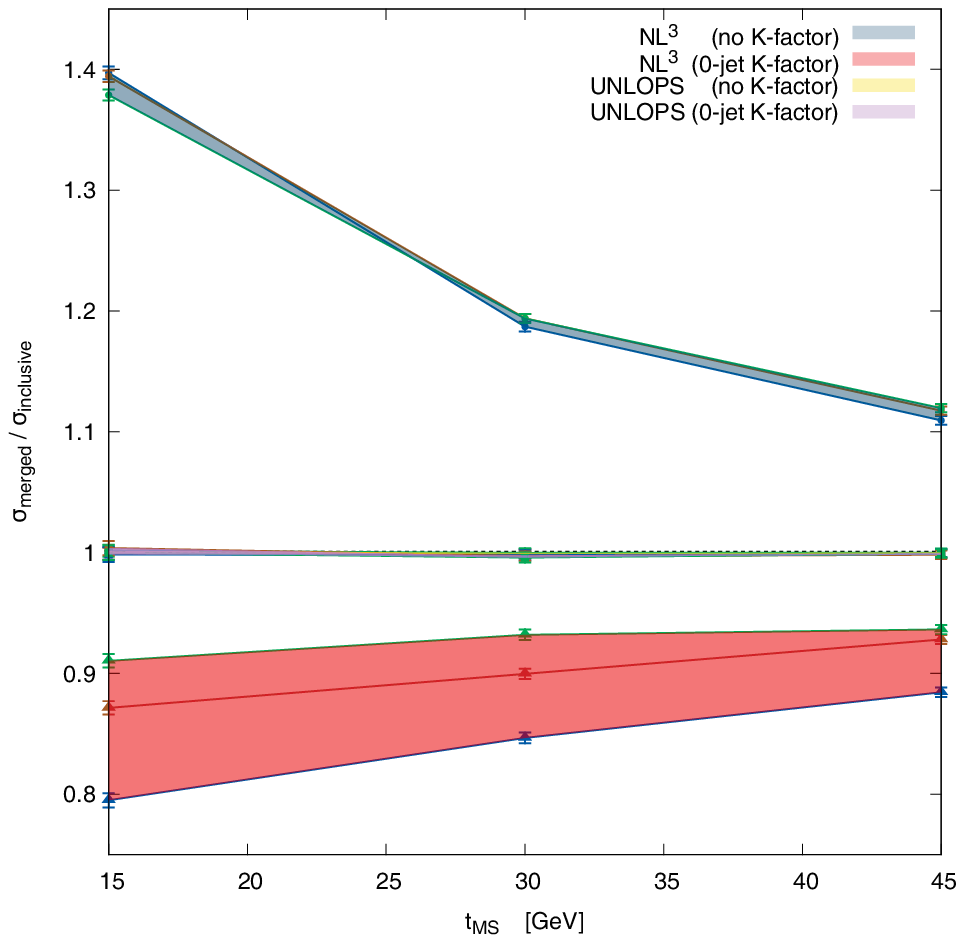}
    \end{center}
  \end{minipage}
  \caption{\label{fig:xsections} Comparison of inclusive cross sections for 
    $\Higgs$-production in gluon fusion $\p\p$ collisions at $\ECM=7000$ GeV, 
    for NL$^3$- and UNLOPS merging, when merging up to two additional partons 
    at LO, and zero and one additional parton at NLO. The results
    of the merging procedure (labelled $\sigma_{\textnormal{merged}}$) are 
    normalised to the zero-jet inclusive NLO cross section 
    $\sigma_{\textnormal{inclusive}}$. The coloured bands indicate the 
    variation from choosing $\muf$ in $\left[~(\frac{1}{2}\mh)^2,\dots (2\mh)^2~\right]$ 
    and $\mur$ in $\left[~(\frac{1}{2}\mz)^2,\dots (2\mz)^2~\right]$.
    The error bars represent only
    the statistical error on the merged cross section. The results of the NLO 
    merging procedure are presented for two different \Kf-factor treatments.
    The result of using \emph{no} \Kf-factor is labelled "no K-factor". 
    Usage of the zero-jet \Kf-factor is indicated by the label 
    "0-jet K-factor". The same variations of NL$^3$ and UNLOPS are plotted,
    although the variance in the UNLOPS results is hardly visible.}
}

In Figure \ref{fig:h-pT1-vs-powheg}, we compare the variation of NLO merged
results with the scale variation in the $\Higgs+$jet NLO calculation of the
default \powhegbox program. The transverse momentum spectrum of the hardest 
jet is softer in NLO results than in \pytppp. We found the same behaviour 
in tree-level merging as well. Interestingly, a similar effect was observed in
pure QCD dijet production, which might indicate that \pytppp overestimates the
hardness of radiation from initial state gluons.

We consider Figure \ref{fig:h-pT1-vs-powheg} a cautionary 
tale. Let us examine the  the upper row first. The merging 
scale variation in $p_{\perp 1}$ is very small. However, when including
renormalisation- and factorisation-scale dependence, the uncertainty of the 
NLO merged results is larger than the variation in the $\Higgs+$jet NLO 
calculation. This is explained by our choice of \Kf-factor. As discussed in
appendix \ref{sec:weight-generation}, this rescaling affects only the ``higher
orders", since the effect on n-jet observables is removed to 
$\Oasof{n+1}{\mur}$. 
Different choices lead to no visible 
effects in $\W$-boson production, since \eg\ a change of 
$\Kf_0=\frac{\int \Bbarev{0}}{\int \Bornev{0}} = 1.16 \approx 1$ will only
result in changing tree-level samples slightly, and $p_{\perp 1}$ is dominated
by the one-jet NLO contribution. 
However, this does not apply to $\Higgs$ production in gluon 
fusion, where $\Kf_0\gtrsim 2$ leads to a significantly larger two-jet 
tree-level contribution. Enhancing the two-jet tree-level contribution will
make the leading-order scale variation of this sample more visible, thus 
leading to an overall larger variation\footnote{The same might naively be 
true for the \powheg result, since two-jet contributions in $\Higgs+$jet in 
\powheg also carry a (much more complex, phase-space dependent) 
\Kf-factor $\Kf = \frac{\Bbarev{1}}{\Bornev{1}}$. This ``one-jet 
\Kf-factor" increases with increasing $\mu_{F,R}$ and counteracts the 
decrease in $\Higgs+$ 2 jet cross section with increasing scales. This leads
-- among other improvements -- to a small scale variation in the \powhegbox 
calculation for $\Higgs+$jet.}. 

Imposing a leading-order scale uncertainty on NLO observables is very 
conservative, and it seems prudent avoid artificial increases due to 
\Kf-factors that rescale higher orders. The lower row of 
\ref{fig:h-pT1-vs-powheg} shows the result of not using any \Kf-factors at all.
The agreement with \powhegbox is reassuring, and the scale variation is 
small. The NL$^3$ result however exhibits major merging scale variations, 
which are mainly induced by an increased cross section in the results
for $\tms=15$ GeV. This unitarity violation was previously ``masked" by a 
large \Kf-factor.

To further illustrate the effect of including \Kf-factors, we
show the merged prediction of the total inclusive Higgs cross section as a 
function of the merging scale in Figure \ref{fig:xsections}. In this figure,
we divide the NLO merged results for scales $\mur$/$\muf$ by the input NLO 
cross section with the same $\mur$/$\muf$ choices. The ideal result should be
unity, without scale uncertainties. This is true to a high degree 
for UNLOPS, which shows that the unitary nature of that method really works 
as expected. For NL$^3$, however, we see that when using \Kf-factors we get 
a large scale variation with a non-negligible merging scale dependence. Removing
the \Kf-factors decreases the scale variations, but on the other hand
increases significantly the merging scale dependence.

The \Kf-factor dependence is a major uncertainty in the NLO merged results for
Higgs production in gluon fusion. We would like to stress that the current 
publication is intended as a technical summary, and not aimed at making 
binding predictions. Rather, we will use this as guidance when
improving the implementation further.

\subsection{$\W$-boson production compared to data}
\label{sec:w-results-data}

\FIGURE[t]{
\centering
  \includegraphics[width=0.47\textwidth]{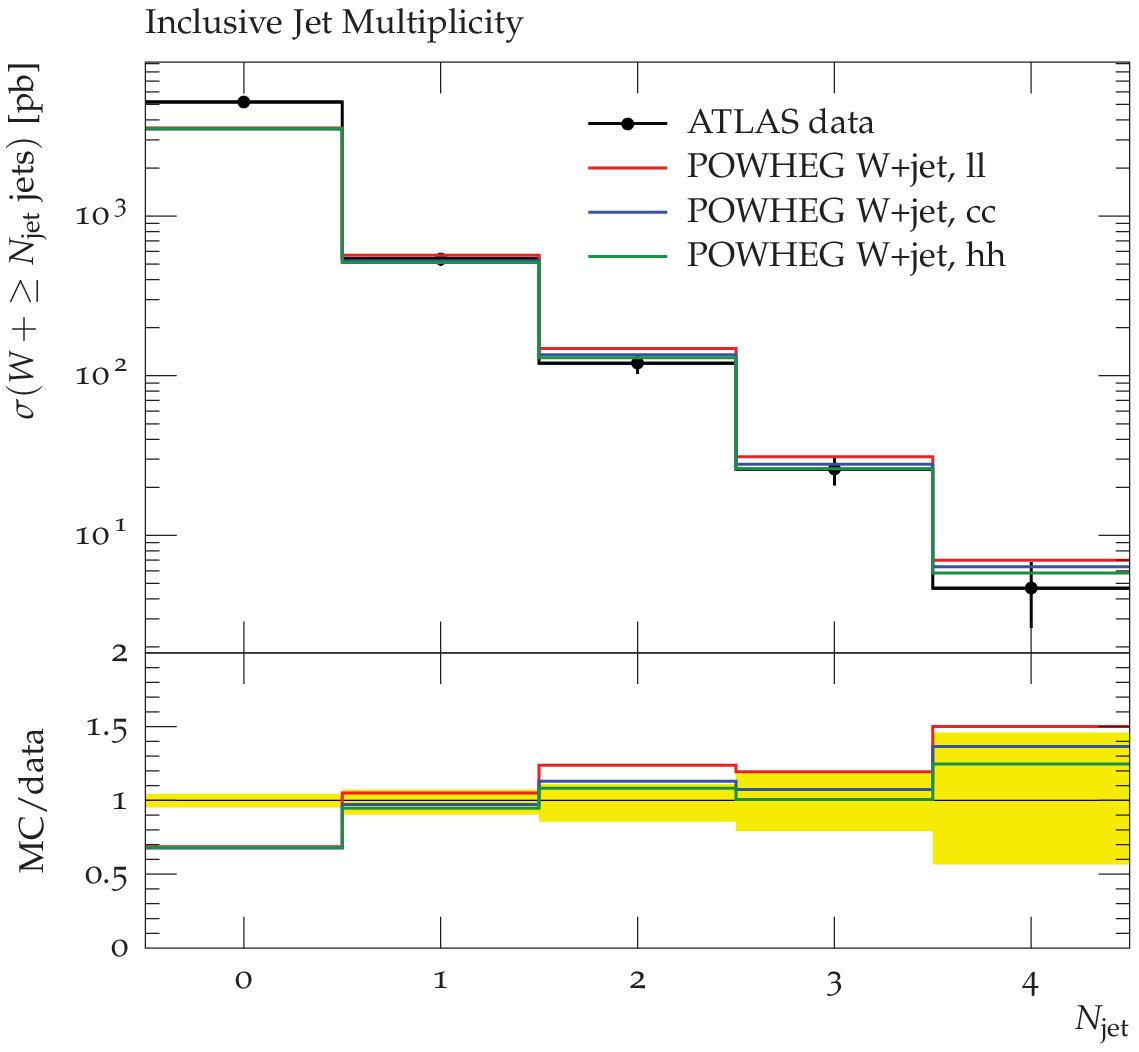}
  \includegraphics[width=0.47\textwidth]{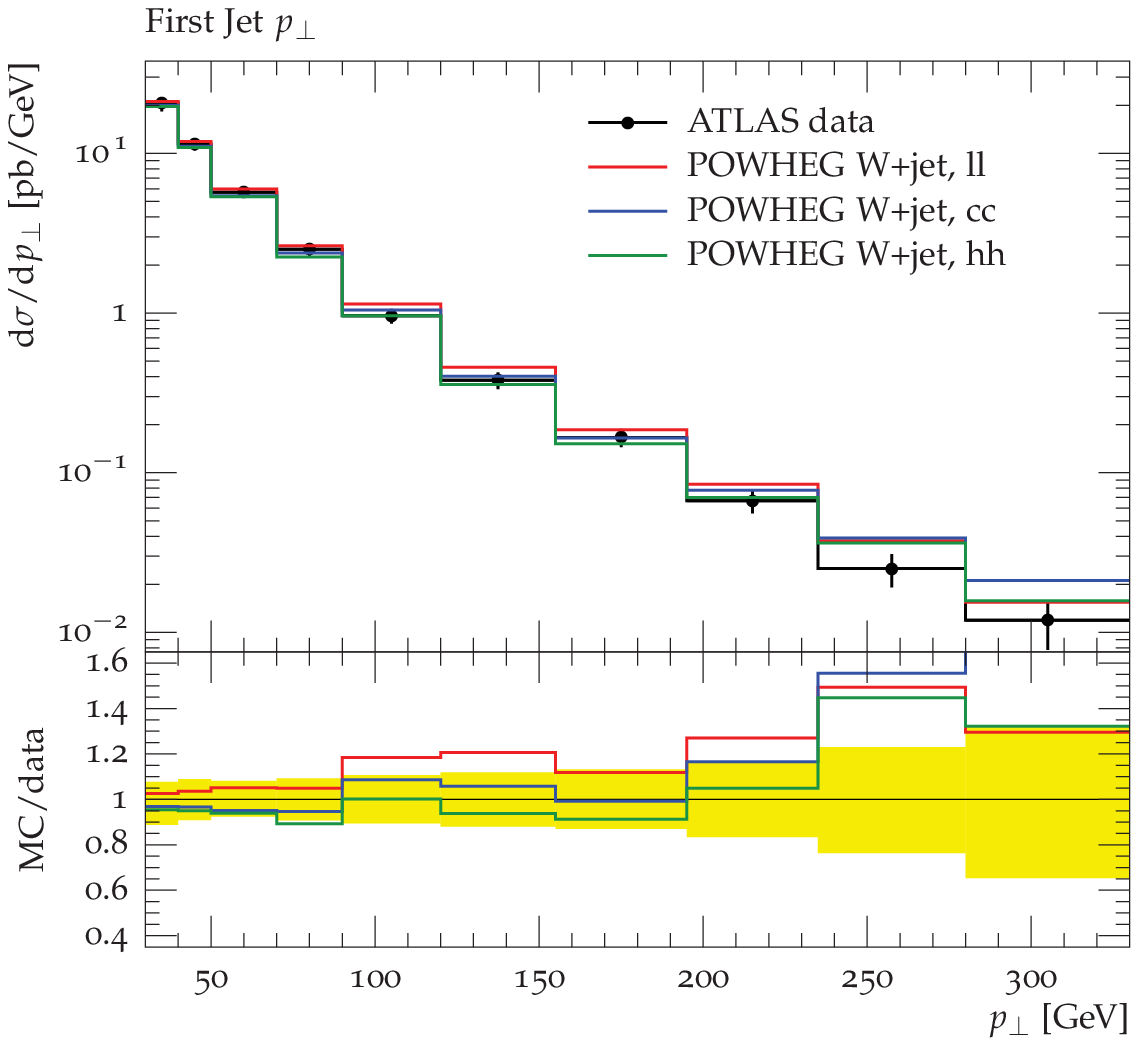}
  \caption{\label{fig:atlas-w-pt-powheg}Jet multiplicity and
    transverse momentum of the hardest jet in $\W$-boson production,
    as measured by ATLAS \cite{Aad:2012en}. MC results taken from the
    \powhegbox program, with three different
    renormalisation/factorisation scales. Effects of partons showers,
    multiple scatterings and hadronisation are included.}
}

In this section, we would like to show NLO merged predictions in
comparison to data. We would like to point out that we have fixed
$\as(\mz)$ in the PS to $\as(\mz)=0.118$, and use CTEQ6M parton
distributions throughout. Please consult appendix \ref{sec:mpi-in-nlo}
for a discussion of multiparton interactions. This means that the
results do not correspond to a tuned version of the \pytppp
shower. Conclusive results can of course only be presented after the
uncertainty of PS tuning has been assessed.

\FIGURE[t]{
  \centering
  \begin{minipage}{1.0\linewidth}
    \includegraphics[width=0.47\textwidth]{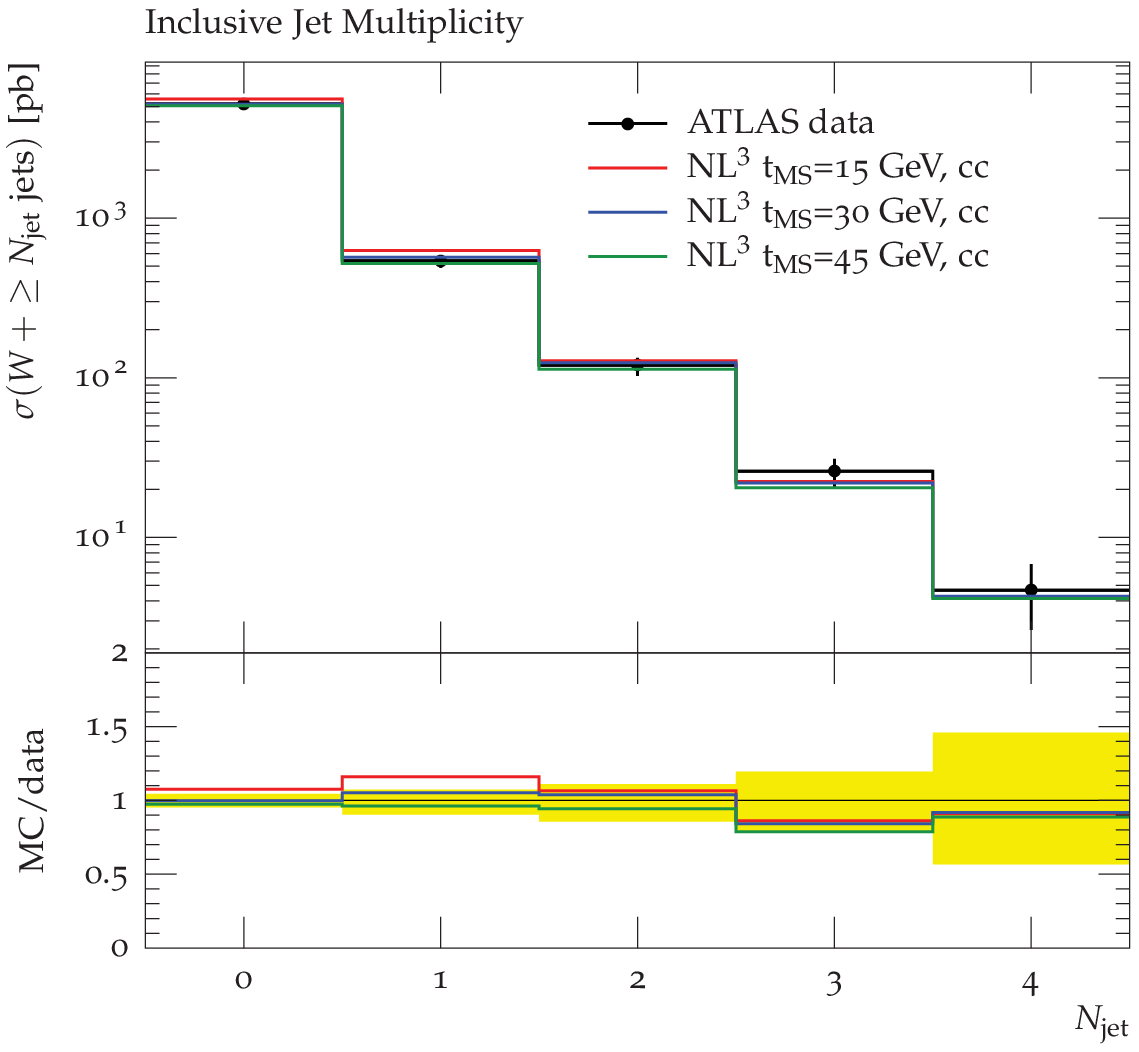}
    \includegraphics[width=0.47\textwidth]{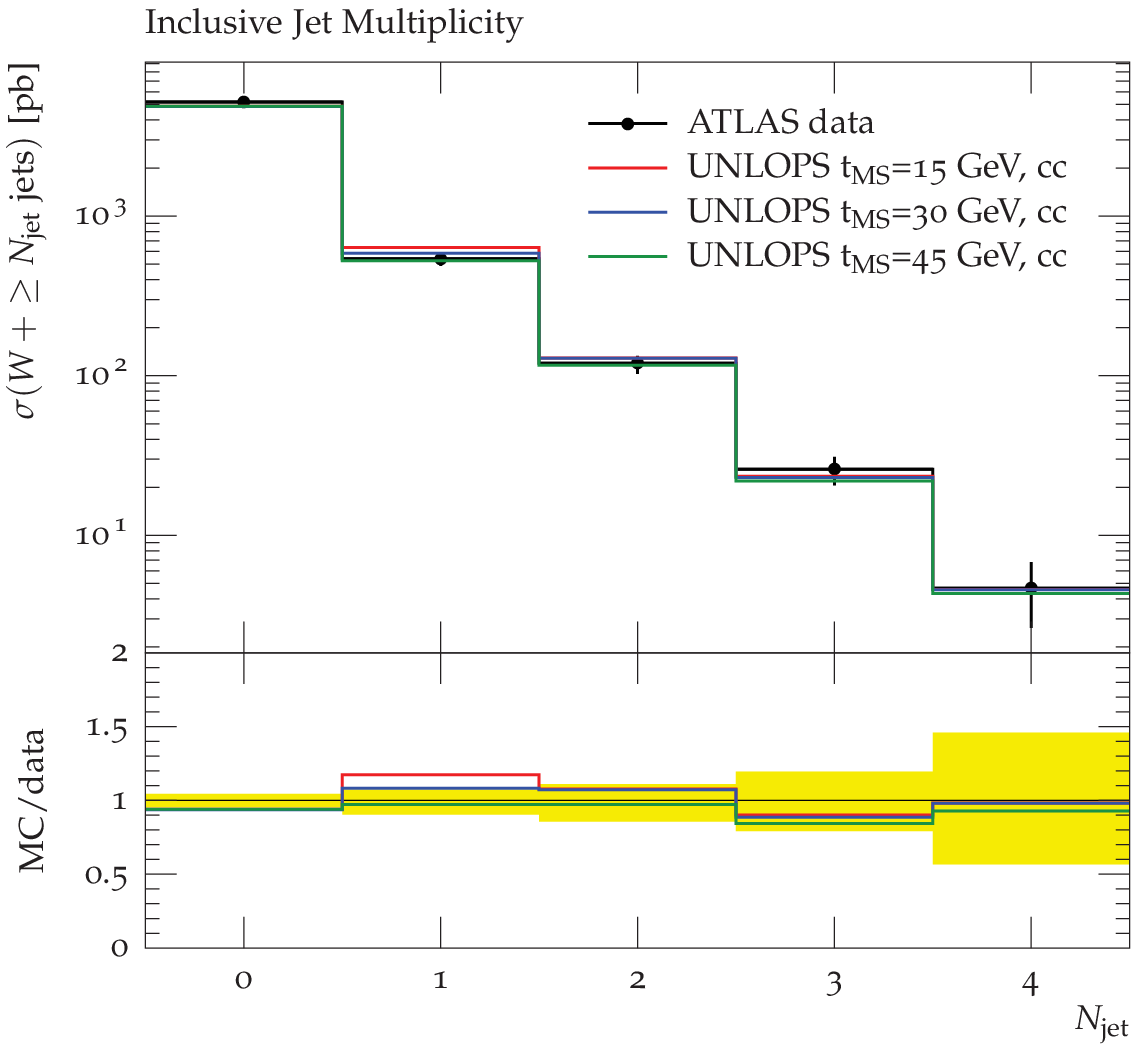}
    
    \includegraphics[width=0.47\textwidth]{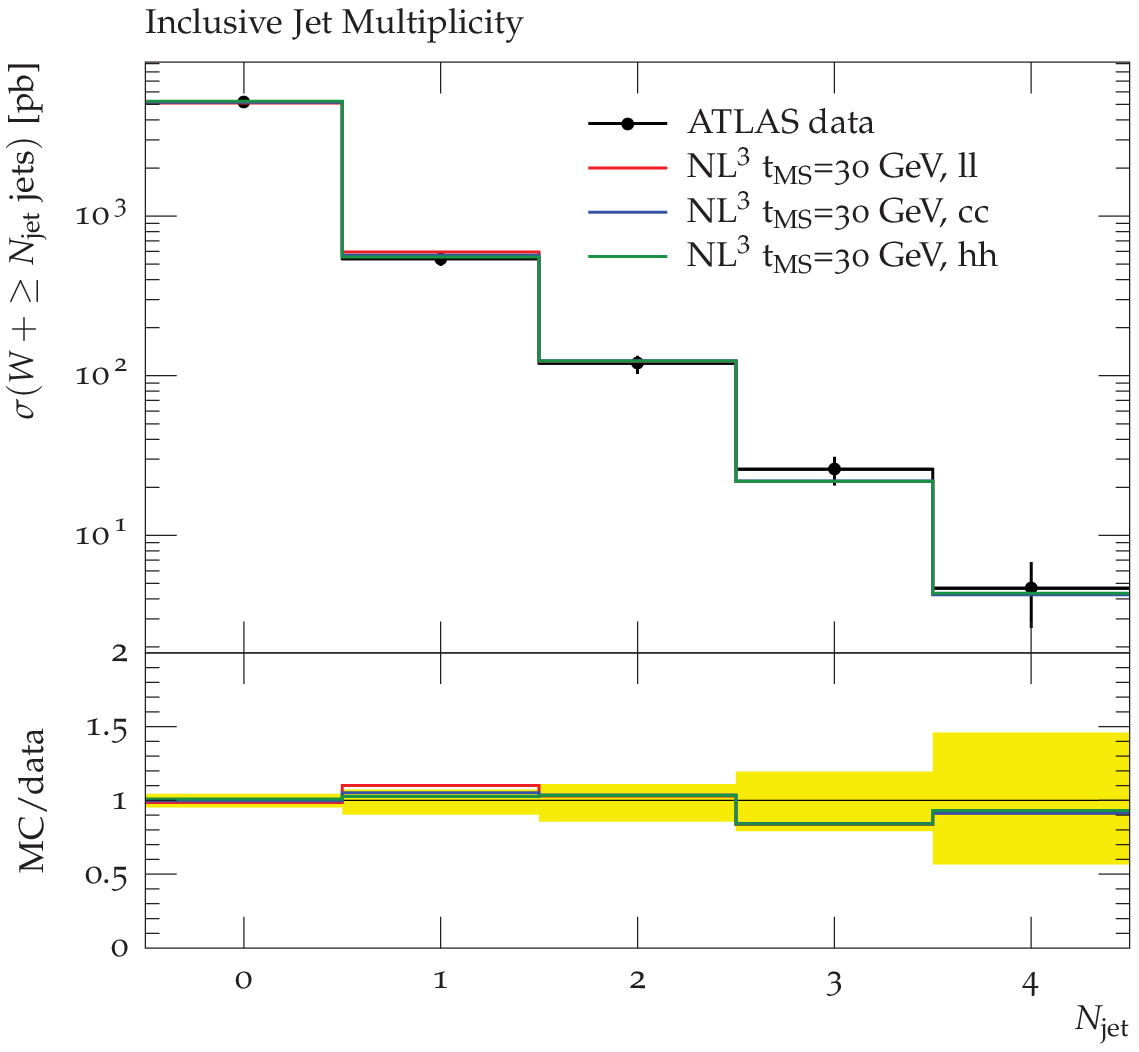}
    \includegraphics[width=0.47\textwidth]{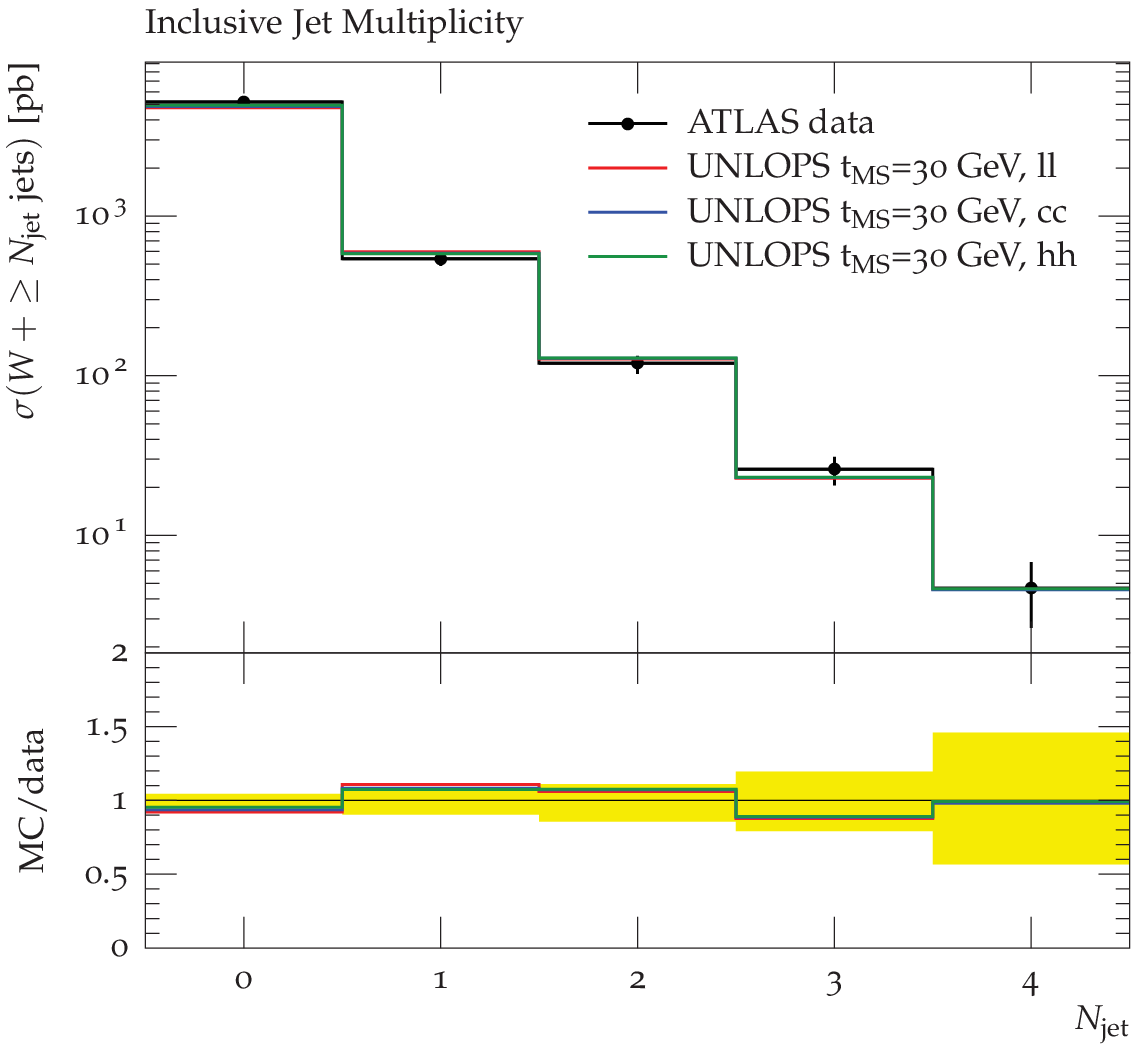}
  \end{minipage}
  \caption{\label{fig:atlas-w-njet-nlo-tmsvar}Jet multiplicity in
    $\W$-boson production, as measured by ATLAS \cite{Aad:2012en}. The
    MC results were obtained by merging up to two additional partons
    at LO, and zero and one additional parton at NLO. MC results are
    shown for three different merging scales (top panels) and for
    three different renormalisation/factorisation scales (bottom
    panels). Effects of multiple scatterings and hadronisation are
    included.  Left panels: Results of NL$^3$.  Right panels: Results
    of UNLOPS.}
}

In figure \ref{fig:atlas-w-njet-nlo-tmsvar}, we show that the jet
multiplicity is well under control in NLO merged predictions. The left
panel of Figure \ref{fig:atlas-w-pt-powheg} shows that, as expected,
it is not possible to describe the number of zero-jet events with a
$\W+$jet NLO calculation. This is of course exactly the strength of
merged calculations: Observables with different jet multiplicities can
be described in a single inclusive sample.

\FIGURE[t]{
  \centering
  \begin{minipage}{1.0\linewidth}
    \includegraphics[width=0.47\textwidth]{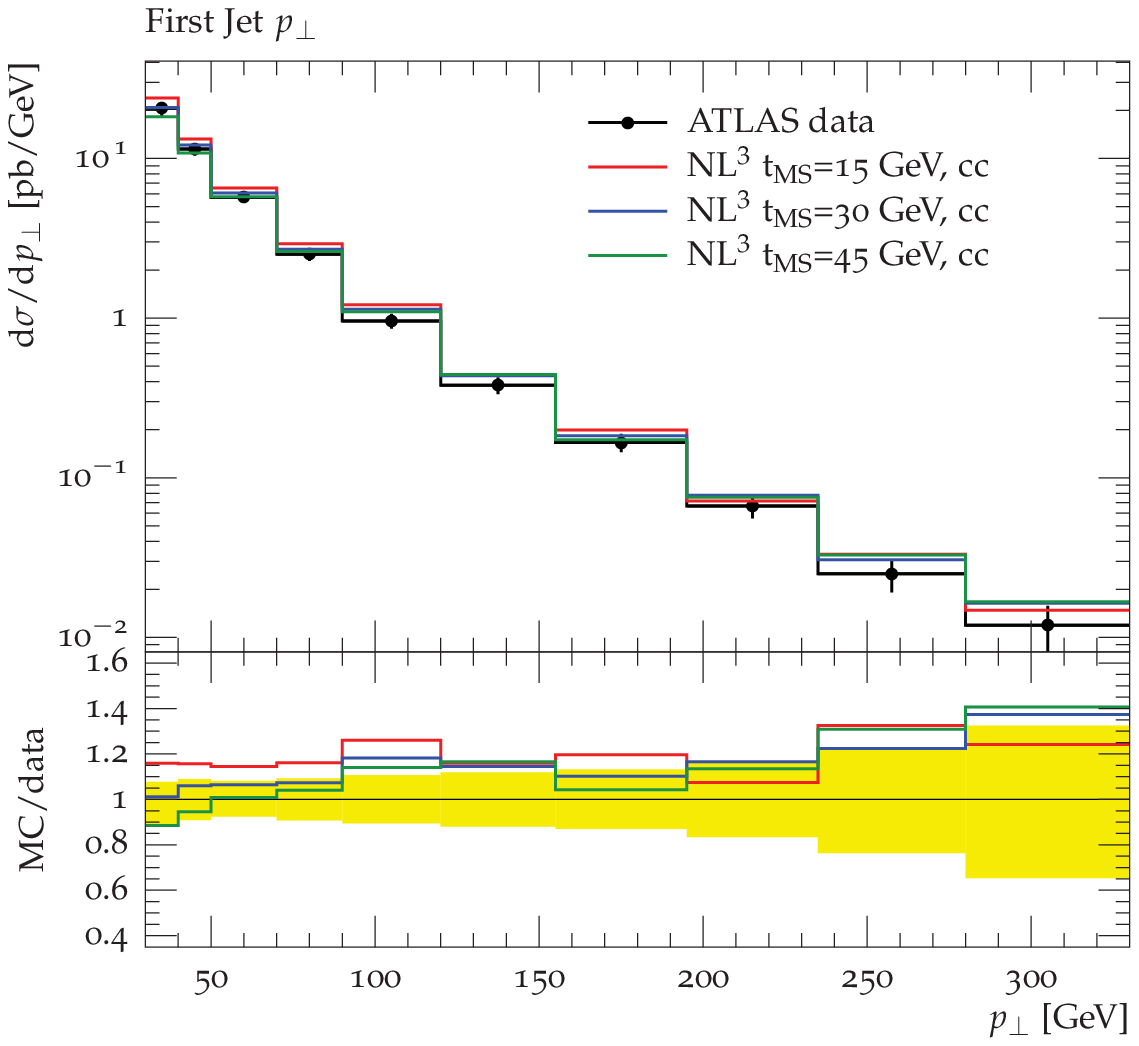}
    \includegraphics[width=0.47\textwidth]{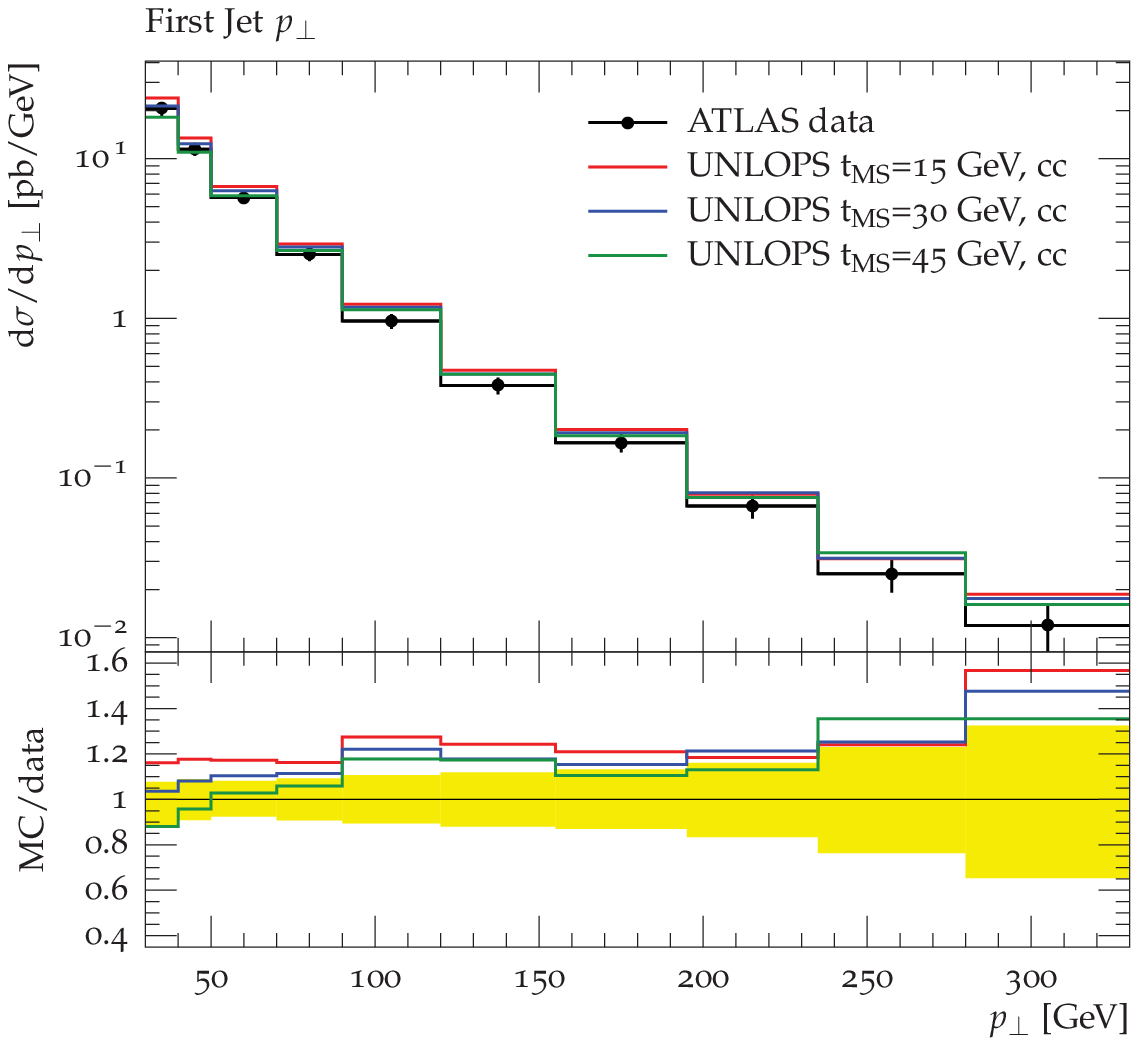}
    
    \includegraphics[width=0.47\textwidth]{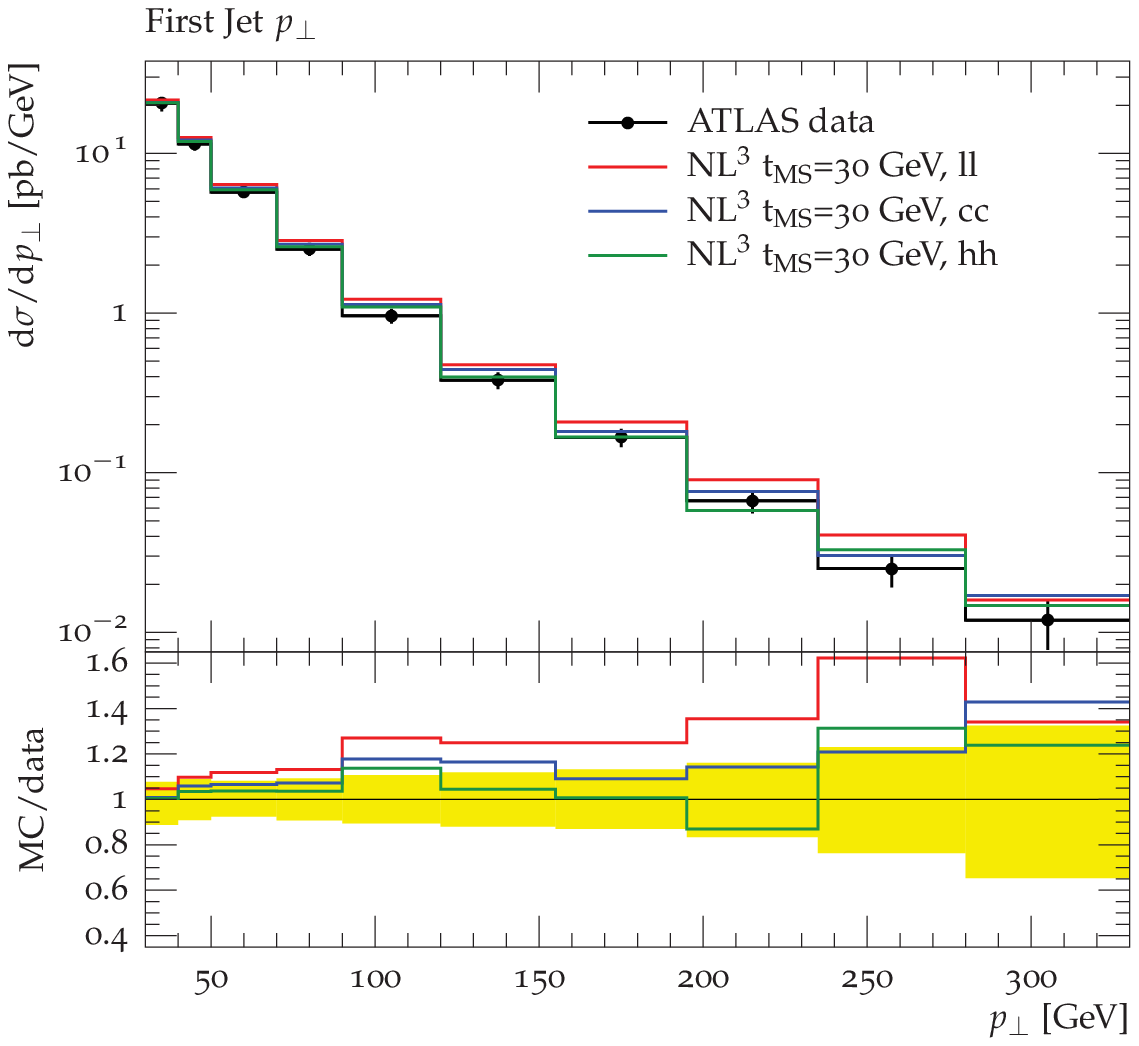}
    \includegraphics[width=0.47\textwidth]{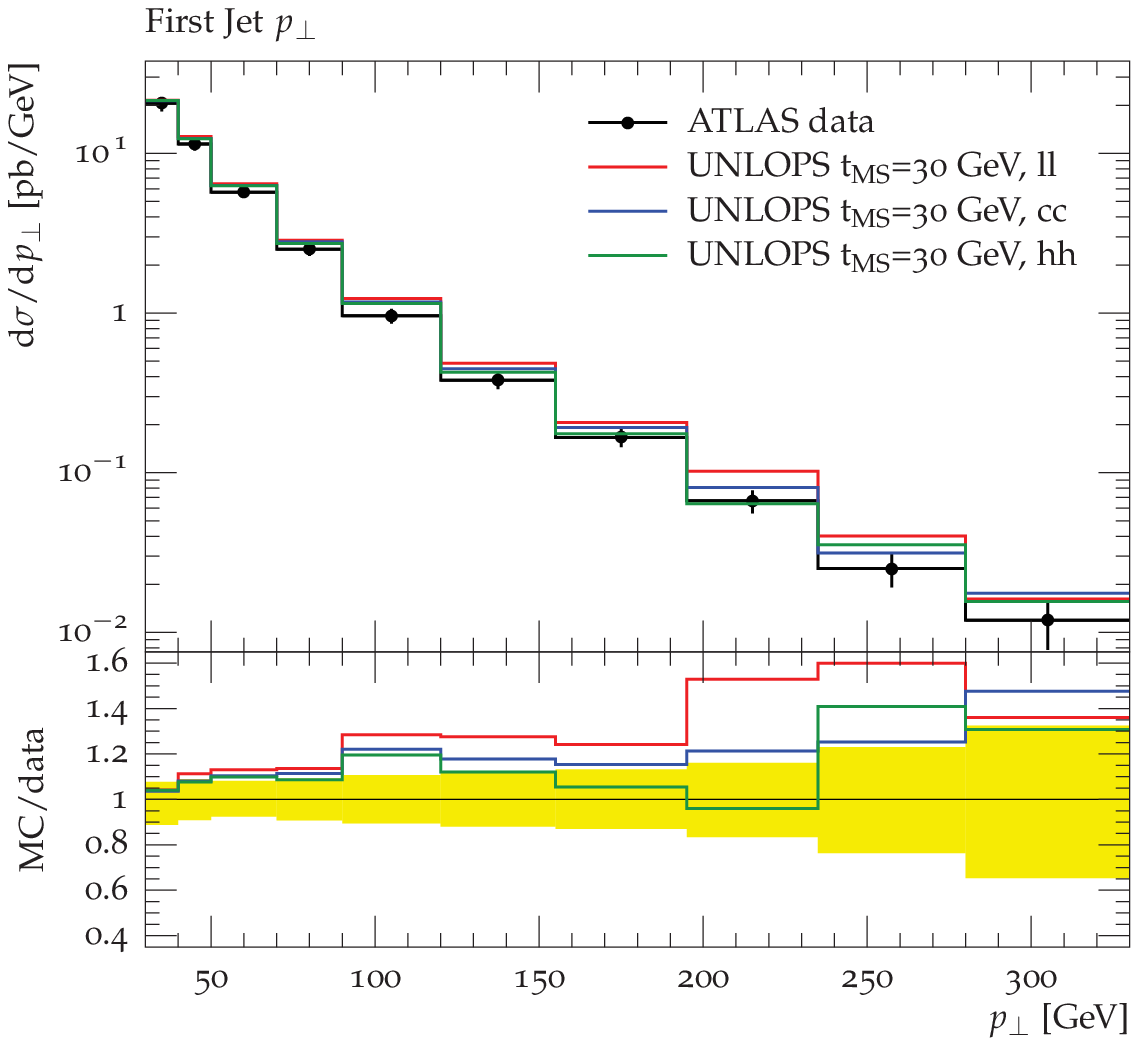}
  \end{minipage}
  \caption{\label{fig:atlas-w-pt-nlo-tmsvar}Transverse momentum of the
    hardest jet in $\W$-boson production, as measured by ATLAS
    \cite{Aad:2012en}. The MC results were obtained by merging up to
    two additional partons at LO, and zero and one additional parton
    at NLO. MC results are shown for three different merging scales
    (top panels) and for three different renormalisation/factorisation
    scales (bottom panels). Effects of multiple scatterings and
    hadronisation are included.  Left panels: Results of NL$^3$.
    Right panels: Results of UNLOPS.}
}

The transverse momentum of the hardest jet in association with a $\W$-boson
is shown in figure \ref{fig:atlas-w-pt-nlo-tmsvar}
and the right panel of Figure \ref{fig:atlas-w-pt-powheg}. It is clear that 
the NLO merged results do not agree with data. We have chosen this particular 
observable because it our exhibits the most unsatisfactory description of data
that we have encountered while testing our NLO merging methods. 
The reason for this disagreement is multifold. First,
we have already mentioned that correcting for inclusive NLO input produces 
harder $p_{\perp1}$ tails. The two-jet sample will eventually dominate the 
tail. We have chosen to rescale the two-jet contribution with a \Kf-factor 
above unity. It could also be argued that the \powhegbox result 
in Figure \ref{fig:atlas-w-pt-powheg} has slight tendency to overshoot. This might indicate that some
part of the ``giant \Kf-factor effect" due to enhancements of 
$\mathcal{O}\left(\as\ln \frac{p_{\perp1}^2}{\mw^2}\right)$ is developing in the 
$\W+$jet NLO calculation of $p_{\perp1}$ because of soft/collinear $\W$-bosons.
The last two points are correlated, since two-jet configurations
have a major impact on the $p_{\perp1}$-dependence of the NLO result, and 
increasing the two-jet contribution can enhance the visibility of giant
\Kf-factors. 

The NL$^3$ and UNLOPS descriptions of data exhibit high similarity. We have already 
noted the semblance of both methods in section 
\ref{sec:w-results}. This observation is specific to $\W$-boson production,
and does not hold for other processes, as for instance illustrated in section 
\ref{sec:h-results}.

\section{Discussion and conclusions}
\label{sec:conclusions}

In this article, we have presented two new methods for combining
multiple next-to-leading order calculations consistently with the
\pytppp parton shower. The NL$^3$ method is a generalisation of the
CKKW-L scheme, while the UNLOPS prescription accomplishes the same for
UMEPS. Both methods achieve a description of zero-, one-, \ldots, $n$-jet
observables simultaneously at NLO accuracy in one inclusive sample,
provided input event files at NLO accuracy for up to $n$ additional
jets are supplied. We would like to point the interested reader to
appendix \ref{sec:unlops-nnlo}, in which we argue out that it is
feasible to extend the UNLOPS method to a NNLO matching scheme.

Two distinct NLO merging schemes were presented to estimate the magnitude of
issues related to sub-leading logarithmic enhancements. 
Although the UNLOPS method can be considered theoretically preferable, no 
large differences between NL$^3$ and UNLOPS have been observed when merging
multiple NLO calculations for $\W$-boson production in association with jets.
This leads us to conclude that for the observables that
were investigated, and the merging scale values that were used, sub-leading 
logarithmic enhancements are sub-dominant.
For $\Higgs-$boson production, differences are visible, with UNLOPS 
delivering a more reliable solution.
 
This article is intended to give a comprehensive description of the 
choices that can be made in deriving and implementing an NLO merging method.
We hope that this publication provides enough information about the actual 
implementation to allow the reader to form clear judgements of the rather
intricate details.  We have tried to remain as general as possible in our 
choice of inputs. It has been shown that different inputs can, due to 
mismatches in phase space mappings, have visible, systematic effects. When 
confronted with such effects, it is clearly preferable to reach an agreement 
over inputs, and we hope that the current publication can contribute to a 
discussion.

We have shown that the merging scale dependence in $\W+$jets is small, and 
contained in the scale variation band of the $\W+$jet NLO calculation. This 
also means that the description of data is governed by the input NLO 
calculation. 

The merging scale dependence in Higgs-boson production in gluon fusion is 
very small. In this case, we highlighted that the dominant uncertainty of 
the algorithms is given by the choice of the \Kf-factor rescaling higher 
orders -- which is beyond the control of the NLO calculation. This is a 
manifestation of the magnitude of \Kf-factors in gluon fusion and the
scale variation of the cross sections. We would like to 
stress that this uncertainty is present because we try to be as general as 
possible, and that the introduction of \Kf-factors does in principle not 
jeopardise NLO accuracy, or degrade the PS approximation. However, if \Kf-factors
are not necessary and instead produce large variations, the removal of 
\Kf-factors should be considered.  

Although we have presented some comparisons to data in this article,
we do not attempt to make any definite predictions. To do this, a further
investigation of the uncertainties has to be performed -- a task we will
return to in future publications. We end this article by listing the
main issues that need to be addressed.

Our methods require events generated according to the exclusive NLO
cross section. There are currently no standard programs that will
produce such events, and instead we have used inclusive NLO cross
sections and subtracted explicit counter events by integrating
tree-level matrix element events over the radiative phase space, using
the mapping of the \pytppp parton shower. We have also ``hacked''
\powhegbox to directly produce the exclusive cross section event, and
have found some differences, due to the different phase space mapping
used there. Modifying the internals of other programs is, of course,
not a viable long-term solution, and we hope that the introduction of
our algorithm may inspire authors of NLO matrix element generators to include 
the generation of exclusive cross sections as an option in their programs.

We have allowed the use of \Kf-factors in the underlying tree-level
merging in the hope that the inclusion of NLO corrections will then
lead to less merging scale variations. Although this can be done
without modifying the formal accuracy of our methods, we see clear
differences compared to the case where \Kf-actors are omitted, in the
case of Higgs-production, where these \Kf-factors are large. We find
indications of reduced factorisation and renormalisation scale
uncertainties in the absence of \Kf-factor, but also note larger
merging scale variations in the NL$^3$ case. This needs to
be investigated further. Other options, e.g.\ including
multiplicity-dependent \Kf-factors, should also be considered to
understand uncertainties.

At very high transverse momenta we expect logarithms of the form
$\ln\left(\frac{p_{\perp \textnormal{jet}}}{\mw}\right)$ to arise,
resulting in so-called ``giant \Kf-factors''
\cite{Rubin:2010xp,Campanario:2012fk}. These logarithms can in principle be 
resummed to all orders, and an inclusion of such resummation is planned for 
the parton shower in \pytppp. We are confident that our methods can be
extended also to deal with this full electro-weak shower, but
meanwhile we need to understand better the uncertainties arising from
these logarithms.

Finally, before we can be confident enough to make precise predictions
with our new methods, a re-tuning of the shower (including MPI) of
\pytppp must be carried out. The currently available tunes have all
been obtained without higher order matrix elements merging, and it is
clear that some of the resulting parameters have been obtained from
trying to fit distributions where we do not expect an uncorrected
parton shower to do a reasonable job. In particular, this applies to
the tuning of the scale factor in \as\ (see \eqref{eq:asscaleb}) in
the shower, and we expect this to change significantly when tuning the
ME corrected shower. This will then also directly influence the MPI,
which also need to be re-tuned. Needless to say, such a tuning as a
major undertaking.

\section*{Note added}

While finishing this article, it came to our attention that an
approach that is similar to UNLOPS has been developed in parallel by
Pl\"{a}tzer \cite{Platzer:2012bs}. Also, on the day of submission, we
noted that Aioli et al.\ \cite{Alioli:2012fc} presented their work on
NLO-matching.

\section*{Acknowledgements}

Work supported in part by the Swedish research council (contracts
621-2009-4076 and 621-2010-3326). We are grateful to Nils Lavesson and
Oluseyi Latunde-Dada for collaboration at the early stages of
developing the NL$^3$ method. We would also like to thank Simon
Pl\"{a}tzer and Keith Hamilton for helpful discussions.


\section*{Appendices}
\begin{appendix}


\section{NLO prerequisites}
\label{sec:nlo-prereq}

This appendix is intended to introduce the merging scale definition used 
throughout this article (see section \ref{sec:pythia-evolution-pt}), discuss
the prerequisites on NLO input (section \ref{sec:exclusive-cross-sections}),
introduce the notation we employ (section \ref{sec:notation}) and finally 
illustrate how the \powhegbox program can be used to generate the input 
necessary for NLO merging (section \ref{sec:powheg-usage}).

\subsection{The Pythia jet algorithm}
\label{sec:pythia-evolution-pt}

Throughout this paper, we use cuts on the minimal \pytppp evolution 
$p_{\perp,evol}$, to disentangle regions of phase space. Since 
$p_{\perp,evol}$ defines a relative $p_\perp$ distance \cite{Sjostrand:2004ef}, we think of 
$p_{\perp,evol}$ as an inter-parton separation criterion. To avoid confusion
with other $p_\perp-$definitions, we will use the symbol $\ord$ for 
$p_{\perp,evol}$.

The phase space regions in which we believe fixed-order calculations to 
dominate is separated from the resummation region by a cut value $\ordms$, 
defined in a parton separation $t \sim \min\{\ord\}$. 
This minimal 
separation is constructed by finding the minimal $\ord$ for any triplet of 
partons 
$e,r,s$, where $e$ is a final state ``emitted" parton, $r$ is a radiating
parton and $s$ is a spectator. All triplets, irrespectively of flavour 
(or colour) constraints are included. In a dipole picture, the radiator $r$
can be thought of as the dipole end whose momentum changes most when splitting
the dipole $(r',s')$ into two dipoles $(r,e),~(e,s)$  while $s$ is the dipole 
end that absorbs the (small) recoil. The functional definition of this parton
separation criterion is
\begin{align}
\label{eq:min-pythia-pt-def}
&    t ~=~ \min
        \left[ \ord_{\{i,j,k\}} \right]
    \qquad
    \textnormal{where $i \in$ \{final state partons\},}\\
& \qquad\qquad\qquad\qquad\qquad\!
    \textnormal{and $j \in$ \{final and initial state partons\},}\nonumber\\
& \qquad\qquad\qquad\qquad\qquad\!
    \textnormal{and $k \in$ \{final and initial state partons\}}\nonumber
\end{align}
where the separation of $i$ for a fixed triplet $(i,j,k)$ of partons with 
momenta $p_i,p_j,p_k$ is
\begin{eqnarray}
\label{eq:pythia-pt-def}
    \ord_{ijk} ~=~ p^2_{\perp,evol,ijk} ~=~
    \begin{cases}
      z_{ijk}(1-z_{ijk})Q_{ij}^2 &
         \textnormal{if the radiator $j$ is a final state parton, and}\\
      &  Q_{ij}^2 = (p_i + p_j)^2 
         ~, \quad
         z_{ijk} = \frac{x_{i,jk}}{x_{i,jk}+x_{j,ik}} \\
      &  x_{i,jk} = \frac{2 p_i(p_i+p_j+p_k)}{(p_i+p_j+p_k)^2}\\
      (1-z_{ijk})Q_{ij}^2 &
         \textnormal{if the radiator $j$ is an initial state parton, and}\\
      &  Q_{ij}^2 = -(p_i - p_j)^2 
         ~, \quad
         z_{ijk} = \frac{(p_i - p_j + p_k)^2}{(p_i+p_k)^2}\\
    \end{cases}
\end{eqnarray}
The cut value $\ordms$ is called merging scale. If all $\ord_{ijk}$ for a 
particular final state parton $i$ are larger than $\ordms$, we call this 
parton a resolved jet. Conversely, if any $\ord_{ijk}$ is below $\ordms$, we
call $i$ an unresolved jet.
We say that a phase space point is in the matrix element region if 
$t > \ordms$, \ie\ all minimal parton separations are larger than the cut. In
other words, a phase space point is in the matrix element region if it only
contains resolved jets.
The parton shower region is disjoint: If any jet separation falls below
$\ordms$, be believe that parton shower resummation is appropriate. 

Using \eqref{eq:min-pythia-pt-def} and \eqref{eq:pythia-pt-def} as merging 
scale definition does not exactly correspond to separating the matrix 
element- and parton shower regions in $p_{\perp,evol}$.
In parton shower algorithms, the resolution scale attributed to a state is
given by the scale of the last splitting. This is just one number, since a 
splitting is generated by a winner-takes-it all strategy: If a splitting is
chosen, all scales attributed to splittings of other partons are considered 
higher. Such a merging scale definition can only be constructed if we know
(all) parton shower histories of an input event.

For now, we use \eqref{eq:min-pythia-pt-def} and \eqref{eq:pythia-pt-def} as 
merging scale definitions, and are content with the fact that $\ordms$ does 
still correspond to a single $p_{\perp,evol}-$value. This means that vetoing
shower emissions that would result in an additional resolved jet will not 
introduce no-emission probabilities above $\ordms$.

We have to point out that fixing the merging scale 
definition is necessary in the NLO merging methods illustrated in this 
article. Otherwise, it would be mandatory to reweight NLO corrections with 
no-emission probabilities for merging-scale-unordered emissions, which would
fundamentally degrade the higher-order description we aim to achieve\footnote{
See appendix \ref{sec:nl3-derivation} for details}. One benefit of CKKW-L 
tree-level merging is that the method allows for a wide class of merging 
scale definitions. Because of the treatment of emissions that are 
unordered in the merging scale, however, the merging scale effectively has to 
define a hardness-measure, since otherwise, only small portions of phase space 
will be endowed with ME corrections \cite{Lonnblad:2011xx}. Different choices
of hardness definitions for different processes in CKKW-L can be helpful 
for efficiently correcting phase space. The current implementation of CKKW-L
in \pytppp allows for both $\min\{\ord\}$ and $\min\{k_\perp\}$ as merging
scales. No major efficacy differences between these merging scale 
definitions has been found so far, leading us to conclude that in practise, 
defining the merging scale in $\min\{\ord\}$ is reasonable.


\subsection{Exclusive cross sections}
\label{sec:exclusive-cross-sections}

In this section, we would like to introduce the concepts of exclusive and 
inclusive NLO cross sections, and comment on how inclusive NLO cross sections
can be made exclusive by the inclusion of a phase space subtraction sample. 

\FIGURE{
\centering
  \includegraphics[width=0.47\textwidth]{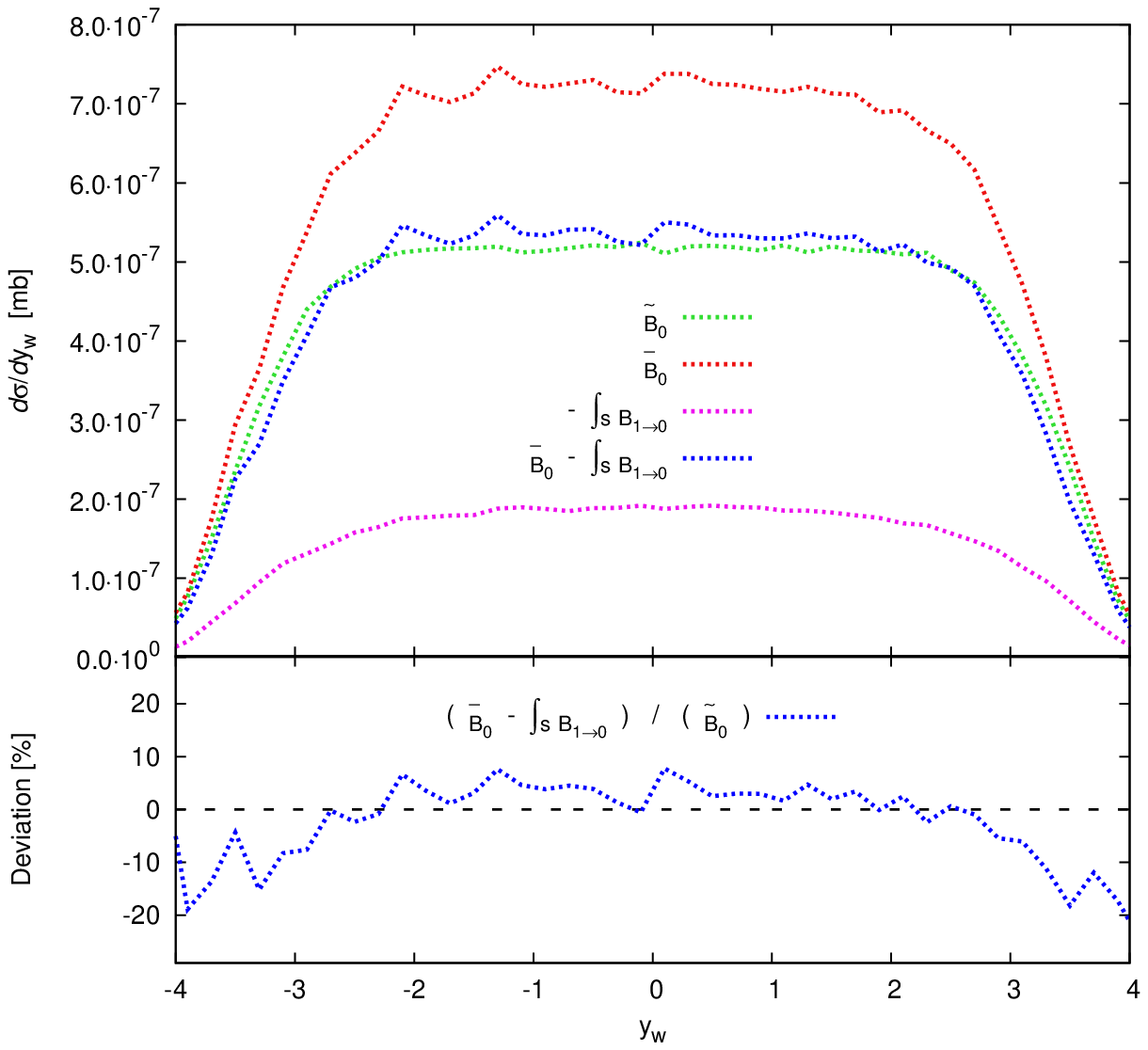}
  \includegraphics[width=0.47\textwidth]{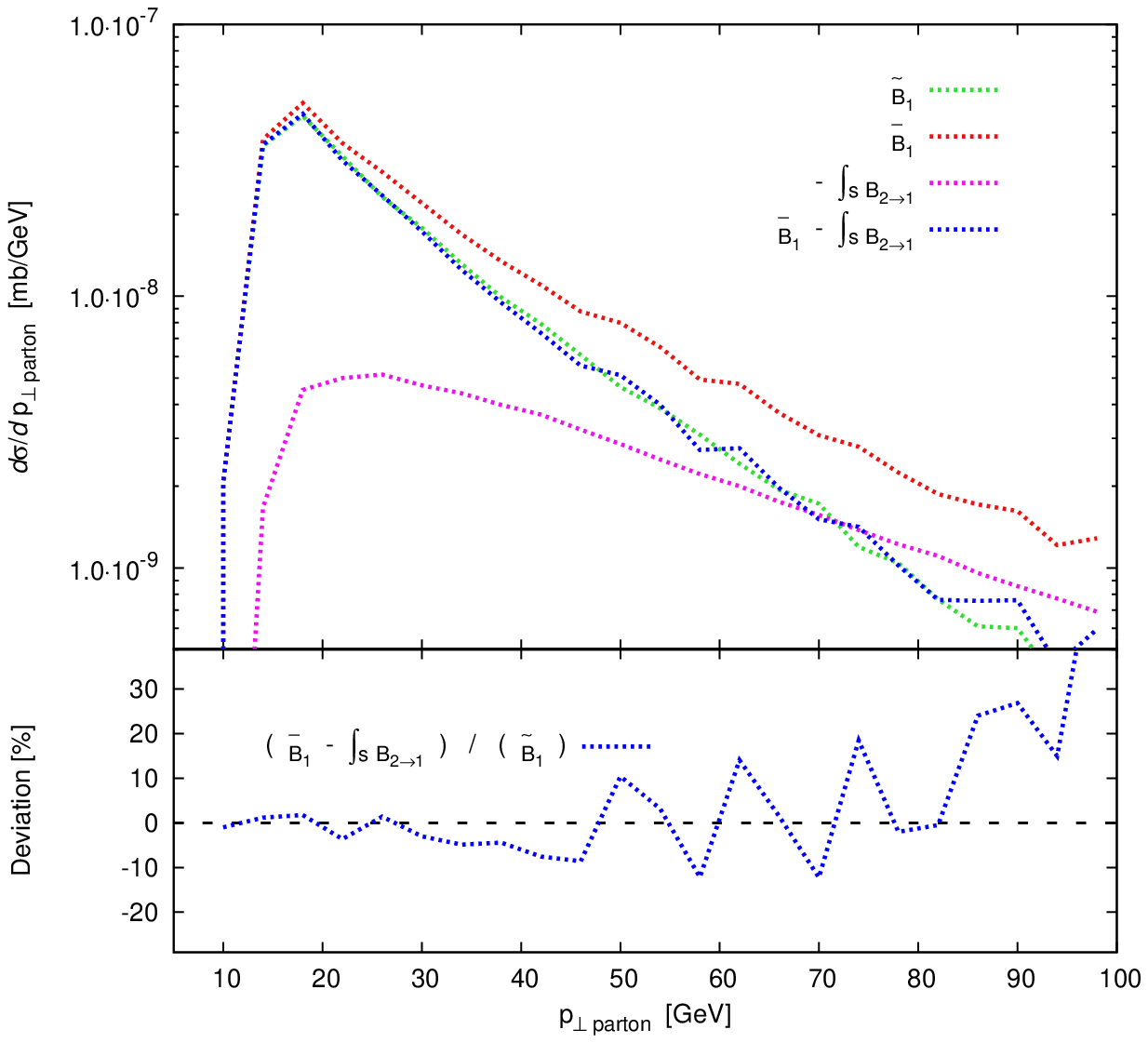}
\caption{\label{fig:exc-xsec-btilde}Comparisons of two way of generating 
  exclusive NLO cross sections. $\Bbarev{i}$ events are 
  calculated without changing the \powhegbox program, while for $\Btilev{i}$,
  we have explicitly introduced the necessary cuts in \powhegbox.
  Left panel: Comparison for zero-jet exclusive NLO cross section, as function
  of the $\W$-boson rapidity. Right panel: Comparison for one-jet exclusive 
  NLO cross section, as function the kinematical transverse momentum of the
  parton.
  }
}

We think of matrix element merging as a two-step measurement.
We first measure the number of resolved jets in the input event, by applying 
a cut. Then, we calculate an interesting observable on events that have been
classified as $n-$jet events. To make the second step independent of 
the choices in the first measurement, we need to sum over all possible jet
multiplicities.

Throughout this publication, we will define jets by the \pytppp evolution 
$p_{\perp}$ jet separation criterion, as discussed in
appendix \ref{sec:pythia-evolution-pt}. Resolved jets are defined as partons
whose separation to any other partons in the state is above the value
$\ordms$. We call $\ordms$ the merging scale.

The output of a tree-level calculations for $k$ final partons can contain
$l=0,\dots,k$ partons with soft or collinear momenta. The result of the 
calculation will diverge as soon as any parton in the calculation becomes 
soft or collinear. We can remove these regions of phase space, if we 
enforce that the output only contains exactly $k$ partons with jet separation 
above $\ordms$, \ie\ $k$ resolved jets. If the jet definition is infrared and
collinear safe, this jet cut will render tree-level calculations finite.

The result of using a next-to-leading order calculation for $k$ partons to 
describe an observable $\mathcal{O}$ can schematically be written as
\begin{eqnarray}
\label{eq:nlo-calculation-fluffy}\!\!\!
\langle \mathcal{O} \rangle
&=&
      \int d^{k}\phi~ \mathcal{O}(\phi_k)~ d\sigma_{Born}
    + \int d^{k}\phi~ \mathcal{O}(\phi_k)~ d\sigma_{Virtual}
    + \int d^{k+1}\phi~ \mathcal{O}(\phi_{k+1})~ d\sigma_{Real}~,\qquad
\end{eqnarray}
where $\int d^{k}\phi$ indicates an integration over the $k-$parton phase space,
$d\sigma_{Born}$ is the tree-level cross section, $d\sigma_{Virtual}$ the 
virtual correction term and $d\sigma_{Real}$ the real emission part.
Equation \ref{eq:nlo-calculation-fluffy} allows contributions of any number 
of $l=0,\dots,k+1$ resolved (or unresolved) jets. Since the tree-level part 
still diverges if any of the $k$ partons approach the soft and collinear 
regions, we require that 
$\int_k \mathcal{O}(\phi_k) d\sigma_{Born}$ always contains exactly $k$ 
resolved jets, which immediately means having the same requirement in $\int_k 
\mathcal{O}(\phi_k) d\sigma_{Virtual}$. This then means that $\int_{k+1} 
\mathcal{O}(\phi_{k+1}) d\sigma_{Real}$ has to be constrained, since 
otherwise, the NLO calculation could include real-emission corrections to
non-existent ``underlying" Born configurations with less than $k$ resolved 
jets. The \powheg method \cite{Nason:2004rx,Frixione:2007vw} eliminates this issue by evaluating using one phase
space point $\overline{\phi}_n$ for tree-level and virtual parts, and 
$\phi_{n+1}$ phase space points that can be projected exactly onto 
$\overline{\phi}_n$ in real-emission terms. We will assume that in neither of
the terms in \ref{eq:nlo-calculation-fluffy}, less than $k$ resolved jets are
included. Further, the observable $\mathcal{O}$ receives -- through real 
corrections -- contributions from $k+1$ (resolved or unresolved) jets. 
Measurements explicitly depending on the kinematics of $k+1$ resolved jets 
are only accurate to tree-level approximation, while contributions for an
unresolved additional jet are necessary to cancel divergences\footnote{In
numerical implementations, the singularities of virtual corrections and 
real-emission contributions are cancelled separately by regularisation terms.}

In multi-jet merging, $(k+1)-$jet contributions enter through explicitly 
adding a reweighted $(k+1)-$jet sample. To merge multiple NLO calculations, we
also need a clean cut, that makes the classification of the input in 
terms of jet multiplicities possible. We define that the ``NLO part" of an
$k-$jet NLO calculation should contain $k$ resolved jets, and at most one 
unresolved jet, while contributions with $k+1$ resolved jets are regarded
leading-order parts
\begin{eqnarray}
\label{eq:nlo-calculation-fluffy-again}
\langle \mathcal{O} \rangle
&=&
   \underbrace{
      \int d^{k}\phi~ \mathcal{O}(\phi_k)~
       \left\{
     ~ d\sigma_{Born}
    ~+~  d\sigma_{Virtual}
    ~+~  \int^{\ordms}d\phi~ d\sigma_{Real} \right\}
    }_{\textnormal{NLO part}}\\
&&   +
   \underbrace{
   \int_{\ordms} d^{k+1}\phi~ \mathcal{O}(\phi_{k+1})~ d\sigma_{Real}
    }_{\textnormal{LO part}} \nonumber
\end{eqnarray}
We call the NLO part of such a calculation the \emph{exclusive} NLO cross 
section. 

Merging schemes naturally act on exclusive cross sections. Technically, we 
assume that all unresolved real emission parts are projected onto $n$-parton
phase space points. In \powheg, this is facilitated by performing the 
$d\phi$ integration in the real-emission term explicitly. If an NLO 
calculation would yield only phase space points with exactly $n$ partons in
$n$ resolved jets, and weight these points with\footnote{We give more 
precise definitions of the notation in the next section.}
\begin{eqnarray}
\Btilev{n} = d\sigma_{Born}
    ~+~  d\sigma_{Virtual}
    ~+~  \int^{\ordms}d\phi~ d\sigma_{Real}~,
\end{eqnarray}
then this calculation could be used immediately for NLO merging. However, 
$\Btilev{n}$ might not be accessible without changing the NLO matrix element
generator, since it might not possible to split the calculation into 
``NLO parts" and ``LO parts" as desired. We instead choose to write
\begin{eqnarray}
\Btilev{n} = d\sigma_{Born}
    ~+~  d\sigma_{Virtual}
    ~+~  \int d\phi~ d\sigma_{Real}
    ~-~  \int_{\ordms} d\phi~ d\sigma_{Real}
\end{eqnarray}
and to use the two samples
\begin{eqnarray}
\Bbarev{n} &=& ~d\sigma_{Born}
    ~+~  d\sigma_{Virtual}
    ~+~  \int d\phi~ d\sigma_{Real}\\
- \int_s \Bornev{n+1\rightarrow n} &=& ~-~  \int_{\ordms} d\phi~ d\sigma_{Real}
\end{eqnarray}
separately. We call $\Bbarev{n}$ the inclusive NLO cross section, and
refer to $\int_s \Bornev{n+1\rightarrow n}$ as phase space subtraction term. 
Adding the inclusive NLO cross section and the subtraction term, we retrieve 
the exclusive NLO cross section. It is possible to formulate NLO merging
acting on exclusive cross sections, or acting on inclusive cross
sections and additional phase space subtraction samples.

We would like to comment on our framework to generate exclusive 
cross sections from inclusive NLO input. It has already been demonstrated in 
\cite{Lonnblad:2012ng} that extracting integrated states that had been 
constructed as intermediate states in the parton shower history does indeed
give the expected results (see Figure 2 of \cite{Lonnblad:2012ng}). However, 
these comparisons were performed between the shower approximation and the
integrated matrix element. To correct for the use of an inclusive NLO cross
section, we have to use integrated tree-level events as phase space 
subtraction for \powhegbox events. Figure \ref{fig:exc-xsec-btilde}
shows how the exclusive cross section generated through explicitly changing
the \powhegbox program code compares to the exclusive cross section produced
by a-posteriori phase space subtraction. The differences between the two
prescriptions stems from a different phase space mapping in \powhegbox and
\pytppp. Such effects are beyond the accuracy of the NLO merging
methods presented in this article. It should be noted that an
a-posteriori phase space subtraction using the \pytppp phase space mapping 
produces a harder $p_\perp$-spectrum and less forward $\W$-bosons. Ideally, we
would like generate the exclusive NLO cross section with an NLO generator 
which allows for the necessary cuts, so that explicit phase space subtraction
become unnecessary. For this publication however, we will use subtractions,
since this allows us to perform merging scale variations without continuously
having to re-generate LHEF output with the \powhegbox program.


\subsection{Notation}
\label{sec:notation}

Formulating NLO merging is unfortunately a fairly notation-heavy task.
In this section, we would like to carefully introduce the symbols we will use 
throughout this article in a tabular style, for easy reference. Let us define 
the lingo

\vspace*{1ex}
\begin{longtable}{p{0.3\textwidth}p{0.65\textwidth}}
$\phi_{n}$: &     Phase space point with $n$ additional resolved jets.
                  This means that $\phi_{n}$ can contain $p$ 
                  lowest-multiplicity particles (\eg\ $\eplus$ and $\eminus$
                  for Drell-Yan production), and $n$ additional partons.
                  Each phase space point has a fixed momentum, flavour and
                  colour configuration.\\
Underlying 
configuration: &   Phase space point $\phi_n$ which can be to constructed from
                  $\phi_{n+1}$ by removing one emission, meaning integrating 
                  over the one-particle phase space of the emission, and 
                  recombining flavours and colours. 
                  We will use the terms underlying momentum configuration,
                  underlying flavour configuration and underlying colour
                  configuration when explicitly emphasising one aspect of
                  the underlying configuration.\\
$\Phi_{\textnormal{rad}}:$
              &   The radiative phase space, meaning that in
\begin{eqnarray}
\phi_{n+1} \approx \phi_n \Phi_{\textnormal{rad}}
\end{eqnarray}
                  $\Phi_{\textnormal{rad}}$ plays the role of the one-particle
                  phase space of the additional parton, while $\phi_n$ is the
                  underlying configuration.\\
Resolved jet: &    Parton, for which all possible jet separations to other 
                  partons in this phase space point are larger than the value
                  $\ordms$.\\
Unresolved jet: &  Parton, for which at least one jet separations to one other 
                  partons in this phase space point is lower than the value
                  $\ordms$. 
\end{longtable}

\noindent
We always assume that matrix elements for $n$ outgoing partons are only 
integrated over phase space regions with exactly $n$ resolved jets, unless 
explicitly stated otherwise. In cases where we integrate over emissions, we
assume that in the ME input, all partons were resolved jets, and that after
the integration, all partons are resolved jets.\\

\noindent
Please note that the methods presented in this publication do not rely on a 
particular regularisation scheme in the NLO calculation, as long as the 
dependence on the regularisation scheme is cancelled locally, \ie\ in the 
weight of each phase space points separately. We will choose a rather symbolic 
notation for parts of NLO calculations, and hope that this will make the 
formulae in this article more accessible. In the following, we would like to 
introduce the shorthands:

\vspace*{1ex}
\begin{longtable}{p{0.1\textwidth}p{0.83\textwidth}}

$\Bornev{n}$: &   Tree-level matrix element with $n$ partons, \ie\ 
\begin{eqnarray}
\Bornev{n} &=& f_n^+(x_n^+,\muf)f_n^-(x_n^-,\muf) \mesqof{n,0}{\muf,\mur}~,
\end{eqnarray}
                  where the first subscript on $\mesqof{n,0}{\muf,\mur}$ 
                  indicates the number of jets, while the second counts loop 
                  integrations. Readers more familiar with the
                  notation of \cite{Frixione:2007vw} should note
\begin{eqnarray*}
\Bornev{n} &=& \left[~ B(\Phi_{n})~ \right]_{f_b}
\end{eqnarray*}
                  For brevity, we always suppress flavour 
                  indication on $\Bornev{n}$. \\
$\Bornev{n+1|n}$: & Sum of tree-level configurations with $n+1$ partons, for
                  which the underlying Born configuration is $\phi_n$.
                  Loosely, we think of $\Bornev{n+1|n}$ as the sum of 
                  matrix elements $\Bornev{n}$, multiplied by splitting 
                  kernels (with $\Bornev{n}$ being evaluated using the 
                  underlying momentum, flavour and colour configuration, while
                  the splitting kernels depend on the radiative phase space).
                  Translating to the notation of \cite{Frixione:2007vw}, this 
                  means
\begin{eqnarray*}
 \int \drad \Bornev{n+1|n}
 ~=~ \sum_{\alpha_r \in \{\alpha_r|f_b\}}~
     \int~ \left[~ d\Phi_{rad}~ R(\Phi_{n+1})~ \right]^{\overline{\Phi}^{\alpha_r}_n = \Phi_n}_{\alpha_r}
\end{eqnarray*}
                  We choose the more symbolic indexing with
                  $n+1|n$, rather than the more rigorous $\alpha_r$-notation
                  of \cite{Frixione:2007vw}, in order to not obfuscate 
                  formulae with details that are not essential for the 
                  discussion.\\
$\int_s\Bornev{n\rightarrow m}$: & Sum of tree-level cross sections with
                  $n$ resolved jets in the input ME events,
                  after integration over the phase space of $n-m$ partons.        
                  Symbolically:
\begin{eqnarray}
 \int_s\Bornev{n\rightarrow m} ~=~ \sum_n \int_s d^{n-m}\phi~\Bornev{n|n-m}
\end{eqnarray}
                  We think of $\phi_n$ being produced from $\phi_{n-m}$ by $m$
                  consecutive splittings. The index $s$ denotes that the
                  integration is accomplished by explicitly removing $m$ 
                  partons from the $n-$parton phase space, meaning that we 
                  substitute the state $\state{n}$ by the state 
                  $\state{n-m}$, as introduced in \cite{Lonnblad:2012ng}.
                  The symbol $\int_s\Bornev{n\rightarrow m}$ also indicates 
                  that the state $\state{m}$ only has resolved jets, and that
                  possibly more than one 
                  integrations had to be performed in case all of the states 
                  $\state{n-1},\dots,\state{m+1}$ contained unresolved jets.\\
$\int_s\Bornev{n\rightarrow m}^{\uparrow}$: & Sum of tree-level cross 
                  sections with $n$ resolved jets in the input ME events,
                  after integration over the phase space of $n-m$ partons.
                  However, in contrast to the symbol 
                  $\int_s\Bornev{n\rightarrow m}$, we explicitly require the 
                  states $\state{n-1}$ to contain $n-1$ resolved jets, and 
                  still perform this second integration.
                  This is indicated by the upward-pointing arrow.  
                  All further integrations only have to be performed because
                  the states $\state{n-2},\dots,\state{m+1}$ contained 
                  unresolved jets.
                  $\int_s\Bornev{3\rightarrow 0}^{\uparrow}$ for example
                  means that we first replace the state $\state{3}$ by
                  $\state{2}$ (with two resolved jets), then demand another 
                  integration, giving $\state{1}$. Then, we
                  find that $\state{1}$ contains an unresolved jet, so that
                  we integrate once more. The last step would not be necessary
                  for the term $\int_s\Bornev{3\rightarrow 1}^{\uparrow}$.\\
$\Virtev{n}$: &   Virtual correction matrix element with $n$ partons above 
                  $\ordms$:
\begin{eqnarray}
\Virtev{n} &=& f_n^+(x_n^+,\muf)f_n^-(x_n^-,\muf) \mesqof{n,1}{\muf,\mur}~,
\end{eqnarray}
                  We assume that all ultraviolet divergences have already been
                  removed.\\
$\Dipev{n+1|n}$:  & Sum of infrared regularisation terms with $n$ partons 
                  above $\ordms$. As above, we indicate these terms can be
                  projected onto underlying Born configurations by the index 
                  $n+1|n$. For simplicity, we may think of these regulators as 
                  Catani-Seymour dipoles, and \emph{very} symbolically put
\begin{eqnarray}
\Dipev{n+1|n} &\sim& \sum_{i'j'k} 
                     f_n^+(x_n^+,\muf)f_n^-(x_n^-,\muf)
                     \mesqof{n,0}{\muf,\mur,\bar{\phi_n}}\\
&&\qquad \qquad
                     \otimes~ D_{ij\rightarrow i'j'k}
                     \left( \Phi_{\textnormal{rad}} \right)\nonumber ~,
\end{eqnarray}
                  where $i$ and $j$ are partons of the underlying
                  configuration $\bar{\phi_n}$, while $i',j'$ and $k$ are 
                  partons of $\phi_{n+1}$.
                  As long as all dependence on the regularisation is 
                  contained in the inclusive (exclusive) NLO cross sections,
                  our method will not depend on the actual form of
                  these terms -- all numerical NLO subtraction schemes are 
                  equally valid. The notation $n+1|n$ is to be understood 
                  in the same way as for $\Bornev{n+1|n}$.\\
$\Insev{n+1|n}$:  & Sum of integrated infrared regularisation terms with 
                  $n$ partons above $\ordms$. Remainders due
                  to initial state partons being collinear with identified
                  initial hadrons are included in $\Insev{n+1|n}$. 
                  As above, we indicate that the terms in $\Insev{n+1|n}$ can 
                  be projected unto underlying configurations by the index 
                  $n+1|n$. Schematically
\begin{eqnarray}
 \Insev{n+1|n} &\sim& \int \drad \Dipev{n+1|n}
\end{eqnarray}
                  In this case, the one-parton phase space integration is 
                  commonly performed analytically. The integration here 
                  covers the complete radiative phase space.
                  Integration variables can change for different types of
                  dipoles $D_{ij\rightarrow i'j'k}$.\\
$\Bbarev{n}$: &   Inclusive NLO weight of $n-$parton phase space points.

\begin{eqnarray}
\Bbarev{n} &=&
         \Bornev{n} + \Virtev{n} + \Insev{n+1|n}\nonumber\\
   &&  + \int \drad
         \left( \Bornev{n+1|n} - \Dipev{n+1|n} \right) ~.
\end{eqnarray}
                 Note that we will assume that this gives a NLO weight for
                 the phase space point $\phi_n$, meaning that we have to 
                 evaluate $\Dipev{n+1|n}$ and $\Insev{n+1|n}$ with $\phi_n$ 
                 rather than $\bar{\phi_n}$.
                 This is the standard procedure in
                 the \powheg and MC@NLO methods.
                 The integration over $\Phi_{\textnormal{rad}}$ covers the
                 full radiative phase space.
                 Real emission terms can give contributions with an additional
                 resolved jet, which are here included into the weight of 
                 $n$-jet phase space points $\phi_n$. As discussed above, the
                 description of an observable at NLO receives $n$-jet and
                 a $n+1$-jet contributions, so that projecting all 
                 $n+1$-jet configurations onto $\phi_n$ (and leaving no
                 $n+1$ events) seems problematic. However, in matrix-element
                 merging, $n+1$-jet events will be included though the 
                 next-higher multiplicity sample. We then need to ensure that
                 the contribution of resolved $n+1$-jet events to the cross 
                 section is not double-counted. This is solved by the 
                 introduction of exclusive NLO jet cross sections.
                 The inclusive cross section is closely related to the 
                 definition of 
                 $\Bbar$ in the \powheg method. Indeed, if no Sudakov factors 
                 are applied in the weight of Born-type phase space points 
                 in \powheg, and all radiative events are projected unto
                 Born configurations, this exactly produces our definition of 
                 $\Bbarev{n}$.\\
$\Btilev{n}$: &  Exclusive NLO weight of $n-$parton phase space points,
\begin{eqnarray}
\Btilev{n} &=&
         \Bornev{n} + \Virtev{n} + \Insev{n+1|n} \nonumber\\
   &&  + \int^{\ordms}\!\!\!\! \drad
         \left( \Bornev{n+1|n} - \Dipev{n+1|n} \right)\nonumber\\
 &=& \Bbarev{n}
    - \int_{\ordms} \drad
      \Bornev{n+1|n}\nonumber\\
 &=& \Bbarev{n} - \int_s\Bornev{n+1\rightarrow n}~.\label{eq:btilde-explicit}
\end{eqnarray}
                 It is clear from the last equality that an 
                 exclusive NLO $n-$jet cross section can be constructed
                 from the inclusive case by explicitly subtracting the phase 
                 space points with an additional resolved jet.\\
$\int_s\Bbarev{n\rightarrow m}$: & Inclusive NLO cross section with
               $n$ resolved jets in the input ME events,
               after integration over the phase space of $n-m$ partons.
               The symbol $\int_s\Bbarev{n\rightarrow m}$ as always also
               indicates that more than one 
               integrations had to be performed because all of the states 
               $\state{n-1},\dots,\state{m+1}$ contained unresolved jets.\\
$\int_s\Btilev{n\rightarrow m}$: & Exclusive NLO cross sections with
               $n$ resolved jets in the input ME events,
               after integration over the phase space of $n-m$ partons.
               The symbol $\int_s\Btilev{n\rightarrow m}$ as always also
               indicates that more than one 
               integrations had to be performed because all of the states 
               $\state{n-1},\dots,\state{m+1}$ contained unresolved jets.
\end{longtable}
\vspace*{1ex}

\noindent
We will use several different event samples as input for multi-jet merging. An 
event consists of a phase space point with an associated weight, and can thus 
be considered completely differential. Only if necessary will we talk about 
predictions for an observable. We will use ``cross section" and ``event" 
interchangeably, and also make little distinction between the terms 
``(phase space) weight" and ``matrix element".

Exclusive cross sections are the basic building blocks needed for multi-jet
merging. Tree-level merging uses phase space points weighted with 
exclusive tree-level matrix elements $\Bornev{n}$ as input. The NLO multi-jet
merging prescriptions advocated in this publication analogously require 
phase space points weighted by exclusive NLO weights as input. If no exclusive
calculation is available, it is possible to extend the algorithm to include 
the explicit subtraction of \eqref{eq:btilde-explicit}.
To make formulae for NLO merging a bit more transparent, let us introduce the 
short-hands 

\vspace*{1ex}
\begin{longtable}{lp{0.8\textwidth}}
$\Tev{n}$:    & UMEPS-reweighted tree-level cross sections with $n$ resolved 
                jets in the input ME events.\\
$\Iev{n}{m}$: & UMEPS-reweighted tree-level cross sections with
               $n$ resolved jets in the input ME events,
               after integration over the phase space of $n-m$ partons. 
               The symbol $\Iev{n}{m}$ also indicates that more than one 
               integrations had to be performed because all of the states 
               $\state{n-1},\dots,\state{m+1}$ contained unresolved jets.\\
$\termX{A}{-a,b}$: & Contribution $A$, with terms of powers $\as^a$ and $\as^b$ 
               removed. The resulting terms are calculated with fixed scales
               $\mur$ and $\muf$.\\
$\termX{A}{c,d}$: & Contribution $A$, with only terms of power $\as^c$ and 
               $\as^d$ retained, calculated with fixed scales
               $\mur$ and $\muf$.\\
\end{longtable}
\vspace*{1ex}

\noindent
The last two short-hands are particularly useful when trying to summarise
terms in the expansion of the tree-level merging weights. For example, the
sum of the second and third term in curly brackets in
\begin{eqnarray*}
\termX{\Bornev{2}\wckkwl{2}}{-2,3}
 &=& \Bornev{2}
     \left\{
     \wumeps{2}
   - \termX{\wumeps{2}}{0}
   - \termX{\wumeps{2}}{1} \right\}
\end{eqnarray*}
is given by \eqref{eq:full-as-expansion} below.


\subsection{Powheg-Box usage}
\label{sec:powheg-usage}

This section is intended to give guidelines on how to use the \powhegbox 
program \cite{Alioli:2010xd} in order to produce the inclusive NLO cross sections \cite{Alioli:2008gx,Alioli:2008tz,Alioli:2010qp,Campbell:2012am} needed for NLO
merging in \pytppp. Ideally, this should suffice as tutorial on how to set up 
the desired \powhegbox outputs. We rely on knowledge on \powhegbox input 
manipulations. With new versions of \powhegbox, the names of the inputs
might change, so that the settings advocated here come without guarantees.

For a given process with $n$ partons in the underlying Born configuration, 
\powhegbox by default generates output Les Houches events with $n-$ and 
$(n+1)-$parton kinematics. $n-$parton phase space points are weighted with 
the matrix element weights
\begin{eqnarray}
&& \Bbarev{n}~\Delta\left(p_{\perp,min}\right)\nonumber\\
&=&\Bigg\{\Bornev{n}\left(\phi_n\right)
     + \Virtev{n}\left(\phi_n\right)
     + \Insev{n+1|n}\left(\phi_n\right)\\
&&\qquad
     +  \int\!\!\drad
        \left[ \Bornev{n+1|n}\left(\phi_n\,\phi_{rad}\right)
             - \Dipev{n+1|n}\left(\phi_n\,\phi_{rad}\right)
        \right]
  \Bigg\}~\Delta\left(p_{\perp,min}\right) ~,
\end{eqnarray}
where the integration $\int\drad$ contains the complete radiative phase space,
and the Sudakov factor is given by
\begin{eqnarray}
\label{eq:powheg-real-events}
\Delta\left(p_{\perp}\right) &=&
 \exp\left\{ - \int_{p_{\perp}}\!\! \drad^\prime 
              \frac{\Bornev{n+1|n}\left(\phi_{n}, \phi_{rad}^\prime\right)}
                   {\Bornev{n}\left(\phi_{n}\right)}
              \right\}~.
\end{eqnarray}
$(n+1)-$parton phase space points are weighted with
\begin{eqnarray}
&& \Bbarev{n}
~\Delta\left(p_{\perp}\right)
~\frac{\Bornev{n+1|n}\left(\phi_{n+1}\right)}
      {\Bornev{n}\left(\phi_{n}\right)}
  \Theta\left(p_{\perp} - p_{\perp,min}\right).
\end{eqnarray}
The program decides if a radiative (\ie\ $(n+1)-$parton) phase space point
is generated by comparing the $p_{\perp}$ of the proposed configuration 
$\phi_{n+1}$ against $p_{\perp,min}$. No radiative events are produced if 
$p_{\perp,min}$ is set to the kinematical limit. Furthermore, we have
\begin{eqnarray}
\left\{
\Delta\left(p_{\perp,min} \right)
\right\}_{p_{\perp,min} \rightarrow \infty} = 1~.
\end{eqnarray}
Thus, using $p_{\perp,min} \rightarrow \infty$, we find
\begin{eqnarray}
\left\{
\Bbarev{n}~\Delta\left(p_{\perp,min}\right)
\right\}_{p_{\perp,min} \rightarrow \infty}
 &=&
\Bornev{n}\left(\phi_n\right)
     + \Virtev{n}\left(\phi_n\right)
     + \Insev{n+1|n}\left(\phi_n\right)\\
&&\qquad
     +  \int\!\!\drad
        \left[ \Bornev{n+1|n}\left(\phi_n\,\phi_{rad}\right)
             - \Dipev{n+1|n}\left(\phi_n\,\phi_{rad}\right)
        \right]\nonumber
\end{eqnarray}
This is exactly the inclusive NLO cross section we need to perform NLO 
merging. The NLO merging prescription will include $(n+1)-$parton 
configurations in a CKKW-L style.

In the \powhegbox program, the parameter $p_{\perp,min}$ can be set by 
changing the input variable {\tt ptsqmin}. For example assigning
\begin{equation}
\label{eq:powheg-ptmin-setting}
{\tt ptsqmin = 1d15} 
\end{equation}
will ensure that for LHC energies, the output events of \powhegbox will
contain only $n-$parton kinematics, weighted with the desired inclusive NLO
cross section.
Setting {\tt ptsqmin} will produce only $n-$parton kinematics only   
if every $(n+1)-$parton phase space has an underlying Born configuration.
This is for example not true for $\qc\cbar\to\qu\dbar\Wm$ scattering via an
$s-$channel gluon. Since such processes do not constitute corrections to any
lower-order process, we regard these as true leading-order parts, and 
(internally) neglect these configuration in the \powhegbox output. 
They will be added by including such configurations with incomplete parton
shower histories through the treatment of tree-level matrix elements. 
Numerically, the treatment of incomplete states does not have any impact.

The desired \powhegbox output should be generated with fixed 
factorisation- and renormalisation scale. To be completely certain that this
is the case, set
\begin{eqnarray}
&&{\tt runningscale\quad0 }\\
&&{\tt runningscales\quad0  }\\
&&{\tt btlscalereal\quad1 }\\
&&{\tt btlscalect\quad1 }\\
&&{\tt ckkwscalup\quad0 }
\end{eqnarray}
The \powhegbox program would, upon making the assignment \ref{eq:powheg-ptmin-setting}, attribute a {\tt SCALUP } value
of ${\tt SCALUP } = \sqrt{ {\tt ptsqmin } }$ to the output LH events. 
This number would normally be read by \pytppp and used as factorisation scale 
in the construction of overestimates for initial state splittings. For our 
purposes, the true value of $\muf$ will be an input for \pytppp, so that
the correct choice can be used internally.

After these settings, the \powhegbox output 
file can be used for NLO merging in \pytppp. We will include a detailed 
documentation of this procedure, and a manual how the schemes presented in 
this publication can be used, in the online documentation of an upcoming 
\pytppp release.


\section{Generation of weights}
\label{sec:weight-generation}

The aim of this appendix is to provide a complete description of the 
$\Oasof{1}{\mur}-$terms needed to implement the NL$^3$ and UNLOPS schemes.
This task is split into subsections containing the expansion of factors 
appearing in the weight $\wckkwl{n}$
\begin{eqnarray}
\wckkwl{n} 
&=&  K\cdot 
    \tfrac{x_{n}^+f_{n}^+(x_{n}^+,\ord_n)}{x_{n}^+f_{n}^+(x_{n}^+,\muf)}
     \tfrac{x_{n}^-f_{n}^-(x_{n}^-,\ord_n)}{x_{n}^-f_{n}^-(x_{n}^-,\muf)}
     \prod_{i=1}^{n} \Bigg[\frac{\as(b_i\ord_i)}{\as(\mur)}
     \tfrac{x_{i-1}^+f_{i-1}^+(x_{i-1}^+,\ord_{i-1})}
          {x_{i-1}^+f_{i-1}^+(x_{i-1}^+,\ord_i)}
     \tfrac{x_{i-1}^-f_{i-1}^-(x_{i-1}^-,\ord_{i-1})}
          {x_{i-1}^-f_{i-1}^-(x_{i-1}^-,\ord_i)}\nonumber\\
&&\qquad\qquad\qquad
   \noem{i-1}(x_{i-1},\ord_{i-1},\ord_i)\Bigg]\noem{n}(x_n,\ord_n,\ordms)
  ~,\label{eq:ckkwl-wgt-ref}~.
\end{eqnarray}
The weight applied in UMEPS differs in that the last no-emission probability
$\noem{n}(x_n,\ord_n,\ordms)$ is not included. Note that we have kept the 
parameter $b_i$ discussed in \ref{sec:getting-ready}.
The generation of the tree-level weights for CKKW-L is discussed
in \cite{Lonnblad:2011xx}, and the UMEPS case is treated in \cite{Lonnblad:2012ng}.
The \Kf-factor is generated by dividing the (integrated) inclusive NLO 
zero-jet cross section by the leading-order result
\begin{eqnarray}
K = \frac{\int \Bbarev{0}}{ \int \Bornev{0}}
\end{eqnarray}
It is in principle possible to rescale the tree-level weights for $n$ partons
by \Kf-factors depending on the jet multiplicity. Multiplicity-dependent 
\Kf-factors will lead to different rescaling of PS higher orders, since the 
$\Oasof{1}{\mur}-$contribution from multiplying \Kf-factors will be removed.
This means that different \Kf-factor choices give changes beyond the accuracy
of the methods. It is interesting to observe that MC@NLO and \powheg also 
differ in the \Kf-factor applied to real-emission events: While the radiative
events are not rescaled in MC@NLO, \powheg includes a phase-space dependent 
\Kf-factor (see \eqref{eq:powheg-real-events}). Though formally 
sub-leading, this difference can be large, particularly if the NLO result
is significantly higher than the Born approximation, \eg\ in Higgs-boson 
production in gluon fusion.
An argument against $n-$dependent \Kf-factors is that rescaling every jet 
multiplicity in CKKW-L by different numbers will result in increased merging 
scale dependencies. In \pytppp, we include the possibility for having 
$n-$dependent \Kf-factors
\begin{eqnarray}
K_n = \frac{\int_{\ordms} \Bbarev{n}}{ \int_{\ordms} \Bornev{n}}~.
\end{eqnarray}
These will then be calculated by dividing the sums of inclusive NLO
and LO cross section weights (with no extra reweighting) of events
above the $\ordms$ cut.

Below, we will give a detailed expansion of all factors in the
tree-level weights, which depend on \as, directly and indirectly
through the PDF-ratios. Since we have demanded that the input cross
sections be calculated with fixed $\muf$ and $\mur$, we will only keep
contributions of powers $\as^1$, and fixed scales.  Incoming particles
with positive (negative) momentum component $p_z$ will be indicated by
a superscript $+$ ($-$). Final state partons will be enumerated by the
superscript $k$. We will make use of the short-hands
\begin{eqnarray}
\label{eq:fhat-notation-isr}
\widehat{f}_{i}^\pm(\tfrac{x_i^\pm}{y},\ord)
&=&  \sum_{j \in \{q,\bar q,g \}}
    \widehat{\splitP}_{ij}^\pm(y)f_j^\pm(\tfrac{x_i^\pm}{y},\ord)
\\
\label{eq:ftilde-notation-isr}
\widetilde{f}_{i}^\pm(\tfrac{x_i^\pm}{y},\ord)
&=&  \sum_{j\in \{q,\bar q,g \}}
     \splitP_{ji}^\pm(y)f_j^\pm(\tfrac{x_i^\pm}{y},\ord)
\end{eqnarray}
where $\splitP_{ji}$ are the unregularised Altarelli-Parisi splitting kernel
for an initial state parton changing from $i$ to $j$ (by backward evolution), 
and $\widehat{\splitP}_{ij}$ are the plus-prescription-regularised 
counterparts for forward evolution. For final state splittings, we will 
write
\begin{eqnarray}
\label{eq:ftilde-notation-fsr}
\widetilde{\splitP}_{i}^k(y,\ord)
 &=& \sum_{j\in \{q,\bar q,g \}} \splitP_{ji}^k(y,\ord)\\
\splitP^k_{ji}(y,\ord)
 &=&
\begin{cases}
  & \textnormal{If the PS step from $\state{j}$ to $\state{i}$ was final state}\\
\vphantom{ \min\Big\{ 1,
             \frac{\frac{x^\pm}{y} f^\pm_j(\frac{x^\pm}{y},\ord)}
                  { x^\pm f^\pm_i(x^\pm,\ord)}\Big\}} 
\splitP_{ji}(y)
 & \textnormal{radiation off leg $k$, with a final state recoiler.}
 \\
\vphantom{ \min\Big\{ 1,
             \frac{\frac{x^\pm}{y} f^\pm_j(\frac{x^\pm}{y},\ord)}
                  { x^\pm f^\pm_i(x^\pm,\ord)}\Big\}}
 & \textnormal{If the PS step from $\state{j}$ to $\state{i}$ was final state}\\
 \splitP_{ji}(y)
 \min\Big\{ 1,
             \frac{\frac{x^\pm}{y} f^\pm_j(\frac{x^\pm}{y},\ord)}
                  { x^\pm f^\pm_i(x^\pm,\ord)}\Big\}
& \textnormal{radiation off leg $k$, and involved the incoming}\\
\vphantom{ \min\Big\{ 1,
             \frac{\frac{x^\pm}{y} f^\pm_j(\frac{x^\pm}{y},\ord)}
                  { x^\pm f^\pm_i(x^\pm,\ord)}\Big\}}
 & \textnormal{parton $\pm$ as recoiler.}\nonumber
 \end{cases}\\
\end{eqnarray}
The factor $\min\{ 1, \frac{x^\pm}{y} f^\pm_j(\frac{x^\pm}{y},\ord)
/ x^\pm f^\pm_i(x^\pm,\ord)\}$ is introduced on purely technical grounds,
because the overestimate of final state radiation in \pytppp does not include PDF 
factors, and violations of the overestimate need to be avoided. 
We split the expansion of \eqref{eq:ckkwl-wgt-ref} into subsections
containing detailed expansions of each factor. 
At the end of each subsection, we will give a description of how the 
necessary terms are generated in \pytppp. 

\subsection{Expansion of \Kf-factors and $\as-$ratios}
\label{sec:k-as-expansion}

The weight $\wckkwl{n}$ in \eqref{eq:ckkwl-wgt-ref} contains the factors
\begin{eqnarray*}
K\prod_{i=1}^{n} \frac{\as(b_i\ord_{i})}{\as(\mur)}
\end{eqnarray*}
Note that we have kept the parameters $b_i \in \left\{b_I, b_F\right\}$ 
stemming from different $\as(\mz)-$values in parton shower and fixed-order 
calculation. The factors have simple $\as-$expansions
\begin{eqnarray}
\Kf
 &=& 1 ~+~ \as(\mur)k_1 ~+~ \mathcal{O}(\as^2(\mur))\\
\as(b_i\ord_{i})
 &=& \as(\mur)
      \left\{
     1 +
      \frac{\beta_0}{4\pi} \as(\mur)\ln\left(\frac{\mur}{b_i\ord_{i}}\right)
      \right\}
     ~+~ \mathcal{O}(\as^2(\mur))
\end{eqnarray}
where $\beta_0 = 11 -\tfrac{2}{3}n_f$. Multiplying these series, we get the
expansion of the product of \Kf-factors and $\as-$ratios
\begin{eqnarray}
\label{eq:k-as-expansion}
\Kf\prod_{i=1}^{n} \frac{\as(b\ord_{i})}{\as(\mur)}
    &=& 1
     + \as(\mur)k_1
     + \sum_{i=1}^n\as(\mur)\frac{\beta_0 }{4\pi}
                   \ln \left(\frac{\mur}{b_i\ord_{i}}\right)
     + \mathcal{O}(\as^2(\mur))
\end{eqnarray}
We generate the $k_1-$term by using $k_1 = \Kf -1$. The sum is generated by 
stepping through the chosen PS history, and adding, for each nodal 
state $\state{i}$, the logarithmic terms, evaluated at the reconstructed
splitting scale $\ord_i$. This of course means that we have to construct and
choose a parton shower history first.


\subsection{Expansion of ratios of parton distributions}
\label{sec:pdf-as-expansion}

The expansion of the PDF ratios
\begin{eqnarray}
     \tfrac{x_{n}^+f_{n}^+(x_{n}^+,\ord_n)}{x_{n}^+f_{n}^+(x_{n}^+,\muf)}
     \tfrac{x_{n}^-f_{n}^-(x_{n}^-,\ord_n)}{x_{n}^-f_{n}^-(x_{n}^-,\muf)}
     \prod_{i=1}^{n}
     \tfrac{x_{i-1}^+f_{i-1}^+(x_{i-1}^+,\ord_{i-1})}
          {x_{i-1}^+f_{i-1}^+(x_{i-1}^+,\ord_i)}
     \tfrac{x_{i-1}^-f_{i-1}^-(x_{i-1}^-,\ord_{i-1})}
          {x_{i-1}^-f_{i-1}^-(x_{i-1}^-,\ord_i)}
\end{eqnarray}
is a bit involved. To derive an expansion, we will infer the DGLAP equation
\begin{align}
&\ord\frac{\partial}{\partial\ord}f_{i}^\pm(x_{i}^\pm,\ord)
  = \frac{\as(\ord)}{2\pi} \int_{x_{i}^\pm}^1 \frac{dy}{y}
    \widehat{f}_{i}^\pm(\tfrac{x_{i}}{y},\ord)\\
&\Longrightarrow f_{i}^\pm(x_{i}^\pm,\ord_{i-1}) - f_{i}^\pm(x_{i}^\pm,\mu)
  = \int_{\mu}^{\ord_{i-1}}\frac{d\ord}{\ord}
      \frac{\as(\ord)}{2\pi} \int_{x_{i}^\pm}^1 \frac{dy}{y}
      \widehat{f}_{i}^\pm(\tfrac{x_{i}}{y},\ord) \nonumber\\
&\Longrightarrow f_{i}^\pm(x_{i}^\pm,\ord_{i-1})
  = f_{i}^\pm(x_{i}^\pm,\ord_{i-1} - \delta\ord)
    + \int_{\ord_{i-1} - \delta\ord}^{\ord_{i-1}}\frac{d\ord}{\ord}
      \frac{\as(\ord)}{2\pi} \int_{x_{i}^\pm}^1 \frac{dy}{y}
      \widehat{f}_{i}^\pm(\tfrac{x_{i}}{y},\ord) ~, \nonumber
\end{align}
where we have used the notation of \eqref{eq:fhat-notation-isr}.
So far, no approximation to DGLAP scale dependence is made. This equation has 
contributions from all orders in $\as$, and should be regarded as an expansion 
in the difference of scales $\delta\ord$, rather than an expansion in 
$\as$. Approximating $\widehat{f}_{i}(\tfrac{x_{i}}{y},\ord)$ as the sum of 
products of PDFs and leading-order or next-to-leading order splitting kernels
(indicated by superscripts (0) and (1) respectively), we find
\begin{eqnarray}
    f_{i}^\pm(x_{i}^\pm,\ord_{i-1})
  &=& f_{i}^\pm(x_{i}^\pm,\muf)\\
  && + \int_{\muf}^{\ord_{i-1}}\frac{d\ord}{\ord}
       \int_{x_{i}^\pm}^1 \frac{dy}{y}
       \left\{
      \frac{\as(\ord)}{2\pi}
      \widehat{f}^{\pm, (0)}_{i}(\tfrac{x_{i}}{y},\ord)
   +  \left(\frac{\as(\ord)}{2\pi}\right)^2
      \widehat{f}^{\pm, (1)}_{i}(\tfrac{x_{i}}{y},\ord) \right\}~.\nonumber
\end{eqnarray}
Shifting the scale in $\as$ to $\mur$, and the scale of parton distributions
to $\muf$, this gives
\begin{eqnarray}
\label{eq:oas2-dglap-expansion-full}
    f_{i}^\pm(x_{i}^\pm,\ord_{i-1})
  &=& f_{i}^\pm(x_{i}^\pm,\muf)\\
  &+& \frac{\as(\mur)}{2\pi}
     \int_{\muf}^{\ord_{i-1}}\frac{d\ord}{\ord}
     \int_{x_{i}^\pm}^1 \frac{dy}{y}
     \widehat{f}^{\pm, (0)}_{i}(\tfrac{x_{i}}{y},\muf)\nonumber\\
  &+& \left(\frac{\as(\mur)}{2\pi}\right)^2
     \int_{\muf}^{\ord_{i-1}}\frac{d\ord}{\ord}
     \int_{x_{i}^\pm}^1 \frac{dy}{y}
     \int_{\muf}^{\ord}\frac{d\ord'}{\ord'}
     \int_{x_{i}^\pm/y}^1 \frac{dy'}{y'}
     \overline{\widehat{f}_{i}^{\smaller{\pm, (0)}}}(\tfrac{x_{i}}{yy'},\muf)\nonumber\\
  &+& \left(\frac{\as(\mur)}{2\pi}\right)^2
     \int_{\muf}^{\ord_{i-1}}\frac{d\ord}{\ord}
     \int_{x_{i}^\pm}^1 \frac{dy}{y}
      \widehat{f}^{\pm, (1)}_{i}(\tfrac{x_{i}}{y},\muf)\nonumber\\
  &+&
     \left(\frac{\as(\mur)}{2\pi}\right)^2
     \int_{\muf}^{\ord_{i-1}}\frac{d\ord}{\ord}
     \int_{x_{i}^\pm}^1 \frac{dy}{y}
     \frac{\beta_0 }{2} \ln \left(\frac{\mur}{\ord} \right)
     \widehat{f}^{\pm, (0)}_{i}(\tfrac{x_{i}}{y},\muf)\nonumber\\
  &+& \mathcal{O}(\as^3(\mur))\nonumber
\end{eqnarray}
where $\widehat{f}_{i}^{\pm,(0)}$ is a convolution of parton densities and
leading-order DGLAP splitting kernels (see \eqref{eq:fhat-notation-isr}), and
$\overline{\widehat{f}_{i}^{\pm,(0)}}$ a convolution of 
$\widehat{f}_{i}^{\pm,(0)}$ and splitting kernels.

The weight $\wckkwl{n}$ contains ratios of parton distributions. Since some
of these ratios are the result of rescaling to the lowest-order cross section
(see discussion after \ref{eq:ckkwl-wgt}), it might well be necessary to 
divide next-to-leading order PDFs. Luckily, \eqref{eq:oas2-dglap-expansion-full}
ensures that if we are only interested in the expansion up to 
$\Oasof{1}{\mur}$, we can safely ignore difficulties relating to NLO 
splitting kernels. Using
\begin{eqnarray}
\int_{\muf}^{\ord_{i-1}}\frac{d\ord}{\ord}
&=&
\ln\left(\frac{\ord_{i-1}}{\muf}\right)~,
\end{eqnarray}
and restricting ourselves to $\Oasof{1}{\mur}$, we arrive at
\begin{eqnarray}
\label{eq:oas1-dglap-expansion-full}
    f_{i}^\pm(x_{i}^\pm,\ord_{i-1})
  &=& f_{i}^\pm(x_{i}^\pm,\muf)
  + \frac{\as(\mur)}{2\pi}
     \ln\left(\frac{\ord_{i-1}}{\muf}\right)
     \int_{x_{i}^\pm}^1 \frac{dy}{y}
     \widehat{f}^{(0)}_{i}(\tfrac{x_{i}}{y},\muf)
   + \mathcal{O}(\as^2(\mur))\qquad
\end{eqnarray}
From now on, we will drop the superscript $(0)$. With 
\eqref{eq:oas1-dglap-expansion-full}, the expansion of a ratio of parton 
distributions is given by
\begin{align}
\label{eq:single-pdf-ratio-oas1}
\frac{ x^\pm_{i-1}f^\pm_{i-1}(x^\pm_{i-1}, \ord_{i-1})}
     { x^\pm_{i-1}f^\pm_{i-1}(x^\pm_{i-1}, \ord_{i})} &=&
1 + \frac{\as(\mur)}{2\pi}
    \ln\left\{\frac{\ord_{i-1}}{\ord_{i}}\right\}
  \hugeint_{x^\pm_{i-1}}^1 \frac{dy}{y}
  \frac{ x^\pm_{i-1} \widehat{f}^\pm_{i-1}(\tfrac{x^\pm_{i-1}}{y},\muf) }
       {x^\pm_{i-1}f^\pm_{i-1}(x^\pm_{i-1}, \muf)}
~~ + \mathcal{O}(\as^2(\mur))
\end{align}
With this, we can write the expansion of the product of PDF ratios to 
$\Oasof{1}{\mur}$ as
\begin{eqnarray}
&&    \tfrac{x_{n}^+f_{n}^+(x_{n}^+,\ord_n)}{x_{n}^+f_{n}^+(x_{n}^+,\muf)}
     \tfrac{x_{n}^-f_{n}^-(x_{n}^-,\ord_n)}{x_{n}^-f_{n}^-(x_{n}^-,\muf)}
     \prod_{i=1}^{n}
     \tfrac{x_{i-1}^+f_{i-1}^+(x_{i-1}^+,\ord_{i-1})}
          {x_{i-1}^+f_{i-1}^+(x_{i-1}^+,\ord_i)}
     \tfrac{x_{i-1}^-f_{i-1}^-(x_{i-1}^-,\ord_{i-1})}
          {x_{i-1}^-f_{i-1}^-(x_{i-1}^-,\ord_i)}\nonumber\\
&=&
 1 + \frac{\as(\mur)}{2\pi}~\hugelcurly
   \ln\left\{\frac{\ord_{n}}{\muf}\right\}
   \hugeint_{x_{n}^+}^1
   \frac{dy}{y}
   \frac{x_{n}^+ \widehat{f}_{n}^+(\tfrac{x_{n}^+}{y},\muf) }
        {x_{n}^+f_{n}^+(x_{n}^+, \muf)}\nonumber\\
&&\qquad\qquad ~+~
   \ln\left\{\frac{\ord_{n}}{\muf}\right\}
   \hugeint_{x_{n}^-}^1
   \frac{dy}{y}
   \frac{x_{n}^- \widehat{f}_{n}^-(\tfrac{x_{n}^-}{y},\muf) }
        {x_{n}^-f_{n}^-(x_{n}^-,\muf)}
   \nonumber\\
&&\qquad\qquad ~+~
   \hugesum_{i=1}^n
   \ln\left\{\frac{\ord_{i-1}}{\ord_i}\right\}
   \hugeint_{x_{i-1}^+}^1
   \frac{dy}{y}
   \frac{x_{i-1}^+ \widehat{f}_{i-1}^+(\tfrac{x_{i-1}^+}{y},\muf) }
        {x_{i-1}^+f_{i-1}^+(x_{i-1}^+, \muf)}
   \nonumber\\
&&\qquad\qquad ~+~
   \hugesum_{i=1}^n
   \ln\left\{\frac{\ord_{i-1}}{\ord_i}\right\}
   \hugeint_{x_{i-1}^-}^1
   \frac{dy}{y}
   \frac{x_{i-1}^- \widehat{f}_{i-1}^-(\tfrac{x_{i-1}^-}{y},\muf) }
        {x_{i-1}^-f_{i-1}^-(x_{i-1}^-, \muf)}
   \nonumber\\
&&\qquad\qquad ~+~ \mathcal{O}(\as^2(\mur))
\label{eq:pdf-ratio-as-expansion}
~\hugercurly
\end{eqnarray}
These integrals can be calculated by explicit numerical integration. Remember
that $\widehat{f}$ has been defined with regularised splitting kernels \cite{Ellis:1991qj}
\begin{eqnarray}
\label{eq:splitting-kernel-plus-reg}
\widehat{P}_{\q\q}(z)
 &=& C_F\frac{1+z^2}{\left(1-z\right)_+} + \frac{3}{2}C_F\delta(1-z)
 ~=~ \widehat{P}_{\qbar\qbar}(z)\\
\widehat{P}_{\g\q}(z) 
 &=& C_F\frac{1+(1-z)^2}{z}
 ~=~ \widehat{P}_{\q\q}(1-z)
 ~=~ \widehat{P}_{\g\qbar}(z)\\
\widehat{P}_{\g\g}(z)
 &=& 2C_A\left[ \frac{z}{\left(1-z\right)_+}
              + \frac{1-z}{z} + z(1-z)\right]
  + \frac{1}{6}\left[11C_A - 4n_fT_R\right]\delta(1-z)\\
\widehat{P}_{\q\g}(z)
 &=& T_R\left[z^2 + (1-z)^2\right] ~.
\end{eqnarray}
By using these functions explicitly, and inserting
\begin{eqnarray}
\int_{0}^{x_{i-1}}\frac{dy}{1-y} = - \ln(1-x_{i-1})~,
\end{eqnarray}
we find that in the case that $i-1$ is a quark or antiquark, the integral in 
\eqref{eq:single-pdf-ratio-oas1} becomes 
\begin{eqnarray}
\label{eq:plus-reg-pdf-integration-quark}
&&\frac{\as(\mur)}{2\pi}
    \ln\left\{\frac{\ord_{i-1}}{\ord_{i}}\right\}
  \hugeint_{x^\pm_{i-1}}^1 \frac{dy}{y}
  \frac{ x^\pm_{i-1} \widehat{f}^\pm_{i-1}(\tfrac{x^\pm_{i-1}}{y},\muf) }
       {x^\pm_{i-1}f^\pm_{i-1}(x^\pm_{i-1}, \muf)}\nonumber\\
&&\quad = 
\frac{\as(\mur)}{2\pi}~\ln\left(\frac{\ord_{i-1}}{\ord_{i}}\right)
\hugelcurly~
\hugeint_{x_{i-1}}^1\!\! \frac{dy}{1-y}~
  \left[ C_F (1+y^2)
        \frac{\tfrac{x_{i-1}}{y} f_\q\left(\tfrac{x_{i-1}}{y},\ord_{i}\right)}
             { x_{i-1} f_\q(x_{i-1}, \ord_{i})}
       -2C_F
  \right]\nonumber\\
&&\qquad\qquad +
  \hugeint_{x_{i-1}}^1\!\! dy~
  \left[ T_R (y^2 
      + (1-y)^2) 
        \frac{\tfrac{x_{i-1}}{y}f_\g\left(\tfrac{x_{i-1}}{y},\ord_{i}\right)}
             {x_{i-1} f_\q(x_{i-1}, \ord_{i})}
 ~\right]\nonumber\\
&&\qquad\qquad
   + ~  2C_F \ln(1-x_{i-1}) ~+~ \frac{3}{2}C_F
\hugercurly~.
\end{eqnarray}
To arrive at this form, we have used that the sum over all possible flavours
in $\widehat{f}_{i-1}$ reduces to two terms, since the antiquark (quark) can
only be produced by the evolution of a gluon or an antiquark (quark). 
If the flavour index $i-1$ indicates a gluon, the integral reads
\begin{eqnarray}
\label{eq:plus-reg-pdf-integration-gluon}
&&\frac{\as(\mur)}{2\pi}
    \ln\left\{\frac{\ord_{i-1}}{\ord_{i}}\right\}
  \hugeint_{x^\pm_{i-1}}^1 \frac{dy}{y}
  \frac{ x^\pm_{i-1} \widehat{f}^\pm_{i-1}(\tfrac{x^\pm_{i-1}}{y},\muf) }
       {x^\pm_{i-1}f^\pm_{i-1}(x^\pm_{i-1}, \muf)}\nonumber\\
&&\quad =
\frac{\as(\mur)}{2\pi}~\ln\left(\frac{\ord_{i-1}}{\ord_{i}}\right)
\hugelcurly~
  \hugeint_{x_{i-1}}^1\!\! \frac{dy}{1-y}~ 
  \left[ 2 C_A ~y~ \frac{ \tfrac{x_{i-1}}{y} f_\g\left(\tfrac{x_{i-1}}{y},\ord_{i}\right) }
                        {x_{i-1} f_\g(x_{i-1}, \ord_{i})}
        -2 C_A  ~
  \right]\nonumber\\
&&\qquad
  +~ \hugeint_{x_{i-1}}^1\!\! dy~ 
    2C_A \left[\frac{1-y}{y} + y(1-y) ~ \right]
         \left[ \frac{ \tfrac{x_{i-1}}{y} f_\g\left(\tfrac{x_{i-1}}{y},\ord_{i}\right) }
                     { x_{i-1} f_\g(x_{i-1}, \ord_{i})}
         \right] \nonumber\\
&&\qquad
  +~ \hugeint_{x_{i-1}}^1\!\! dy~
     C_F \left[\frac{1 +(1-y)^2}{y}   ~ \right]
         \left[ \frac{\tfrac{x_{i-1}}{y} f_\q\left(\tfrac{x_{i-1}}{y},\ord_{i}\right)}
                     { x_{i-1} f_\g(x_{i-1}, \ord_{i})}
          ~+~   \frac{\tfrac{x_{i-1}}{y} f_{\qbar}\left(\tfrac{x_{i-1}}{y},\ord_{i}\right)}
                     { x_{i-1} f_\g(x_{i-1}, \ord_{i})}
         \right] \nonumber\\
&&\qquad
  +~ 2C_A \ln(1-x_{i-1})
  ~+~ \frac{1}{6}\left[11C_A - 4n_fT_R~\right]
\hugercurly
\end{eqnarray}
The sum over all flavours is reduces to three terms -- the evolution of a 
gluon, a quark and an antiquark. Since all remaining 
integrands in eqs. \ref{eq:plus-reg-pdf-integration-quark} and 
\ref{eq:plus-reg-pdf-integration-gluon} involve PDFs, we need to perform the 
integrals numerically. 
This numerical integration will be performed by Monte-Carlo integration, 
although we have also implemented the Gaussian adaptive quadrature method for 
this task. We checked that compared to performing a many-point numerical 
integration, it is sufficient to only pick a single integration point for the 
Monte-Carlo integral evaluation, already when averaging over a small number of 
events ($\mathcal{O}(50)$). Thus, we choose the less time-consuming 
Monte-Carlo method as default. For this, we roll a uniformly distributed 
random number $r \in [0,1]$, and pick the integration variable as
\begin{eqnarray}
y_{mc} =
\begin{cases}
x^{r} & \textnormal{if $i-1$ is a gluon}\\
x + r\cdot (1-x) & \textnormal{otherwise}
\end{cases}
\end{eqnarray}
The full weight in \eqref{eq:pdf-ratio-as-expansion} is generated 
by stepping through the chosen PS history, and adding, for each reconstructed
state $\state{i-1}$ $(i-1 < n)$, the $x_{i-1}-$bounded integral for both 
incoming partons, multiplied by the logarithm of the reconstructed scales 
$\ord_{i-1}$ and $\ord_i$. The values $x_{i-1}$, $\ord_{i-1}$ and $\ord_{i}$ 
are easily accessible since the parton shower history contains a complete 
sequence of fully reconstructed states $\state{0},\dots,\state{n}$. For the 
ME state $\state{n}$, we add the $x_n-$bounded integral, multiplied by the 
logarithm of $\ord_n$ and $\muf$, for both incoming partons.


\subsection{Expansion of no-emission probabilities}
\label{sec:noem-as-expansion}

The factors $\noem{i-1}(x_{i-1},\ord_{i-1},\ord_i)$ in 
\eqref{eq:ckkwl-wgt-ref} symbolise the probability of a state $\state{i-1}$ to
evolve from scale $\ord_{i-1}$ to $\ord_i$ without resolving emissions. 
This means that $\noem{i-1}(x_{i-1},\ord_{i-1},\ord_i)$ is a product of 
no-emission probabilities for each dipole in the state:
\begin{eqnarray}
\noem{i-1}(x_{i-1},\ord_{i-1},\ord_i)
&=&
 \noem{i-1}^+(x_{i-1}^+,\ord_{i-1},\ord_i)~
 \noem{i-1}^-(x_{i-1}^-,\ord_{i-1},\ord_i)\\
&&\qquad \prod_{k \in \{\textnormal{final partons}\}}
 \noem{i-1}^k(x_{i-1}^\pm,\ord_{i-1},\ord_i)\nonumber
\end{eqnarray}
Using the notation of \eqref{eq:ftilde-notation-isr} and 
\eqsref{eq:ftilde-notation-fsr}, we can write
\begin{eqnarray}
\label{eq:noem-isr}
 \noem{i-1}^\pm(x_{i-1}^\pm,\ord_{i-1},\ord_i)
&=&
\exp\left\{
 - \hugeint^{\ord_{i-1}}_{\ord_{i}}
   \frac{d\ord}{\ord}
   \hugeint_{\Omega_I}
   \frac{dy}{y}
   \frac{\as(b_I\ord)}{2\pi}
   \frac{x^\pm_{i-1}\widetilde{f}^\pm_{i-1}(\frac{x^\pm_{i-1}}{y},\ord)}
        {x^\pm_{i-1}f^\pm_{i-1}(x^\pm_{i-1},\ord)}
\right\}
\end{eqnarray}
and
\begin{eqnarray}
\label{eq:noem-fsr}
 \noem{i-1}^k(x_{i-1}^\pm,\ord_{i-1},\ord_i)
&=&
\exp\left\{
 - \hugeint^{\ord_{i-1}}_{\ord_{i}}
   \frac{d\ord}{\ord}
   \hugeint_{\Omega_F}
   \frac{dy}{y}
   \frac{\as(b_F\ord)}{2\pi}
   \widetilde{\splitP}_{i-1}^k(y,\ord)
\right\}
\end{eqnarray}
From \eqref{eq:noem-fsr}, we find
\begin{eqnarray}
\label{eq:noem-fsr-exp-1}
 \noem{i-1}^k(x_{i-1}^\pm,\ord_{i-1},\ord_i)
&=& 1 
 - \hugeint^{\ord_{i-1}}_{\ord_{i}}
   \frac{d\ord}{\ord}
   \hugeint_{\Omega_F}
   \frac{dy}{y}
   \frac{\as(b_F\ord)}{2\pi}
   \widetilde{\splitP}_{i-1}^k(y,\ord)
+ \Oasof{2}{\ord}~.
\end{eqnarray}
All NLO contributions included in NLO merging schemes are generated 
with fixed scales $\mur$ and $\muf$. For a NLO-accurate description, we need to
remove the approximate shower contributions for exactly these scales. 
Otherwise, \eg\ if we choose to remove the second term on the right-hand side
of \eqref{eq:noem-fsr-exp-1}, we will remove $\Oasof{2}{\mur}$ terms as 
well, thus degrading the description of higher orders. To remove only 
precisely those parton shower terms that have corresponding contributions in
the NLO calculation, we extract an $\Oasof{}{\mur}-$expansion from
\eqref{eq:noem-fsr-exp-1}: 
\begin{eqnarray}
\label{eq:noem-fsr-exp-2}
 \noem{i-1}^k(x_{i-1}^\pm,\ord_{i-1},\ord_i)
&=& 1 
 - \frac{\as(\mur)}{2\pi}
   \hugeint^{\ord_{i-1}}_{\ord_{i}}
   \frac{d\ord}{\ord}
   \hugeint_{\Omega_F}
   \frac{dy}{y}
   \widetilde{\splitP}_{i-1}^k(y,\muf)
+ \Oasof{2}{\mur}~,
\end{eqnarray}
where we have used that the difference between $\as(b_F\mur)$ and $\as(\mur)$
is of $\Oasof{2}{\mur}$. The PDFs in appearing in 
$\widetilde{\splitP}_{i-1}^k(y)$, for final state radiation with an initial 
state recoiler, should be evaluated at $\muf$.
Expanding the exponential in \eqref{eq:noem-isr}, we find
\begin{eqnarray}
 \noem{i-1}^\pm(x_{i-1}^\pm,\ord_{i-1},\ord_i)
&=&
 1
 - \hugeint^{\ord_{i-1}}_{\ord_{i}}
   \frac{d\ord}{\ord}
   \hugeint_{\Omega_I}
   \frac{dy}{y}
   \frac{\as(b_I\ord)}{2\pi}
   \frac{x^\pm_{i-1}\widetilde{f}^\pm_{i-1}(\frac{x^\pm_{i-1}}{y},\ord)}
        {x^\pm_{i-1}f^\pm_{i-1}(x^\pm_{i-1},\ord)}
+ \Oasof{2}{\ord}~,\qquad\quad
\end{eqnarray}
which can be expanded further to give
\begin{eqnarray}
\label{eq:noem-isr-exp-2}
 \noem{i-1}^\pm(x_{i-1}^\pm,\ord_{i-1},\ord_i)
&=&
 1 -
   \frac{\as(\mur)}{2\pi}
   \hugeint^{\ord_{i-1}}_{\ord_{i}}
   \frac{d\ord}{\ord}
   \hugeint_{\Omega_I}
   \frac{dy}{y}
   \frac{x^\pm_{i-1}\widetilde{f}^\pm_{i-1}(\frac{x^\pm_{i-1}}{y},\muf)}
        {x^\pm_{i-1}f^\pm_{i-1}(x^\pm_{i-1},\muf)}
+ \Oasof{2}{\mur}~.\qquad\quad
\end{eqnarray}
Here, we have used the result \eqref{eq:oas1-dglap-expansion-full}, \ie\ that 
if we are interested in the $\Oasof{1}{\mur}$ parts
in \eqref{eq:noem-isr-exp-2}, we can safely evaluate the ratio of parton 
distributions at $\muf$, rather than $\ord$.

When expanding the CKKW-L weight $\wckkwl{n}$, we also need to discuss the 
``last" no-emission probability
\begin{eqnarray}\!\!\!\!\!\!
\noem{n}(x_{n},\ord_{n},\ordms)
=
 \noem{n}^+(x_{n}^+,\ord_{n},\ordms)~
 \noem{n}^-(x_{n}^-,\ord_{n},\ordms)\!\!\!\!\!\!\!\!
 \prod_{k \in \{\textnormal{final partons}\}}\!\!\!\!\!\!\!\!\!\!\!\!
 \noem{n}^k(x_{n}^\pm,\ord_{n},\ordms)\quad
\end{eqnarray}
with
\begin{eqnarray}
\label{eq:noem-isr-last}\!\!\!\!\!\!
\noem{n}^\pm(x_{n}^\pm,\ord_{n},\ordms)
=
\exp\left\{\!
 - \!\!\!\hugeint^{\ord_{n}}_{\ord_{n+1}}
   \!\!\!
   \frac{d\ord}{\ord}
   \!\!\!
   \hugeint_{\Omega_I}
   \!\!\!
   \frac{dy}{y}
   \frac{\as(b_I\ord)}{2\pi}
   \frac{x^\pm_{n}\widetilde{f}^\pm_{n}(\frac{x^\pm_{n}}{y},\ord)}
        {x^\pm_{n}f^\pm_{n}(x^\pm_{n},\ord)}
   \Theta\left(t(\state{n+1},\ord) - \ordms\right)
\right\}\quad
\end{eqnarray}
and
\begin{eqnarray}
\label{eq:noem-fsr-last}\!\!\!\!\!\!
\noem{n}^k(x_{n}^\pm,\ord_{n},\ordms)
=
\exp\left\{\!
 - \!\!\!\hugeint^{\ord_{n}}_{\ord_{n+1}}
   \!\!\!
   \frac{d\ord}{\ord}
   \!\!\!
   \hugeint_{\Omega_F}
   \!\!\!
   \frac{dy}{y}
   \frac{\as(b_F\ord)}{2\pi}
   \widetilde{\splitP}_{n}^k(y,\ord)
   \Theta\left(t(\state{n+1},\ord) - \ordms\right)
\right\}\quad
\end{eqnarray}
The expansion of these terms carries an additional $\Theta$ function
\begin{eqnarray}
&&\label{eq:noem-isr--last}\\\nonumber\!\!\!\!\!\!
&&\noem{n}^\pm(x_{n}^\pm,\ord_{n},\ordms)
= 1
-
   \frac{\as(\mur)}{2\pi}
   \!\!\!\hugeint^{\ord_{n}}_{\ord_{n+1}}
   \!\!\!
   \frac{d\ord}{\ord}
   \!\!\!
   \hugeint_{\Omega_I}
   \!\!\!
   \frac{dy}{y}
   \frac{x^\pm_{n}\widetilde{f}^\pm_{n}(\frac{x^\pm_{n}}{y},\muf)}
        {x^\pm_{n}f^\pm_{n}(x^\pm_{n},\muf)}
   \Theta\left(t(\state{n+1},\ord) - \ordms\right)
~+~\dots\qquad
\end{eqnarray}
and
\begin{eqnarray}
\label{eq:noem-fsr--last}\!\!\!\!\!\!
\noem{n}^k(x_{n}^\pm,\ord_{n},\ordms)
&=& 1
-
   \frac{\as(\mur)}{2\pi}
   \!\!\!\hugeint^{\ord_{n}}_{\ord_{n+1}}
   \!\!\!
   \frac{d\ord}{\ord}
   \!\!\!
   \hugeint_{\Omega_F}
   \!\!\!
   \frac{dy}{y}
   \widetilde{\splitP}_{n}^k(y,\muf)
   \Theta\left(t(\state{n+1},\ord) - \ordms\right)
~+~\dots\qquad
\end{eqnarray}
The definition of $t$ is given in appendix \ref{sec:pythia-evolution-pt}.
Collecting all terms, we find that the expansion of the product of no-emission
probabilities in the CKKW-L weight $\wckkwl{n}$ is given by
\begin{eqnarray}
\label{eq:noem-as-expansion}
&&   \noem{n}(x_n,\ord_n,\ordms) \prod_{i=1}^{n} 
   \noem{i-1}(x_{i-1},\ord_{i-1},\ord_i)\\
&=&
1
-
\hugesum_{i=1}^n
\hugelcurly
   ~
   \sum_{\pm}
   \frac{\as(\mur)}{2\pi}
   \hugeint^{\ord_{i-1}}_{\ord_{i}}
   \frac{d\ord}{\ord}
   \hugeint_{\Omega_I}
   \frac{dy}{y}
   \frac{x^\pm_{i-1}\widetilde{f}^\pm_{i-1}(\frac{x^\pm_{i-1}}{y},\muf)}
        {x^\pm_{i-1}f^\pm_{i-1}(x^\pm_{i-1},\muf)}\nonumber\\
&&\qquad~~\! + \qquad
   \sum_{k}
   \frac{\as(\mur)}{2\pi}
   \hugeint^{\ord_{i-1}}_{\ord_{i}}
   \frac{d\ord}{\ord}
   \hugeint_{\Omega_F}
   \frac{dy}{y}
   \widetilde{\splitP}_{i-1}^k(y,\muf)~
\hugercurly\nonumber\\
&&\quad -~
   \sum_{\pm}
   \frac{\as(\mur)}{2\pi}
   \hugeint^{\ord_{n}}_{\ord_{n+1}}
   \frac{d\ord}{\ord}
   \hugeint_{\Omega_I}
   \frac{dy}{y}
   \frac{x^\pm_{n}\widetilde{f}^\pm_{n}(\frac{x^\pm_{n}}{y},\muf)}
        {x^\pm_{n}f^\pm_{n}(x^\pm_{n},\muf)}
    \Theta\left(t(\state{n+1},\ord) - \ordms\right)
   \nonumber\\
&&\quad -~
   \sum_{k}
   \frac{\as(\mur)}{2\pi}
   \hugeint^{\ord_{n}}_{\ord_{n+1}}
   \frac{d\ord}{\ord}
   \hugeint_{\Omega_F}
   \frac{dy}{y}
   \widetilde{\splitP}_{n}^k(y,\muf)
   \Theta\left(t(\state{n+1},\ord) - \ordms\right)
~ +~
 \Oasof{2}{\mur}~.\nonumber
\end{eqnarray}
Although this looks fairly complicated, it is in fact easily generated. It is
useful to remember that if the probability for $n$ incidents (\eg\ nuclear 
decays) is given by
\begin{eqnarray}
P_n = \frac{1}{n!} x^n e ^{-x}~,
\end{eqnarray}
then the average number of incidents is 
\begin{eqnarray}
\langle n \rangle
 = \sum_{n=0}^\infty n P_n
 = \sum_{n=1}^\infty n P_n 
 = e^{-x}
   \sum_{n=1}^\infty
   \frac{1}{(n-1)!} x^n 
 = x e^{-x}
   \sum_{n=1}^\infty
   \frac{1}{(n-1)!} x^{n-1} 
 = x ~.\qquad
\end{eqnarray}
The probability $P_n$ has the same form as the probability for $n$ emissions
in a parton shower, and the $\Oasof{1}{\mur}-$terms in 
\eqref{eq:noem-as-expansion} can be identified with the exponents $x$.
That means we can calculate the $\Oasof{1}{\mur}-$terms by
generating, with fixed $\mur$ and $\muf$, an average number of emissions.
In final state radiation off final parton $k$ for example had been generated 
with fixed scales, we can write
\begin{eqnarray}
&&\!\!\!\!\!\!\!\!\!\!\!\!
\langle n_{\textnormal{FSR emissions between $\ord_{i-1}$ and $\ord_i$}} \rangle\nonumber\\
&=& \sum_{n=0}^\infty n \frac{1}{n!}
   \mathop{\mathlarger{\mathlarger{\Bigg(}}}
   \hugeint^{\ord_{i-1}}_{\ord_{i}}
   \frac{d\ord}{\ord}
   \hugeint_{\Omega_F}
   \frac{dy}{y}
   \widetilde{\splitP}_{i-1}^k(y,\muf)~
   \mathop{\mathlarger{\mathlarger{\Bigg)}}}^n
   \exp
   \hugelcurly
   \! - \!\!\!
   \hugeint^{\ord_{i-1}}_{\ord_{i}}
   \frac{d\ord}{\ord}
   \hugeint_{\Omega_F}
   \frac{dy}{y}
   \widetilde{\splitP}_{i-1}^k(y,\muf)~
   \hugercurly\qquad\nonumber\\
&=& 
   \hugeint^{\ord_{i-1}}_{\ord_{i}}
   \frac{d\ord}{\ord}
   \hugeint_{\Omega_F}
   \frac{dy}{y}
   \widetilde{\splitP}_{i-1}^k(y,\muf)~
\end{eqnarray}
The average number of emissions is additive. This means that instead of
averaging over emissions from each leg separately, we can directly average 
\emph{all} emissions: We can simply start the shower off state 
$\state{i-1}$ at $\ord_{i-1}$, count any emission above $\ord_i$, restart the
shower (off $\state{i-1}$ from $\ord_{i-1}$) $N$ times, and average. This 
will give the sum of all contributions to 
one $i$ in \eqref{eq:noem-as-expansion} \footnote{Even if we had an analytic 
way of generating the integrals, implementing the average number of emissions 
is superior, since using trial emissions will capture correlations between
potentially radiating dipoles, and automatically include phase space 
constraints.}.

For the $\Oasof{1}{\mur}-$terms of the expansion of the last no-emission 
probability (\ie\ the last two lines of \eqref{eq:noem-as-expansion}), the
$\Theta$ function is included by only increasing the number of counted 
emissions for a trial emissions is above the merging scale.

In our implementation, we will generate all trial emissions with running 
scales, and count relevant trial emissions with the weight
\begin{eqnarray}
g_{e}
&=&
\frac{\as\left(\mur\right)}{\as\left(\ord_{e}\right)}
g_{\textnormal{pdf},e}\\
g_{\textnormal{pdf},e}
&=&
\begin{cases}
\frac{x_{i-1}f_{i-1}\left( x_{i-1}, \ord_{e} \right)}
     {x_{i-1}f_{i-1}\left( x_{i-1}, \muf \right)}
\frac{x_{i-1}\widehat{f}_{e}\left( \frac{x_{i-1}}{y_{e}}, \muf \right)}
     {x_{i-1}\widehat{f}_{e}\left( \frac{x_{i-1}}{y_{e}}, \ord_{e} \right)}
&
\textnormal{ if the trial emission was produced in ISR,}\\
\vphantom{\frac{x_{i-1}f_{i-1}\left( x_{i-1}, \ord_{e} \right)}
     {x_{i-1}f_{i-1}\left( x_{i-1}, \muf \right)}
\frac{x_{i-1}\widehat{f}_{e}\left( \frac{x_{i-1}}{y_{e}}, \muf \right)}
     {x_{i-1}\widehat{f}_{e}\left( \frac{x_{i-1}}{y_{e}}, \ord_{e} \right)}}
1
&
\textnormal{ if the trial emission was produced in FSR,}\\
\vphantom{\frac{x_{i-1}f_{i-1}\left( x_{i-1}, \ord_{e} \right)}
     {x_{i-1}f_{i-1}\left( x_{i-1}, \muf \right)}
\frac{x_{i-1}\widehat{f}_{e}\left( \frac{x_{i-1}}{y_{e}}, \muf \right)}
     {x_{i-1}\widehat{f}_{e}\left( \frac{x_{i-1}}{y_{e}}, \ord_{e} \right)}}
&
\textnormal{ with final state recoiler,}\\
\frac{x_{i-1}f_{i-1}\left( x_{i-1}, \ord_{e} \right)}
     {x_{i-1}f_{i-1}\left( x_{i-1}, \muf \right)}
\frac{x_{i-1}\widehat{f}_{e}\left( \frac{x_{i-1}}{y_{e}}, \muf \right)}
     {x_{i-1}\widehat{f}_{e}\left( \frac{x_{i-1}}{y_{e}}, \ord_{e} \right)}
&
\textnormal{ if the trial emission was produced in FSR,}\\
\vphantom{\frac{x_{i-1}f_{i-1}\left( x_{i-1}, \ord_{e} \right)}
     {x_{i-1}f_{i-1}\left( x_{i-1}, \muf \right)}
\frac{x_{i-1}\widehat{f}_{e}\left( \frac{x_{i-1}}{y_{e}}, \muf \right)}
     {x_{i-1}\widehat{f}_{e}\left( \frac{x_{i-1}}{y_{e}}, \ord_{e} \right)}}
&
\textnormal{ with initial state recoiler,}
\end{cases}
\end{eqnarray}
where $\ord_e$ is the evolution scale of the trial emission, $y_{e}$ the 
energy splitting, and $e$ the flavour of the initial line after the 
trial emission. This weight will give trial emissions generated with fixed 
scales, as can \eg\ be verified for initial state splittings by using 
\begin{eqnarray}
&&\!\!\!\!\!
   \hugeint^{\ord_{i-1}}_{\ord_{i}}
   \frac{d\ord}{\ord}
   \hugeint_{\Omega_I}
   \frac{dy}{y}
   \frac{\as\left(\ord\right)}{2\pi}
   \frac{x^\pm_{i-1}\widetilde{f}^\pm_{i-1}\left(\frac{x^\pm_{i-1}}{y},\ord\right)}
        {x^\pm_{i-1}f^\pm_{i-1}\left(x^\pm_{i-1},\ord\right)}
= 
 V \frac{1}{N} \hugesum_{l}^{N}
   \frac{\as\left(\ord_{l}\right)}{2\pi}
   \frac{x^\pm_{i-1}\widetilde{f}^\pm_{l}\left(\frac{x^\pm_{i-1}}{y_{l}},\ord_{l}\right)}
        {x^\pm_{i-1}f^\pm_{i-1}\left(x^\pm_{i-1},\ord_{l}\right)}\qquad\\
&&=
 V \frac{1}{N} \hugesum_{l}^{N}
   \frac{\as\left(\mur\right)}{2\pi}
   \frac{x^\pm_{i-1}\widetilde{f}^\pm_{l}\left(\frac{x^\pm_{i-1}}{y_{l}},\muf\right)}
        {x^\pm_{i-1}f^\pm_{i-1}\left(x^\pm_{i-1},\muf\right)} \times g_{l}^{-1}
\qquad\quad \left(\textnormal{where } V = \int^{\ord_{i-1}}_{\ord_{i}} \frac{d\ord}{\ord}
   \int_{\Omega_I}
   \frac{dy}{y}\right) ~.\nonumber
\end{eqnarray}
Weighting every trial emission with $g_{l}$ will exactly cancel the 
$g_{l}^{-1}$ factor in the sum, thus producing the desired fixed-scale terms.

In conclusion, the algorithm to generate all $\Oasof{1}{\mur}-$terms in one
multiplicity $i$ is
\begin{enumerate}
\item Start a trial shower off state $\state{i-1}$ at scale $\ord_{i-1}$.
\item If the trial shower yields an emission, and 
  \begin{enumerate}
  \item $i-1 < n$, count the emissions with weight $g_{e}$
                   if $\ord_{e} > \ord_{i}$
  \item $i-1 = n$, count the emissions with weight $g_{e}$
                   if $t\left(S_{e},\ord_{e}\right) > \ordms$.
  \end{enumerate}
\item If the trial emission has been counted, restart the trial shower off 
      state $\state{i-1}$, with a starting scale $\ord_{e}$.
      Repeat steps 2 and 3.\\
      If $\ord_{e} < \ord_{i}$, or $\ord_{e} < \ordms$, or no trial
      emission has been constructed, set the number of emissions to the sum
      of weights, and exit.
\end{enumerate}
The average is generated by restarting this algorithm $N$ times, and dividing
the sum of the $N$ results by $N$. 
To generate the sum of all $\Oasof{1}{\mur}-$terms in 
\eqref{eq:noem-as-expansion}, we step through the reconstructed PS history,
and subtract, for each reconstructed state $\state{i-1}$ $(i-1 < n)$, the 
average number of emissions between $\ord_{i-1}$ and $\ord_{i}$, as generated 
by the above algorithm.


\subsection{Summary of weight generation}
\label{sec:weight-generation-summary}

This section is intended to collect the results of appendices
\ref{sec:k-as-expansion}, \ref{sec:pdf-as-expansion} and 
\ref{sec:noem-as-expansion}, and to summarise how the necessary weights are 
generated.
In NL$^3$ and UNLOPS, tree-level samples $\tree{m}^\prime$ ($\untree{m}$) are,
respectively, defined as
\begin{eqnarray}
\tree{m}^\prime =
\Bornev{m} \left\{\wckkwl{m}
      - \termX{\wckkwl{m}}{0}
      - \termX{\wckkwl{m}}{1}\right\}~,\qquad\qquad
\untree{m} =
\Bornev{m} \left\{\wumeps{m}
      - \termX{\wumeps{m}}{0}
      - \termX{\wumeps{m}}{1}\right\}\nonumber
\end{eqnarray}
The weights $\wckkwl{n}$ and $\wumeps{n}$ differ in that $\wumeps{n}$ does 
not contain the ``last" no-emission probability ($\noem{n}$) present in 
$\wckkwl{n}$. Once all approximate $\Oasof{0}{\mur}$ and $\Oasof{1}{\mur}$ 
terms in the weight of tree-level samples are removed, we can add NLO events,
and still retain NLO accuracy.
Collecting all $\Oasof{0}{\mur}$ and $\Oasof{1}{\mur}$ terms of appendices
\ref{sec:k-as-expansion}, \ref{sec:pdf-as-expansion} and 
\ref{sec:noem-as-expansion}, we find
\begin{eqnarray}
&&\!\!\!\!\!\!\!\!\!\!\!\!\!\!\!\!\!\!\!
   \termX{\wckkwl{m}}{0}
 + \termX{\wckkwl{m}}{1}
\quad=\quad 1
 + \as(\mur)k_1
 + \sum_{i=1}^n\as(\mur)\frac{\beta_0 }{4\pi}
               \ln \left(\frac{\mur}{b_i\ord_{i}}\right)\nonumber\\
&&+~
   \frac{\as(\mur)}{2\pi}
\hugesum_{i=1}^n
\hugelcurly
   ~
   \sum_{\pm}\hugelsquare
   \ln\left\{\frac{\ord_{i-1}}{\ord_i}\right\}
   \hugeint_{x_{i-1}^\pm}^1
   \frac{dy}{y}
   \frac{x_{i-1}^\pm \widehat{f}_{i-1}^\pm(\tfrac{x_{i-1}^\pm}{y},\muf) }
        {x_{i-1}^\pm f_{i-1}^\pm(x_{i-1}^\pm, \muf)}\nonumber\\
&&\qquad\qquad\qquad\qquad\qquad\qquad
-
   \hugeint^{\ord_{i-1}}_{\ord_{i}}
   \frac{d\ord}{\ord}
   \hugeint_{\Omega_I}
   \frac{dy}{y}
   \frac{x^\pm_{i-1}\widetilde{f}^\pm_{i-1}(\frac{x^\pm_{i-1}}{y},\muf)}
        {x^\pm_{i-1}f^\pm_{i-1}(x^\pm_{i-1},\muf)}\hugersquare\nonumber\\
&&\qquad\qquad\qquad - \quad~~\!
   \sum_{k}
   \hugeint^{\ord_{i-1}}_{\ord_{i}}
   \frac{d\ord}{\ord}
   \hugeint_{\Omega_F}
   \frac{dy}{y}
   \widetilde{\splitP}_{i-1}^k(y,\muf)~
\hugercurly\nonumber\\
&&+~
   \frac{\as(\mur)}{2\pi}
   \sum_{\pm}\hugelsquare
   \ln\left\{\frac{\ord_{n}}{\muf}\right\}
   \hugeint_{x_{n}^\pm}^1
   \frac{dy}{y}
   \frac{x_{n}^\pm \widehat{f}_{n}^\pm(\tfrac{x_{n}^\pm}{y},\muf) }
        {x_{n}^\pm f_{n}^\pm(x_{n}^\pm, \muf)}\nonumber\\
&&\qquad\qquad\qquad\qquad
-
   \hugeint^{\ord_{n}}_{\ord_{n+1}}
   \frac{d\ord}{\ord}
   \hugeint_{\Omega_I}
   \frac{dy}{y}
   \frac{x^\pm_{n}\widetilde{f}^\pm_{n}(\frac{x^\pm_{n}}{y},\muf)}
        {x^\pm_{n}f^\pm_{n}(x^\pm_{n},\muf)}
    \Theta\left(t(\state{n+1},\ord) - \ordms\right)
   \hugersquare
   \nonumber\\
&&-~
   \frac{\as(\mur)}{2\pi}
   \sum_{k}
   \hugeint^{\ord_{n}}_{\ord_{n+1}}
   \frac{d\ord}{\ord}
   \hugeint_{\Omega_F}
   \frac{dy}{y}
   \widetilde{\splitP}_{n}^k(y,\muf)
   \Theta\left(t(\state{n+1},\ord) - \ordms\right)~.\label{eq:full-as-expansion}
\end{eqnarray}
The first terms in the expansion of the UNLOPS weight $\left(
\termX{\wumeps{m}}{0} + \termX{\wumeps{m}}{1}\right)$ do not
contain terms proportional to $\Theta$, but are otherwise identical. 
The complete expression can be generated by using the reconstructed
parton shower history of the matrix element input event $\state{n}$. Each 
step in the parton shower history corresponds to a fully reconstructed state
$\state{i}$, and an associated production scale $\ord_i$. Thus, we have enough
information at step $i$ to generate all $\Oasof{1}{\mur}-$terms depending on
the index $i$ in \ref{eq:full-as-expansion}. Without specifying details, the
method to generate the right-hand side of \ref{eq:full-as-expansion} is
\begin{enumerate}
\item Construct all PS histories for $\state{n}$, select one history.
      Set $v=1+\as(\mur)k_1$, and start stepping through the history at the 
      lowest-multiplicity state $\state{0}$. Steps will be counted as $i-1$,
      starting from $i=1$.
\item For each step $i-1$,
      \begin{enumerate}
      \item If $i-1<n$, increase $v$ by the term due to the expansion of 
            $\as-$ratios, calculated with scale $\ord_i$.
      \item Increase $v$ by the term due to the expansion of 
            PDF-ratios containing the index $i-1$. This means adding, at each
            step $i-1$, two numerically integrated terms. 
      \item Decrease $v$ by the term due to the expansion of no-emission 
            probabilities containing the index $i-1$. This means subtracting, 
            at each step $i-1$, the average number of emissions between the
            scales $\ord_{i-1}$ and $\ord_i$. In UNLOPS, only subtract until
            step $i-1=n-1$.
      \end{enumerate}
\item To arrive at $\tree{n}^\prime$ ($\untree{n}$) samples, subtract $v$ from the
      weight $\wckkwl{n}$ ($\wumeps{n}$). Use the weight
      \begin{eqnarray}
      \wckkwl{n}
      - v \qquad\qquad
      \left(\textnormal{ or }~ \wumeps{n}
      - v \right)\nonumber
      \end{eqnarray}
      to fill histograms.
\end{enumerate}
In UNLOPS, there is an additional complication in that the contributions
\begin{align}
&\termX{\Iev{n}{n-1}}{-n,n+1}
 &=& \int \Bornev{n\rightarrow n-1}
     \left\{ \wumeps{n} - \termX{\wumeps{n}}{0} - \termX{\wumeps{n}}{1}
     \right\}\nonumber\\
&\termX{\Iev{n}{n-1}}{-n}
 &=& \int \Bornev{n\rightarrow n-1}
     \left\{ \wumeps{n} - \termX{\wumeps{n}}{0} \right\}
  =  \int \Bornev{n\rightarrow n-1}
     \left\{ \wumeps{n} - 1 \right\}\nonumber
\end{align}
have to be generated. For this, we generate the weights $\wumeps{n}$ and the
necessary subtractions, and afterwards integrate over the phase space of the
$n$'th jet, as outlined in the generation of class C in section 
\ref{sec:unlops-step-by-step}.
Having the weights $\wckkwl{n}$,  $\wumeps{n}$, and the first terms in their 
expansions $\left(\termX{\wckkwl{n}}{0},\termX{\wckkwl{n}}{1},
\termX{\wumeps{n}}{0},\termX{\wumeps{n}}{1} \right)$, at our disposal, 
we can produce NL$^3$ or UNLOPS predictions for merging multiple NLO 
calculations by following
the steps in \ref{sec:nl3-step-by-step} and \ref{sec:unlops-step-by-step}.


\section{Derivation of NL$^3$}
\label{sec:nl3-derivation}

The aim of this section is to give a motivated derivation of the NL$^3$ 
method. Since this derivation will explicitly require NLO-correctness as 
condition, the following can also be seen as a proof of the validity of the
scheme. We will use the notation of sections \ref{sec:nlo-merging-intro} and 
\ref{sec:notation}.

To include the parton shower resummation in a CKKW-L style, corrective 
weights will have to be applied to events. When deriving a NLO merging 
method, the aim must be that the scheme
\begin{enumerate}
\item[(a)] Is correct to next-to-leading order for all exclusive $n-$jet 
          observables;
\item[(b)] Keeps the parton shower (\ie\ CKKW-L) approximation for all higher
          orders.
\item[(c)] Shows small dependence on the separation between matrix element and
          parton shower region, especially in the inclusive cross section.
\end{enumerate}
To handle both inclusive and exclusive NLO cross sections at the same time, 
let us introduce the symbols
\begin{eqnarray}
&&\textnormal{L}_n =
\begin{cases}
\Bbarev{n} & \qquad\qquad\textnormal{ if inclusive NLO cross sections are used,}\\
\Btilev{n} & \qquad\qquad\textnormal{ if exclusive NLO cross sections are used.}
\end{cases}\\
&&\textnormal{S}_n =
\begin{cases}
 -\int_s \Bornev{n+1\rightarrow n} & \textnormal{ if inclusive NLO 
  cross sections are used,}\\
0 & \textnormal{ if exclusive NLO cross sections are used,}
\end{cases}
\end{eqnarray}
where $\Bornev{n}$ are the tree-level cross sections. If all the
corresponding event samples are multiplied with corrective weights,
conditions (a) and (b) read
\begin{eqnarray}
\label{eq:nl3-constraint-full}
 &&  \Bornev{n} w_B 
 ~+~   \textnormal{L}_n w_L
 ~-~   \textnormal{S}_n w_S\\
 =&&
    \Bornev{n}
  + \Virtev{n}
  + \Insev{n+1|n}
  +
    \int^{\ordms}\!\!\!\! \drad
    \left( \Bornev{n+1|n}
         - \Dipev{n+1|n}\right)
 ~+~ \Bornev{n}\sum_{i=2}^\infty \termX{\wckkwl{n}}{i}
\end{eqnarray}
and
\begin{eqnarray}
\label{eq:nl3-constraint-tree}
\Bornev{n+1} \wckkwl{n} = 
\Bornev{n+1} \sum_{i=0}^\infty \termX{\wckkwl{n}}{i}~.
\end{eqnarray}
The last equation is trivially fulfilled if we choose to 
reweight higher-multiplicity tree-level matrix elements as in CKKW-L.
If we do not have control over the NLO calculation, we cannot 
assume that tree-level, virtual and real emission samples are evaluated
at identical $n$-jet phase space points. Thus, without having the actual 
functional form of the matrix element weights, the merging conditions for 
$\Bornev{n}$, $\textnormal{L}_n$ and $\textnormal{S}_n$ have to
decouple, since we cannot allow $w_B, w_L$ and $w_S$ to be functions of 
the matrix elements\footnote{This is not the case for MEPS@NLO in \sherpa, where full 
control of the matrix element functions is available, thus opening other 
avenues for NLO merging \cite{Gehrmann:2012yg,Hoeche:2012yf}.}. To accommodate the merging constraint 
\eqref{eq:nl3-constraint-full}, we make an ansatz
\begin{eqnarray}
\label{eq:wgt-ansatz.v2}
w_B &=& a_{B,0} + \sum_{i=1}^\infty b_{B,i}\as^i
     + \sum_{i=1}^\infty c_{B,i}\left(\frac{1}{\as}\right)^i\\
w_L &=& a_{L,0} + \sum_{i=1}^\infty b_{L,i}\as^i
     + \sum_{i=1}^\infty c_{L,i}\left(\frac{1}{\as}\right)^i\nonumber\\
w_S &=& a_{S,0} + \sum_{i=1}^\infty b_{S,i}\as^i
     + \sum_{i=1}^\infty c_{S,i}\left(\frac{1}{\as}\right)^i\nonumber
\end{eqnarray}
We choose this form to allow for complete generality. Negative powers of 
$\as$ are included to allow for a simple visualisation of weights stemming 
\eg\ from division by an all-order expression, if such factors should be
desirable.

If we insert \eqref{eq:wgt-ansatz.v2} into \eqref{eq:nl3-constraint-full}, 
remember that $\Bornev{n}$ is of $\Oas{n}$, that $\textnormal{L}_n$ contains 
a Born term of $\Oas{n}$ and corrections of $\Oas{n+1}$, and that 
$\textnormal{S}_n$ is of $\Oas{n+1}$, we can read off constraints on the 
coefficients order by order. This leads to weights of the 
form\footnote{If $\textnormal{L}_n$ does not contain an additional Born 
term, we would not have to subtract the term $\termX{\wckkwl{n}}{0}$ in $w_B$, 
which would have the benefit of fewer negative weights.}
\begin{eqnarray}
w_B &=& \wckkwl{n} - \termX{\wckkwl{n}}{0} - \termX{\wckkwl{n}}{1}
     + \sum_{i=1}^\infty c_{B,i}\left(\frac{1}{\as}\right)^i\\
w_L &=& 1 + \sum_{i=2}^\infty c_{L,i}\left(\frac{1}{\as}\right)^i\\
w_S &=& 1 + \sum_{i=2}^\infty c_{S,i}\left(\frac{1}{\as}\right)^i
\end{eqnarray}
So far, we have allowed the coefficients $c$ to be non-vanishing. If we
naively do so, we allow changes of $\Oas{n-i}$ to the $n-$jet cross section. 
Since the exact result (in $\Oas{n}$ and $\Oas{n+1}$) should not be changed 
by numerically large terms, we are bound to enforce the conditions
\begin{eqnarray}
&&c_{B,1} = 0\\
&&c_{B,i} + c_{L,i} + c_{S,i} = 0 \qquad (i\geq 2)
\end{eqnarray}
\emph{One} way to include this condition is by replacing $c_{B,i}$ in the 
tree-level weight, which gives allowed weights of the form
\begin{eqnarray}
w_B &=& \wckkwl{n}  - \termX{\wckkwl{n}}{0} - \termX{\wckkwl{n}}{1}
     - \sum_{i=2}^\infty \left[c_{V,i} + c_{R,i}\right]
                         \left(\frac{1}{\as}\right)^i
\end{eqnarray}
Finally, if we choose to use inclusive NLO cross sections for 
$\textnormal{L}_n$, there are non-trivial cancellations between 
$\textnormal{L}_n$ and $\textnormal{S}_n$, since $\textnormal{S}_n$ was 
introduced as an explicit phase space subtraction. If we choose $w_L$ and 
$w_S$ differently, these cancellations are jeopardised in higher orders.
We thus think it reasonable to only allow the weights
\begin{eqnarray}
\label{eq:nl3-general-weights-tree}
w_B &=& \wckkwl{n} - \termX{\wckkwl{n}}{0} - \termX{\wckkwl{n}}{1}
     - \sum_{i=2}^\infty 2c_{L,i}\left(\frac{1}{\as}\right)^i\\
\label{eq:nl3-general-weights-virt}
w_L &=& w_S = 1 + \sum_{i=2}^\infty c_{L,i}\left(\frac{1}{\as}\right)^i
\end{eqnarray}
This still allows for some arbitrariness, since $c_{L,i}$ is not 
fixed. 
We choose a pragmatic approach, and exclude weights that are not easily
generated by the \pytppp shower\footnote{We would like to point out that the 
approach of Pl\"{a}tzer \cite{Platzer:2012bs} does indeed use a smart choice 
to generate different
weights, which contain Sudakov-form-factor denominators.}.
An example for such are weights with negative
$\as$ order. Thus, we set
\begin{eqnarray}
\label{eq:nl3-simple-weights-tree}
w_B &=& \wckkwl{n} -  \termX{\wckkwl{n}}{0} - \termX{\wckkwl{n}}{1}\\
\label{eq:nl3-simple-weights-virt}
w_L &=& w_S = 1
\end{eqnarray}
We will not reweight any $\Oas{n+1}-$terms. This immediately implies that the
merging scale should be defined in the parton shower evolution variable, 
since otherwise, Sudakov factors would have to be multiplied in regions of 
$\tms$-unordered splittings \cite{Lavesson:2008ah}. Sudakov factors can be represented
by a power series in positive powers of $\as$, so that even in 
\eqref{eq:nl3-general-weights-virt}, we could not easily accommodate such factors. The main
constraint on $w_L$ and $w_S$ is condition (b), which can also be interpreted
as the statement that only tree-level samples are allowed ``seeds" for higher 
order contributions.

So far, we have only been concerned with conditions (a) and (b). Condition (c)
becomes important when combining different jet multiplicities. In NL$^3$, this
combination is constructed by simply summing all reweighted $n-$jet samples.
Let $M$ be number of additional jets in the highest-multiplicity NLO 
calculation, and let us use
\begin{eqnarray}
\tree{n}
&=&
\begin{cases}
\Bornev{n}w_B &\qquad\qquad\qquad~ (n\leq M)\\
\Bornev{n}\wckkwl{n} &\qquad\qquad\qquad~  (n > M)
\end{cases}
\\
\virt{n} &=& \textnormal{L}_n = \Bbarev{n} \qquad\qquad\qquad\qquad\!\!\!\!\! (n\leq M)
\\
\subt{n} &=& \textnormal{S}_n = -\int_s \Bornev{n+1\rightarrow n}\qquad\quad (n\leq M)
\end{eqnarray}
The NL$^3$ method then sums reweighted event samples:
\begin{itemize}
\item If $n\leq M$, include the samples $\Bornev{n}$, 
      $\textnormal{L}_n$ and $\textnormal{S}_n$, reweighted according to 
      \eqref{eq:nl3-simple-weights-tree} or 
      \eqref{eq:nl3-simple-weights-virt}. Remember that
      $\textnormal{S}_n$ has a negative sign. This will produce 
      $\tree{n}^{\prime}$, $\virt{n}$ and $\subt{n}$.
\item If $n > M$, reweight $\Bornev{n}$ as in CKKW-L. This will produce 
      $\tree{n}$.
\end{itemize}
For an observable $\mathcal{O}$, this produces the prediction
\begin{eqnarray}
\langle \mathcal{O} \rangle
 &=& \sum_{m=0}^M \int d\phi_0 \int\!\cdots\!\int 
     \mathcal{O}(\state{mj})
     ~\Bigg\{ \tree{m}^{\prime} + \virt{m} + \subt{m} ~\Bigg\}
\nonumber\\
&+&
  \sum_{n=M+1}^N
  \int d\phi_0 \int\!\cdots\!\int 
  \mathcal{O}(\state{nj})
  \tree{n}\nonumber\\
 &=& \sum_{m=0}^M \int d\phi_0 \int\!\cdots\!\int 
     \mathcal{O}(\state{mj})
     ~\Bigg\{\nonumber\\
&&\qquad
   \Bornev{m} \left\{\wckkwl{m}
        - \termX{\wckkwl{m}}{0}
        - \termX{\wckkwl{m}}{1}
    \right\}
 + \Bornev{m} + \Virtev{m} + \Insev{m+1|m}\nonumber\\
&&\qquad
 + \int \drad \left( \Bornev{m+1|m} - \Dipev{m+1|m} \right)
  -\int_s \Bornev{m+1\rightarrow m} \quad\Bigg\}\nonumber\\
&+&
  \sum_{n=M+1}^N
  \int d\phi_0 \int\!\cdots\!\int 
  \mathcal{O}(\state{nj})~
  \Bornev{n} \wckkwl{n}
\end{eqnarray}
Let us briefly investigate how this method changes the inclusive cross
section in the special case of merging zero- and one-jet NLO calculations.
At $\Oasof{1}{\mur}$, the cross section is given by the full NLO 
result by construction, while at $\Oasof{2}{\mur}$, we find
\begin{eqnarray}
\termX{\langle \mathcal{O} \rangle}{2}
 &=& \int d\phi_0
     \mathcal{O}(\state{0j}) ~
     \Bornev{0} \termX{\wckkwl{0}}{2} \nonumber\\
 &&+ \int d\phi_0 \int
     \mathcal{O}(\state{1j})
     ~\Bigg\{~
      \Virtev{1} + \Insev{1|0}
 +   \int \drad \left( \Bornev{1|0} - \Dipev{1|0} \right)
 -   \int_s \Bornev{2\rightarrow 1} \quad\Bigg\}\nonumber\\
 &&+
  \int d\phi_0 \int\!\!\int 
  \mathcal{O}(\state{2j})~
  \Bornev{2}
\end{eqnarray}
It is useful to compare this to the result of CKKW-L
\begin{eqnarray}
\termX{\langle \mathcal{O} \rangle}{2}
 &=& \int d\phi_0
     \mathcal{O}(\state{0j}) ~
     \Bornev{0} \termX{\wckkwl{0}}{2} \nonumber\\
 &&+ \int d\phi_0 \int
     \mathcal{O}(\state{1j})~
     \Bornev{1} \termX{\wckkwl{1}}{1}\nonumber\\
 &&+
  \int d\phi_0 \int\!\!\int 
  \mathcal{O}(\state{2j})~
  \Bornev{2}~.
\end{eqnarray}
For very low merging scales, logarithmic contributions of a 
single jet in $\Bornev{2}$ approaching the soft/collinear limit should be 
cancelled to better accuracy in NL$^3$, since the one-jet NLO results should 
contain the complete logarithmic structure. However, since the zero-jet 
description is given by the CKKW-L result, enhancements for the one-jet NLO 
contributions approaching the phase space boundary are not fully cancelled. 
For $\W-$boson production, this means that NL$^3$ does not fully 
compensate contributions of 
$\mathcal{O}\left(\as^2\ln^2\left\{\frac{\muf}{\ordms}\right\} \right)$. 
Enhancements due to both jets in $\Bornev{2}$ stretching into the infrared
are unchecked in CKKW-L, but should cancel some of the one-jet NLO terms in 
NL$^3$.

We find it difficult to assess if CKKW-L or NL$^3$ is more problematic. The 
merging scale value is chosen to separate the parton shower phase space from 
the hard matrix element region. Seeing that multiparton interactions at the 
LHC certainly play a role already at scales of 
$\mathcal{O}(10~\textnormal{GeV})$, it is common practise to set the 
merging scale to a slightly higher value. In the particular case of 
$\W-$boson production, with a merging scale of $\ordms \gtrsim 10$ GeV, 
double logarithms are considerably smaller than the next higher order in 
$\as$, so that it is difficult to isolate the questionable terms. 
We think these important issues nevertheless, and address them, in the 
context of the UNLOPS method, in section \ref{sec:unlops-math}.\\

\noindent
For completeness, we add the NL$^3$ result for exclusive NLO input, which 
simply does not contain phase space subtraction samples:
\begin{eqnarray}
\langle \mathcal{O} \rangle
 &=& \sum_{m=0}^M \int d\phi_0 \int\!\cdots\!\int 
     \mathcal{O}(\state{mj})
     ~\Bigg\{ \tree{m}^\prime + \virt{m} ~\Bigg\}
~+~
  \sum_{n=M+1}^N
  \int d\phi_0 \int\!\cdots\!\int 
  \mathcal{O}(\state{nj})
  \tree{n}\nonumber\\
 &=& \sum_{m=0}^M \int d\phi_0 \int\!\cdots\!\int 
     \mathcal{O}(\state{mj})
     ~\Bigg\{~
   \Bornev{m} \left\{\wckkwl{m}
        - \termX{\wckkwl{m}}{0}
        - \termX{\wckkwl{m}}{1}
    \right\}
 + \Bornev{m} + \Virtev{m} + \Insev{m+1|m}\nonumber\\
&&\qquad
 + \int \drad \left( \Bornev{m+1|m} - \Dipev{m+1|m} \right)
 ~\Bigg\}\nonumber\\
&+&
  \sum_{n=M+1}^N
  \int d\phi_0 \int\!\cdots\!\int 
  \mathcal{O}(\state{nj})~
  \Bornev{n} \wckkwl{n}
\end{eqnarray}
This ends the derivation and discussion of the NL$^3$ method. The main 
conclusion of this section is that the allowed weights for samples in 
NLO merging are restricted by merging conditions. The constraints
apply to other CKKW-L inspired NLO merging schemes as well. If, for example, 
NLO accuracy has to be safeguarded, it is mandatory to remove the 
$\Oasof{0}{\mur}$- and $\Oasof{1}{\mur}$-parts of the weight of tree-level 
events.


\section{Derivation of UNLOPS}

\label{sec:unlops-derivation}

In this part, we aim to give a step-by-step derivation of the UNLOPS method.
We will use the notation defined in section \ref{sec:nlo-merging-intro}, and 
start with the UMEPS prediction for incorporating up to three additional jets:
$\mathcal{O}$ as
\begin{eqnarray}
\langle \mathcal{O} \rangle
 &=& \int d\phi_0\Bigg\{ \mathcal{O}({\state{0j}})\left( \Tev{0}
 ~-~ \Iev{1}{0}
 ~-~ \Iev{2}{0}
 ~-~ \Iev{3}{0}~\right)\nonumber\\
&&\qquad\quad
 + \int \mathcal{O}({\state{1j}}) \left(
     \Tev{1}
 ~-~ \Iev{2}{1}
 ~-~ \Iev{3}{1}~\right)\nonumber\\
&&\qquad\quad
 + \int\!\!\!\int \mathcal{O}({\state{2j}}) \left(
     \Tev{2}
 ~-~ \Iev{3}{2}~\right)
\nonumber\\
&&\qquad\quad
 + \int\!\!\int\!\!\int \mathcal{O}({\state{3j}})~\Tev{3} ~\Bigg\}
\end{eqnarray}
Our method will be to identify which prediction for an exclusive $n$-jet
observable we want, replace the UMEPS approximation by these terms, 
find the difference between the improved result and the UMEPS prediction,
and remove this difference from the next-lower jet multiplicity.

As a warm-up exercise, let us include a NLO calculation for zero-jet 
observables. We then want zero-jet observables to be described by the sum of
Born, virtual and unresolved real terms, and also include the PS resummation.
Keeping in mind that we do not want to introduce approximate $\Oas{0}$- or 
$\Oas{1}$-terms\footnote{We use the intuitive result in 
\eqref{eq:nl3-simple-weights-tree}: If we
want to include the $n$-jet NLO result, we have to subtract the $\Oas{0}$- and 
$\Oas{1}$-terms of the weight for $n$-jet tree-level events to ensure 
NLO accuracy.},
the zero-jet exclusive cross section should read
\begin{eqnarray}
\Btilev{0} + \termX{\Tev{0}}{-0,1}
 ~-~ \termX{\Iev{1}{0}}{-1}
 ~-~ \Iev{2}{0}
 ~-~ \Iev{3}{0}\nonumber\\
=
\Btilev{0}
 ~-~ \termX{\Iev{1}{0}}{-1}
 ~-~ \Iev{2}{0}
 ~-~ \Iev{3}{0}~.
\end{eqnarray}
We would thus need to remove the terms 
\begin{eqnarray}
\int \Btilev{0} + \int\Tev{0}
\end{eqnarray}
from the next-lower multiplicity. Since there is no next-lower multiplicity, 
we simply get
\begin{eqnarray}
\langle \mathcal{O} \rangle
 &=& \int d\phi_0\Bigg\{ \mathcal{O}({\state{0j}})\left( \Btilev{0}
 ~-~ \termX{\Iev{1}{0}}{-1}
 ~-~ \Iev{2}{0}
 ~-~ \Iev{3}{0}~\right)\nonumber\\
&&\qquad\quad
 + \int \mathcal{O}({\state{1j}}) \left(
     \Tev{1}
 ~-~ \Iev{2}{1}
 ~-~ \Iev{3}{1}~\right)\nonumber\\
&&\qquad\quad
 + \int\!\!\int \mathcal{O}({\state{2j}}) \left(
     \Tev{2}
 ~-~ \Iev{3}{2}~\right)
\nonumber\\
&&\qquad\quad
 + \int\!\!\int\!\!\int \mathcal{O}({\state{3j}})~\Tev{3} ~\Bigg\}
\end{eqnarray}
or, in terms of the inclusive NLO cross section
\begin{eqnarray}
\label{eq:unlops-0}
\langle \mathcal{O} \rangle
 &=& \int d\phi_0\Bigg\{ \mathcal{O}({\state{0j}})\left( \Bbarev{0}
 ~-~ \Iev{1}{0}
 ~-~ \Iev{2}{0}
 ~-~ \Iev{3}{0}~\right)\nonumber\\
&&\qquad\quad
 + \int \mathcal{O}({\state{1j}}) \left(
     \Tev{1}
 ~-~ \Iev{2}{1}
 ~-~ \Iev{3}{1}~\right)\nonumber\\
&&\qquad\quad
 + \int\!\!\int \mathcal{O}({\state{2j}}) \left(
     \Tev{2}
 ~-~ \Iev{3}{2}~\right)
\nonumber\\
&&\qquad\quad
 + \int\!\!\int\!\!\int \mathcal{O}({\state{3j}})~\Tev{3} ~\Bigg\}
\end{eqnarray}
Thus, we can promote UMEPS to a tree-level multi-jet merged, 
lowest-multiplicity NLO corrected calculation by simply replacing the 
zero-jet Born cross section with the inclusive zero-jet NLO cross section. 
Such a scheme is often called MENLOPS \cite{Hamilton:2010wh,Hoche:2010kg,Alioli:2011nr}.
Note that the the total cross section is conserved, because we started
from the UMEPS prediction, rescaled with a factor 
$K = \int \Bbarev{0}~ / \int \Bornev{0}$.

The derivation of a multi-jet merging scheme with NLO accuracy for
any $n-$jet observable is organised as follows. First, we will extend the 
MENLOPS result \eqref{eq:unlops-0} to simultaneously include a one-jet NLO
calculation. Then, we add the two jet NLO result. After a short discussion, 
we present the general case.

To including one-jet NLO predictions, we need to replace the UMEPS one-jet
result by
\begin{eqnarray}
\Btilev{1} + \termX{\Tev{1}}{-1,2}
 ~-~ \termX{\Iev{2}{1}}{-2}
 ~-~ \Iev{3}{1}
\end{eqnarray}
Further we have to subtract the difference of the new one-jet prediction and
the UMEPS case from the zero-jet part. The difference is given by
\begin{eqnarray}
 -~ \left( -~ \Btilev{1} 
 ~+~ \termX{\Tev{1}}{1,2}
 ~-~ \termX{\Iev{2}{1}}{2}~
 \right)~,
\end{eqnarray}
so that the zero-jet contribution becomes
\begin{eqnarray}
\langle \mathcal{O} \rangle_0
 &=& \int d\phi_0 \mathcal{O}({\state{0j}}) \Bigg\{ \Btilev{0}
 ~-~ \termX{\Iev{1}{0}}{-1}
 ~-~ \Iev{2}{0}
 ~-~ \Iev{3}{0}\nonumber\\
&&
 ~-~ \int_s \Btilev{1\rightarrow 0}
 ~+~ \termX{\Iev{1}{0}}{1,2}
 ~-~ \termX{\Iev{2}{0}^{\uparrow}}{2} \Bigg\}\nonumber\\
 &=& \int d\phi_0 \mathcal{O}({\state{0j}}) \Bigg\{ \Btilev{0}
 ~-~ \int_s \Btilev{1\rightarrow 0} 
 ~+~ \int_s \Bornev{1 \rightarrow 0}
 ~-~ \termX{\Iev{1}{0}}{-1,2}\nonumber\\
&&
 ~-~ \int_s \Bornev{2 \rightarrow 0}^{\uparrow}
 ~-~ \Iev{2}{0}
 ~-~ \Iev{3}{0} \Bigg\}
\end{eqnarray}
Note that the integrated two-jet contribution 
\begin{eqnarray}
\termX{\Iev{2}{0}^{\uparrow}}{2} = \int_s \Bornev{2 \rightarrow 0}^{\uparrow}
\end{eqnarray}
is integrated twice, even though the result of the first integration 
($\state{1}$) contains only resolved jets. Putting the pieces together, we 
arrive at
\begin{eqnarray}
\label{eq:unlops-01-exclusive}
\langle \mathcal{O} \rangle
 &=& \int d\phi_0 \Bigg\{ \mathcal{O}({\state{0j}}) \left( \Btilev{0}
 ~-~ \int_s \Btilev{1\rightarrow 0} 
 ~+~ \int_s \Bornev{1 \rightarrow 0}
 ~-~ \termX{\Iev{1}{0}}{-1,2}\right.\nonumber\\
&& \qquad\qquad\qquad\left.
 ~-~ \int_s \Bornev{2 \rightarrow 0}^{\uparrow}
 ~-~ \Iev{2}{0}
 ~-~ \Iev{3}{0} ~\right)\nonumber\\
&&\qquad\quad
 + \int \mathcal{O}({\state{1j}}) \left(
\Btilev{1} + \termX{\Tev{1}}{-1,2}
 ~-~ \termX{\Iev{2}{1}}{-2}
 ~-~ \Iev{3}{1}~\right)\nonumber\\
&&\qquad\quad
 + \int\!\!\int \mathcal{O}({\state{2j}}) \left(
     \Tev{2}
 ~-~ \Iev{3}{2}~\right)
\nonumber\\
&&\qquad\quad
 + \int\!\!\int\!\!\int \mathcal{O}({\state{3j}})~\Tev{3} ~\Bigg\}
\end{eqnarray}
Let us rewrite this in terms of inclusive NLO calculations:
\begin{eqnarray}
\label{eq:unlops-01}
\langle \mathcal{O} \rangle
 &=& \int d\phi_0 \Bigg\{ \mathcal{O}({\state{0j}}) \left( \Bbarev{0}
 ~-~ \int_s \Bbarev{1\rightarrow 0} 
 ~-~ \termX{\Iev{1}{0}}{-1,2}
 -~ \Iev{2}{0}
 ~-~ \Iev{3}{0} ~\right)\nonumber\\
&&\qquad\quad
 + \int \mathcal{O}({\state{1j}}) \left(
\Bbarev{1} + \termX{\Tev{1}}{-1,2}
 -~ \Iev{2}{1}
 ~-~ \Iev{3}{1}~\right)\nonumber\\
&&\qquad\quad
 + \int\!\!\int \mathcal{O}({\state{2j}}) \left(
     \Tev{2}
 ~-~ \Iev{3}{2}~\right)
\nonumber\\
&&\qquad\quad
 + \int\!\!\int\!\!\int \mathcal{O}({\state{3j}})~\Tev{3} ~\Bigg\}
\end{eqnarray}
This result will be used in section \ref{sec:unlops-math}, and is discussed 
there.

Before presenting a general master formula, we will take yet another 
intermediate step, and generalise \eqref{eq:unlops-01} to further include 
two-jet NLO predictions. Two-jet observables should be described by
\begin{eqnarray}
\Btilev{2} + \termX{\Tev{2}}{-2,3}
 ~-~ \termX{\Iev{3}{2}}{-3}~.
\end{eqnarray}
To conserve unitarity, we then have to subtract the difference of the new 
result and the UMEPS contributions, \ie\
\begin{eqnarray}
 -~ \left( -~ \Btilev{2} + \termX{\Tev{2}}{2,3}
 ~-~ \int_s \Bornev{3\rightarrow2}\right)~,
\end{eqnarray}
from the one-jet case. This gives the new one-jet contributions
\begin{eqnarray}
   \Btilev{1}
 - \int_s \Btilev{2\rightarrow1}
 + \termX{\Tev{1}}{-12}
 - \termX{\Iev{2}{1}}{-2,3}
 + \int_s \Bornev{2\rightarrow1}
 - \int_s \Bornev{3\rightarrow1}^{\uparrow}
 - \Iev{3}{1}~.
\end{eqnarray}
In the case that $\Btilev{2\rightarrow1}$, $\int_s \Bornev{3\rightarrow1}^{\uparrow}$
or $\termX{\Iev{2}{1}}{2,3}$ does not result in a state with one jet above the 
merging scale, we choose to integrate twice, as in the case of UMEPS. 
Thus, the UNLOPS prediction for simultaneously merging zero, 
one and two jets at next-to-leading order is given by
\begin{eqnarray}
\label{eq:unlops-012}
\langle \mathcal{O} \rangle
 &=& \int d\phi_0 \Bigg\{ \mathcal{O}({\state{0j}}) \left( \Btilev{0}
 ~-~ \int_s \Btilev{1\rightarrow 0} 
 ~-~ \int_s \Btilev{2\rightarrow 0}
 ~+~ \int_s \Bornev{1 \rightarrow 0}
 ~-~ \termX{\Iev{1}{0}}{-1,2}\right.\nonumber\\
&& \qquad\qquad\qquad\left.
 ~-~ \int_s \Bornev{2 \rightarrow 0}^{\uparrow}
 ~-~ \int_s \Bornev{3 \rightarrow 0}^{\uparrow}
 ~-~ \termX{\Iev{2}{0}}{-2,3}
 ~-~ \Iev{3}{0} ~\right)\nonumber\\
&&\qquad\quad
 + \int \mathcal{O}({\state{1j}}) \left(
   \Btilev{1}
 - \int_s \Btilev{2\rightarrow1}
 + \termX{\Tev{1}}{-1,2}
 - \termX{\Iev{2}{1}}{-2,3}\right.\nonumber\\
&& \qquad\qquad\qquad\quad\left.
 + \int_s \Bornev{2\rightarrow1}
 - \int_s \Bornev{3\rightarrow1}^{\uparrow}
 - \Iev{3}{1}
~\right)\nonumber\\
&&\qquad\quad
 + \int\!\!\int \mathcal{O}({\state{2j}}) \left(
    \Btilev{2}
 + \termX{\Tev{2}}{-2,3}
 -~ \termX{\Iev{3}{2}}{-3}
~\right)
\nonumber\\
&&\qquad\quad
 + \int\!\!\int\!\!\int \mathcal{O}({\state{3j}})~\Tev{3} ~\Bigg\}
\end{eqnarray}
Again expressing this in terms of NLO inclusive cross sections, we find
\begin{eqnarray}
\langle \mathcal{O} \rangle
 &=& \int d\phi_0 \Bigg\{ \mathcal{O}({\state{0j}}) \left( \Bbarev{0}
 ~-~ \int_s \Bbarev{1\rightarrow 0} 
 ~-~ \int_s \Bbarev{2\rightarrow 0} 
 ~-~ \termX{\Iev{1}{0}}{-1,2}\right.\nonumber\\
&& \qquad\qquad\qquad\left.
 ~-~ \termX{\Iev{2}{0}}{-2,3} 
 ~-~ \Iev{3}{0} ~\right)\nonumber\\
&&\qquad\quad
 + \int \mathcal{O}({\state{1j}}) \left(
   \Bbarev{1}
 - \int_s \Bbarev{2\rightarrow1}
 + \termX{\Tev{1}}{-1,2}
 - \termX{\Iev{2}{1}}{-2,3}
 - \Iev{3}{1}
~\right)\nonumber\\
&&\qquad\quad
 + \int\!\!\int \mathcal{O}({\state{2j}}) \left(
    \Bbarev{2}
 + \termX{\Tev{2}}{-23}
 -~ \Iev{3}{2}
~\right)
\nonumber\\
&&\qquad\quad
 + \int\!\!\int\!\!\int \mathcal{O}({\state{3j}})~\Tev{3} ~\Bigg\}
\end{eqnarray}
Let us have a closer look at \eqref{eq:unlops-012}, and in particular how the
$\Oasof{3}{\mur}-$term of the one-jet descriptions:
\begin{eqnarray}
\label{eq:unlops-012-1j-oas3}
\termX{\langle \mathcal{O} \rangle_1}{3}
 =
   \termX{ \Tev{1}}{3}
 - \termX{ \int_s \Btilev{2\rightarrow1}}{3}
 - \termX{ \int_s \Bornev{3\rightarrow1}^{\uparrow}}{3}
 - \termX{ \Iev{3}{1} }{3}
\end{eqnarray}
The first term is the parton shower approximation of unresolved emissions in
the underlying zero-jet configurations. The second and third terms give an
approximation of unresolved contributions in one-jet states. Compared to the
UMEPS result $\left( \int_s \Bornev{2\rightarrow1}\termX{\wumeps{2}}{1}
\right)$, this should give an improved description. The last term in 
\eqref{eq:unlops-012-1j-oas3} is unchanged compared to UMEPS, and should not 
induce logarithmic terms in $\ordms$.

Coming back to the first term in \eqref{eq:unlops-012-1j-oas3}, it is natural
to ask if reweighting the $\Oas{2}-$part of $\Btilev{1}$ would not give a 
better description. Since reweighting exclusive NLO events (which contain both
$\Oas{n}$ and $\Oas{n+1}$ parts) will mix higher-order terms in a difficult
way, this interesting possibility is not examined here. We hope to come back 
to this issue for comparisons between different NLO merging 
prescriptions\footnote{We in particular think about comparisons with 
MEPS@NLO, were $\Btilev{}-$events are reweighted.}.

We hope it is clear that the UNLOPS method preserves NLO accuracy by 
construction, and improves the higher-order description of UMEPS further. 
The formal accuracy of exclusive $n-$jet 
observables, however, is not better than next-to-leading order combined with 
PS resummation. Only for a limited set of observables do parton showers 
capture more than leading logarithmic enhancements. 
We have presented UNLOPS both for exclusive and inclusive NLO cross
sections. This is motivated by trying to accommodate NLO calculations while 
requiring only minor -- or ideally no -- changes to the actual NLO 
implementation. 

After explicitly deriving the UNLOPS scheme for describing up to two jet 
observables next-to-leading order accuracy, we now give the UNLOPS master 
formula, when using exclusive NLO samples
\begin{eqnarray}
\label{eq:unlops-full-exclusive-appendix}
\langle \mathcal{O} \rangle
 &=&
 \sum_{m=0}^{M-1}~ \int d\phi_0 \int\!\cdots\!\int 
     \mathcal{O}(\state{mj})
     ~\Bigg\{~
   \Btilev{m}
 + \termX{\Tev{m}}{-m,m+1}
 + \int_s\Bornev{m+1\rightarrow m}
\nonumber\\
&&\qquad\qquad\qquad~
 - \sum_{i=m+1}^{M} \int_s\Btilev{i\rightarrow m}
 - \sum_{i=m+1}^{M} \termX{\Iev{i}{m}}{-i,i+1}
 - \sum_{i=m+1}^{M} \int_s\Bornev{i+1\rightarrow m}^{\uparrow}
\nonumber\\
&&\qquad\qquad\qquad~
 - \sum_{i=M+1}^{N} \Iev{i}{m}
~\Bigg\}\nonumber\\
&+&
   \int d\phi_0 \int\!\cdots\!\int 
   \mathcal{O}(\state{Mj})\Bigg\{~
   \Btilev{M}
 + \termX{\Tev{M}}{-M,M+1}
 - \termX{\Iev{M+1}{M}}{-M}\nonumber\\
&&\qquad\qquad\qquad~
 - \sum_{i=M+1}^{N} \Iev{i+1}{M}~
   \Bigg\}\nonumber\\
&+&
  \sum_{n=M+1}^N
  \int d\phi_0 \int\!\cdots\!\int 
  \mathcal{O}(\state{nj})~\left\{ \Tev{n} - \sum_{i=n+1}^{N} \Iev{i}{n} ~\right\}
\end{eqnarray}
Furthermore, since most results in this publication are produced using 
inclusive NLO samples, we also give the UNLOPS prediction for inclusive input
\begin{eqnarray}
\label{eq:unlops-full-inclusive-appendix}
\langle \mathcal{O} \rangle
 &=&
 \sum_{m=0}^{M-1}~ \int d\phi_0 \int\!\cdots\!\int 
     \mathcal{O}(\state{mj})
     ~\Bigg\{
   \Bbarev{m}
 + \termX{\Tev{m}}{-m,m+1}
\nonumber\\
&&\qquad\qquad\qquad\quad~
 - \sum_{i=m+1}^{M} \int_s\Bbarev{i\rightarrow m}
 - \sum_{i=m+1}^{M} \termX{\Iev{i}{m}}{-i,i+1}
 - \sum_{i=M+1}^{N} \Iev{i}{m}
~\Bigg\}\nonumber\\
&+&
   \int d\phi_0 \int\!\cdots\!\int 
   \mathcal{O}(\state{Mj})\Bigg\{~
   \Bbarev{M}
 + \termX{\Tev{M}}{-M,M+1}
 - \sum_{i=M+1}^{N} \Iev{i}{M}~
   \Bigg\}\nonumber\\
&+&
  \sum_{n=M+1}^N
  \int d\phi_0 \int\!\cdots\!\int 
  \mathcal{O}(\state{nj})~\left\{ \Tev{n} - \sum_{i=n+1}^{N} \Iev{i}{n} ~\right\}
\end{eqnarray}
Although the number of contributions in UNLOPS becomes somewhat unwieldy, 
we still only require $M+N$ input event files, since some files can be reused
-- as in NL$^3$. The procedure (including processing some input events 
multiple times) has been implemented in \pytppp, and will become available in the 
near future.


\subsection{Upgrading one-jet UNLOPS to a NNLO matching scheme}
\label{sec:unlops-nnlo}

The UNLOPS scheme has the advantage that the lowest-multiplicity cross section
is not reweighted. This makes replacements of this term with more accurate 
calculations relatively easy. Here, we would like to hint at how UNLOPS could
be shaped into a NNLO matching scheme.
The starting point is again UMEPS, but instead of multiplying every UMEPS 
contribution with a NLO \Kf-factor, we rescale with a NNLO \Kf-factor 
$K^\prime$. Apart from this change, we directly move to the UNLOPS 
prescription including zero- and one-jet NLO calculations. 
Then, we assume that an exclusive zero-jet NNLO 
calculation is available, producing phase space points with the weight 
$\widetilde{\widetilde{\textnormal{B}}}_0$. The weight 
$\widetilde{\widetilde{\textnormal{B}}}_0$ should be the sum of the Born 
approximation, one-loop corrections, unresolved single real corrections, 
two-loop corrections, one-loop corrections with an additional unresolved jet,
and double unresolved double real radiation contributions.

We replace $\Btilev{0}$ in
\eqref{eq:unlops-01-main-text} by $\widetilde{\widetilde{\textnormal{B}}}_0$,
and remove all other $\Oasof{2}{\mur}-$terms in the zero-jet part:
\begin{eqnarray}
\label{eq:unlops-01--exclusive}
\langle \mathcal{O} \rangle
 &=& \int d\phi_0 \Bigg\{ \mathcal{O}({\state{0j}}) \left(
     \widetilde{\widetilde{\textnormal{B}}}_0
 ~-~ \termX{\Iev{1}{0}}{-1,2}
 ~-~ \termX{\Iev{2}{0}}{-2} ~\right)\nonumber\\
&&\qquad\quad
 + \int \mathcal{O}({\state{1j}}) \left(
\Btilev{1} + \termX{\Tev{1}}{-1,2}
 ~-~ \termX{\Iev{2}{1}}{-2} ~\right)\nonumber\\
&&\qquad\quad
 + \int\!\!\int \mathcal{O}({\state{2j}}) ~
     \Tev{2} ~\Bigg\}
\end{eqnarray}
We see that the inclusive cross section is given by
\begin{eqnarray}
\int d\phi_0 \mathcal{O}({\state{0j}}) 
     \widetilde{\widetilde{\textnormal{B}}}_0
 ~+~ \int d\phi_0\int \mathcal{O}({\state{1j}})~ \Btilev{1}
\end{eqnarray}
Zero-jet observables are correct to $\Oasof{2}{\mur}$, as is the
description of one- and two-jet observables. It is of course possible
to improve higher orders by including additional matrix elements in
UMEPS-fashion.  The major obstacle for implementing this method is the
lack of available ME generators generating phase space points
according to $\widetilde{\widetilde{\textnormal{B}}}_0$.


\section{NLO merging and multiparton interactions}
\label{sec:mpi-in-nlo}

Observable jets at hadron colliders are not only produced in a single 
energetic interactions, but also emerge from additional scatterings
of other proton constituents.
Multiparton interactions (MPI) models are essential in describing 
hadron collider data which include these ``underlying events'' \cite{Akesson:1986iv,Abe:1997bp,Aad:2010fh}. 
When trying to describe the underlying event at the LHC, one is in practise 
still largely forced to use phenomenological models, although efforts are 
under way to construct more solid theoretical foundations (see
\cite{Bartalini:2011jp} for a recent review). Due to continuous development, 
it is valid to say that current phenomenological models offer a good 
description of a wide range of experimental data.

A sophisticated MPI machinery has always been a cornerstone of \pytppp \cite{Sjostrand:2006za,Corke:2009pm,Corke:2009tk,Corke:2011yy}. 
Multiparton interactions in \pytppp are modelled by including QCD 
$2\rightarrow2$ scatterings in addition to the hard process. It is reasonable
to assume that energetic secondary scatterings induce constraints on how
much beam energy would be left for further initial state radiation. \pytppp
incorporates such phase space constraints by interleaving MPI with parton 
showering: An energetic secondary scattering is produced \emph{before} soft
radiation. This is achieved by generating initial state radiation, final 
state radiation, and MPI in one decreasing sequence of evolution scales. One 
benefit of this method is that for a high shower starting scale, more 
jet-like MPI are produced, which increases the underlying event activity for 
increasing hardness of the core scattering -- a phenomenon called pedestal 
effect. 

From the point of precision QCD, we have to recognise that for observables 
which are influenced by multiple interactions, the formal accuracy of any 
merging method will be governed by MPI. Only if the influence of MPI is 
negligible are statements about the formal accuracy of the result reasonable.
However, suppressing hard multiparton interactions leads to an inferior data 
description.
Following the philosophy of \cite{Lonnblad:2011xx} and \cite{Lonnblad:2012ng}, we will sacrifice
the formal accuracy label of the NLO merging method in regions where MPI are 
important. This does not mean that we undo any improvements of our method, 
but only that we can no longer claim a particular accuracy, even in the 
presence of improvements.

We include MPI in NL$^3$ and UNLOPS in the same way as was previously
done for UMEPS in \cite{Lonnblad:2012ng} and we refer to that publication and
\cite{Lonnblad:2011xx} for more background. First we amend the normal
PS no-emission probabilities with no-MPI factors,
$\noem{n}\sup{MPI}$. This means that all event samples with $n$
partons ($\untree{n}$, $\virt{n}$, $\unsubt{n}$, and
$\unvirt{n}$)
are multiplied with the no-MPI probabilities
\begin{equation}
  \label{eq:no-mpi}
  \prod_{i=0}^{m-1}\noem{i}\sup{MPI}(\ord_i,\ord_{i+1}), 
\end{equation}
which are easily incorporated in the trial showers. Note, however, that
MPI emissions are not taken into account when calculating the
$\Oas{1}-$term of no-emission probabilities (which is natural since
MPI's are of $\Oas{2}$). Then, when the shower is started from the
reweighted (and possibly reclustered) states, using $\ord_n$ as
starting scale, MPI's are included. As before, for $n<N$, any parton
emission above \ordms\ are vetoed, but if a MPI is generated, it is
always accepted, and no emissions are vetoed in the subsequent
showering.

As in the UMEPS and CKKW-L methods, this means that any event where
the $n\le N$ hardest jets are bove the merging scale and are from the
primary interaction a, these will be described by the corresponding
tree-level ME. In addition if $n\le M$ these jets will be described by
the corresponding NLO ME. In both cases, the higher order \as-terms
will be resummed to the precision of the shower. Again we note that
the NLO-prediction will be modified by the inclusion of MPI. The
modification is beyond the ``leading twist'' approximation of the NLO
calculation, but may nevertheless be large, especially for jets with
low transverse momenta.

Before concluding, we would also point out that in this article, we
have used {\tt{CTEQ6M}} PDFs in the generation of secondary
scatterings. This is not advisable, since the dominant contribution to
the underlying event stem from soft secondary scatterings. For such
scatterings, PDFs are evaluated at low scales $\mathcal{O}\left(
  1~\textnormal{GeV}\right)$, and very small $x-$values, \ie\ in a
region where NLO PDFs are poorly constrained and need not even be
positive definite, which clearly is problematic in the
probabilistic MPI picture.  When developing a future tune to be used
together with NLO merged predictions, we will utilise NLO parton
distributions for the hard interaction, while
employing leading-order PDFs in the multiple interactions and parton showers.

\end{appendix}

\bibliographystyle{utcaps}  
\bibliography{references} 

\end{document}